\documentclass[10pt]{article}
\usepackage[left=3cm,right=3cm,top=3cm,bottom=3cm]{geometry}
\usepackage[bookmarks, bookmarksnumbered]{hyperref}
\usepackage{graphicx}
\usepackage{amsmath,amsfonts,amssymb}
\numberwithin{equation}{section}
\numberwithin{figure}{section}

\renewcommand{\k}{\kappa}
\newcommand{\w}{\omega}
\newcommand{\W}{\Omega}

\newcommand{\g}{\gamma}

\renewcommand{\r}{\rho}
\newcommand{\s}{\sigma}
\newcommand{\sbar}{\bar{\sigma}}
\renewcommand{\th}{\theta}

\newcommand{\cK}{{\mathcal K}}
\newcommand{\cM}{{\mathcal M}}
\newcommand{\cN}{{\mathcal N}}
\newcommand{\cNHH}{{\mathcal N_{\rm HH}}}
\newcommand{\NHH}{N_{\rm HH}}
\newcommand{\cD}{{\mathcal D}}
\newcommand{\cZ}{{\mathcal Z}}
\newcommand{\abs}[1] {\left\lvert #1 \right\rvert} 
\newcommand{\norm}[2] {\left\lVert #1 \right\rVert_{#2}} 
\newcommand{\half}{\frac{1}{2}}
\newcommand{\thalf}{\tfrac{1}{2}}
\DeclareMathOperator{\sech}{sech}
\DeclareMathOperator{\sgn}{sgn}

\newcommand{\RR}{{\mathbb R}}
\newcommand{\ZZ}{{\mathbb Z}}

\newcommand{\innerProd}[2]{\langle #1, #2 \rangle}
\newcommand{\fN}{{\mathfrak N}}
\newcommand{\mfs}{{\mathfrak s}}
\newcommand{\bsl}{{\Big[}}
\newcommand{\bsr}{{\Big]}}
\newcommand{\mywidth}{1.75in}
\DeclareMathOperator{\Or}{O}

\title{Hamiltonian Hopf bifurcations and chaos of NLS/GP standing-wave modes}
\author{Roy Goodman\\ 
Department of Mathematical Sciences\\ 
New Jersey Institute of Technology\\
Newark, NJ 07102\\
goodman@njit.edu}

\date{}
\begin{document}
\maketitle
\begin{abstract}
We examine the dynamics of solutions to nonlinear Schr\"odinger/Gross-Pitaevskii equations that arise due to Hamiltonian Hopf (HH) bifurcations--the collision of pairs of eigenvalues on the imaginary axis.  To this end, we use inverse scattering to construct localized potentials for this model which lead to HH bifurcations in a predictable manner.  We perform a formal reduction from the partial differential equations (PDE) to a small system of ordinary differential equations (ODE).  We show numerically that the behavior of the PDE is well-approximated by that of the ODE and that both display Hamiltonian chaos. We analyze the ODE to derive conditions for the HH bifurcation and use averaging to explain certain features of the dynamics that we observe numerically.
\end{abstract}

\section{Introduction}

In the dynamical systems approach to mathematical physics, an important and physically-motivated approach is to consider the behavior of special solutions: fixed points, periodic orbits, and the like.  In particular, one often wants to know whether a given solution is stable, i.e.\ whether it can be destroyed by introducing a small perturbation to the initial conditions.  Moreover, the study of bifurcations has shown that instabilities can, in general, occur in a finite numbers of ways.  Oscillatory instabilities in Hamiltonian systems arise due to Hamiltonian Hopf (HH) bifurcations and have been seen in a great number of analytical and numerical studies, as outlined later in the introduction.  The aim of this paper is to study in detail the \emph{nonlinear} dynamics that occur in one such system that arises in various applications as a way to get a handle on this phenomenon in general.

The nonlinear Schr\"{o}dinger/Gross-Pitaevskii equation (NLS/GP) 
\begin{equation}
\label{NLS}
i \partial_\zeta \psi = H \psi -|\psi|^2 \psi; \; H= -\partial_x^2 + V(x),
\end{equation}
is important in mathematical physics in (at least) two main contexts.  In nonlinear optics, it arises in the paraxial approximation for light propagating in a thin waveguide constructed in a material with Kerr nonlinearity~\cite{Boyd:2008,Newell:2003}. In a Kerr material, the refractive index of light takes the form $n = n_0 + n_2(\abs{E}^2)$ where $E$ represents the electric field.  In particular, the electric field is given by 
$$
E(x,z,t) = \Re({e^{i(kz-\w t)} \psi(x,\zeta)}).
$$
Here $z$ is the direction of propagation along the waveguide, $x$ the direction transverse, and $t$ is time. The waveguide is assumed to be thin in the $y$ direction, and the variation in this direction can safely be ignored. The potential $V(x)$ represents the contribution due to the geometry of the waveguide, and which we assume to be smooth, negative and exponentially localized. The variable $\zeta$ represents a scaled version of the propagation distance $z$.  An effective equation for the envelope $\psi(x,\zeta)$ can be derived by the method of multiple scales, and we assume that the independent and dependent variables in this equation can be rescaled to obtain the simple form of~\eqref{NLS} dependent on as few parameters as possible.  \emph{Despite the physical meaning of the variable $\zeta$, for the remainder of the paper, we shall call this variable $t$ to remind us that it is the independent variable of evolution.}

When the sign on the nonlinear term of~\eqref{NLS} is reversed (and $\zeta$ now genuinely represents time), the equation describes the state of a Bose-Einstein condensate, a state of matter achievable at extreme low temperatures where atoms lose their individual identities and are described by a common wavefunction~\cite{Pitaevskii:2003}.  For equation~\eqref{NLS} to hold, the three-dimensional condensate must be strongly confined by a steep potential in the two transverse directions $y$ and $z$ so that it assumes a ``cigar'' shape.  The term $V(x)$ then represents a less steeply confining potential in the third spatial dimension.

In both these systems, a fundamental object of study is the nonlinear bound state, i.e.\ a localized solution to~\eqref{NLS} of the form 
$$
\psi(x,t) = e^{-i\W t}\Psi(x).
$$ 
A solution consists of $\Psi(x)$, a sufficiently rapidly decaying real-valued function, and two real numbers $\W$ and $\cN$ that satisfy
\begin{equation}\begin{split}
\label{stationary_NLS}
\W \Psi & = H \Psi -\Psi^3;\\
\int_{-\infty}^{\infty} \Psi^2(x) dx &= ||\Psi||_2^2 = \cN
\end{split}\end{equation}
The parameter $\cN>0$, the square of the $L^2$-norm, represents the number of particles of a BEC or the total intensity of the light in optics.  This solution may be thought of as a nonlinear generalization of an eigenfunction of a linear Schr\"odinger equation, although, of course, the principle of superposition fails to apply in this instance.  We expect, and find, continuous families of solutions that are indexed by the intensity $\cN$. In fact, as $\cN\to 0$, some of these solutions approach, in shape, the eigenpairs of the linear system.

Nonlinear bound states, or standing waves, represent coherent and simple states that might be observable in a laboratory experiment.  Such bound states may be found numerically, or, for specially constructed potentials $V(x)$, might be easily computed exactly in the linear limit and approximately for $\cN \neq 0$.  In order for such states to be observable in experiments, they would have to be stable, i.e.\ if  a solution to equation~\eqref{NLS} is initialized at $t=0$ with value close to, but not equal to, a solution to system~\eqref{stationary_NLS}, then it must stay in a neighborhood of that solution for all $t>0$.

Much work, of course, has gone into studying the stability, especially the spectral stability, of solutions, i.e.\  the presence of of unstable modes (corresponding to spectrum with positive real part) in the linearization of~\eqref{NLS} about a given solution.   In particular, we may think of $\cN$ as a bifurcation parameter.  Since it is usually the case that there exist families of solutions to~\eqref{stationary_NLS} continuously parameterized by $\cN$, we may ask for what values of $\cN$ the solution is stable.   

The stability of a standing wave is not, however, the whole story. Bifurcation theory dictates that there is a relatively small set of scenarios (bifurcation types) that may be observed in the transition from stability to instability, and in each of these scenarios certain types of solutions and dynamics may be observed. System~\eqref{NLS} is Hamiltonian, and this fact further restricts the types of behaviors that can be seen near a bifurcation. 

Several recent studies have focused on the types of bifurcations observable in system~\eqref{NLS} and related systems and we review a few of them here, in order to motivate the current study.  In addition to the stability of a solution changing as a parameter is varied, a bifurcation may create new solutions.
Kirr et al., for example,  have demonstrated that solutions to~\eqref{stationary_NLS} with a double-well potential
\begin{equation}
V^{(2)}_L(x)= \tilde{V}(x-L)+ \tilde{V}(x+L)
\label{2well}
\end{equation}
undergo a symmetry-breaking bifurcation as the parameter $\cN$ is raised from zero~\cite{KirKevShl:08}.  At a critical value $\cN_{\rm SB}$, a symmetric solution to equation~\eqref{stationary_NLS} loses stability and two stable, asymmetric standing wave modes are created. Kapitula, Kevrekids, and Chen~\cite{KapKevChe:06} have shown that for a triple well potential of the form
\begin{equation}
V^{(3)}_L(x)= \tilde{V}(x-L)+ \tilde{V}(x) + \tilde{V}(x+L).
\label{3well}
\end{equation}
that these symmetry-breaking bifurcations are replaced by saddle-node bifurcations.  In the symmetry breaking bifurcations, the new families of standing waves ``branch off'' of the existing families exactly at the location of the bifurcation, while for a saddle-node bifurcation, the new families of solutions are not connected to the the existing families at this point.

Also associated with bifurcations are certain features in the dynamics in a neighborhood of the family of solutions.  The symmetry-breaking bifurcation studied by Kirr et al.\ was shown by Marzuola and Weinstein to display the dynamics typical of such systems.  Below the bifurcation, the ODE system has a single-well potential energy,and thus a one-parameter family, of periodic orbits.  Above the bifurcation, the potential energy has a dual-well shape and thus three topologically distinct families of periodic orbits. This manifests itself in a wobbling of the shape of the asymmetric solutions or a periodic exchange of energy between the two wells~\cite{Marzuola:2010}; see also~\cite{Mayteevarunyoo:2008,Pelinovsky:2011}.

One particular type of bifurcation that can give rise to much more complicated dynamics is the HH bifurcation.  While~\cite{KapKevChe:06} concentrates on enumerating all the standing wave states, they also numerically compute the stability of these standing waves, and they do demonstrate a HH bifurcation (figure 6d); see also~\cite{Kapitula:2005}.  The HH bifurcation has also been observed in other NLS-related settings.  Several studies have demonstrated numerically the existence of ``Krein collisions''---defined in section~\ref{sec:HH} below---in discrete wave equations~\cite{Johansson:2004,Kapitula:2001,Morgante:2000,Panda:2005} and in Bose-Einstein condensates (BEC)~\cite{Kevrekidis:2003,Kevrekidis:2005,Li:2005,Nistazakis:2007,Theocharis:2010}.  In these studies, and most others, the bifurcation is discussed only in the context of detecting the instability transition in the linear spectrum, or by performing a small number of numerical solutions to the initial value problem.

The HH bifurcation is often described as an instability resulting from a ``collision of modes'' (i.e.\ frequencies), for example in describing the motion of multiple dark solitons in a quasi-one-dimensional BEC's~\cite[fig. 5c]{Theocharis:2010}, Theocharis et al.\ remark on an instability caused by ``the collision of the second anomalous mode with the quadrupole mode'' in describing dynamics that look remarkably like our figure~\ref{fig:ODEsolutions}b, column 3.  A goal of this paper is to shed light on the origin of such patterns in this and similar numerical simulations.

Kapitula et al.\ have developed rigorous analytical methods for counting the number eigenvalues that might lead to instability in a wide variety of Hamiltonian nonlinear wave equations~\cite{Kapitula:2004,Kapitula:2005a}, and are thus able to rigorously determine the stability of localized solutions of these infinite-dimensional Hamiltonian systems  In that work, they apply this method to investigate the stability of localized solutions to a system of coupled NLS equations.  In~\cite{Kapitula:2007}, they use this machinery to study the stability of rotating matter waves in Bose-Einstein condensates, and demonstrate the presence of HH bifurcations.  They supplement this with well-chosen numerical simulations in order to demonstrate the dynamics that occur when the solution is destabilized.

In related work, Goodman and Weinstein~\cite{Goodman:08} study the linear stability of standing wave modes of the nonlinear coupled mode equations (NLCME). In that paper, several bifurcation scenarios are outlined, including both symmetry-breaking (figure 4.2c) and HH (figure 4.2d).  In extensive numerical studies, they found symmetry-breaking bifurcations, but were unable to locate any HH bifurcations.  Part of the motivation for the construction in the present paper was to engineer potentials where these bifurcations can be observed and understood, first in the simpler and better-known NLS/GP equation.  In forthcoming work parallel to this, we perform similar analysis for NLCME and find largely similar results.

In this paper, we focus the dynamics in the vicinity of a HH bifurcation.  In the following subsection, we summarize the notation used in the paper.  In section~\ref{sec:background}, we discuss the assumptions about the potential under which this bifurcation may be observed and state the main findings of this paper, including a slight reformulation of the problem in section~\ref{sec:alternate}.  In section~\ref{sec:construction}, we sketch the inverse-scattering techniques used to construct the potential, while leaving more of the details to Appendix~\ref{sec:appendix}.  Section~\ref{sec:finiteDim} discusses the elementary properties of the finite-dimensional model. In section~\ref{sec:derivation}, we briefly describe the derivation of a finite-dimensional model~\ref{3modes} for the of the dynamics of equation~\ref{NLS} and in section~\ref{sec:reduction} a further reduction~\ref{complexform} of the dimension based symmetries of the system. Section~\ref{sec:stationary} reviews the known stationary solutions of system~\ref{3modes}.  In section~\ref{sec:stability}, we derive a formula to detect the HH bifurcation.  Section~\ref{sec:numerics} contains numerical confirmation of this formula and numerical explorations of the dynamics of both the PDE~\ref{NLS} and the finite-dimensional model~\eqref{complexform}.  We discuss a further symmetry reduction of the system~\eqref{complexform} in section~\ref{sec:further}, which allow a fuller understanding of the dynamics, and finally conclude in section~\ref{sec:conclusions}. Appendix~\ref{sec:projection_appendix} contains some formulas related to the derivation in section~\ref{sec:finiteDim}.

\subsection{Notation}
\begin{itemize}
\item An overbar, $\bar z$ represents the complex conjugate of $z$.  
\item The expressions $\Re z$ and $\Im z$ represent, respectively, the real and imaginary parts of $z$.
\item We denote the $L^2$ inner product over complex-valued $L^2$ functions of a real argument by $\innerProd{f}{g} = \int_\RR f(x) \bar{g}(x) dx$. 
\end{itemize}

\section{Technical Background}
\label{sec:background}

\subsection{Discrete spectrum of the operator $H$.} 
If $V(x)$ has even symmetry, $V(-x)=-V(x)$, then solutions to the (linear) eigenvalue equation 
\begin{equation}
\W \Psi = H \Psi,
\label{linearEig}
\end{equation}
that is, the $\cN \to 0$ limit of equation~\eqref{stationary_NLS}, will 
will have either odd or even symmetry.  If the NLS system~\eqref{linearEig} possesses two discrete eigenvalues $\W_1 < \W_2 < 0$, then standard Sturm-Liouville theory requires that the associated eigenfunctions $\Psi_1$ and $\Psi_2$ are, respectively, even and odd functions of $x$.  $\Psi_1$ is the minimizer of the associated Hamiltonian and is thus referred to as the ground state.  The mode $\Psi_2$ is referred to as the excited state.  The spectrum of $H$ will, independently of its symmetry, generically consist of a finite number of real discrete eigenvalues $\W_k<0$ and continuous spectrum on the non-negative real axis.

These standing wave modes persist as $\cN$ is increased from zero, with their shapes and frequencies altering as well.  For sufficiently small amplitudes, they will inherit the neutral stability of their linear limits--barring resonances among the the eigenvalues $\W_k$ that we will discuss shortly. 
In~\cite{KirKevShl:08}, Kirr et al.\ \ prove that as the $L^2$ norm of the solution is increased, then at a critical amplitude $$\cN_{\rm SB} \propto \W_2-\W_1,$$ the solution that continues from the ground state loses stability and a new stable solution to~\eqref{stationary_NLS} appears, possessing neither even nor odd symmetry.  That is, there is a symmetry-breaking or (Hamiltonian) supercritical pitchfork bifurcation.  Marzuola and Weinstein have demonstrated for this system in the unstable regime, over a long time period, the dynamics of~\eqref{NLS} are well-approximated by a Duffing oscillator-like dynamics~\cite{Marzuola:2010} when the initial condition is sufficiently close to an elliptic fixed point.   Pelinovsky and Phan have generalized this result to a wider class of initial conditions and provided a proof that relies on simpler estimates~\cite{Pelinovsky:2011}.

The present problem is naturally modeled by a three degree-of-freedom Hamiltonian system, which due to symmetry, as we will discuss, can be reduced to a two degree-of-freedom system.  It should also be noted that the symmetry-breaking bifurcation is non-generic--if $V(x)$ is non-symmetric, the system will generally feature a saddle-node bifurcation instead.
HH bifurcations are not possible in the two-mode system, and to observe them, we must consider a system with an additional degree of freedom.  We demonstrate via formal asymptotics, and observe numerically, that $\Psi_2$, the first excited state, generically becomes unstable in an HH bifurcation when $V(x)$ supports three localized eigenmodes and with eigenfrequencies satisfying the following assumptions.
\begin{samepage}
\subsubsection*{Assumptions}
\begin{enumerate}
\item $\W_1 < \W_2 < \W_3<0$, \label{A1}
\item $\W_2 - \W_1 = \Or{(1)}$,
\item $\W_3 - \W_2 = \Or{(1)}$, 
\item $(\W_3 - \W_2) - (\W_2 - \W_1) \ll 1$, and \label{2ndDifference}
\item $\W_3 = \Or{(1)}$ (i.e.\  a sufficient gap between the three eigenmodes and the band edge).
\end{enumerate}
\end{samepage}
To satisfy assumption~\ref{2ndDifference} in particular, we let
\begin{equation}
\label{nearResonanceAssumption}
\W_2-\W_1 = W - \epsilon \text{\ and\ } \W_3-\W_2 = W  + \epsilon
\end{equation}
where $\epsilon \ll W $ and $W =\Or{(1)}$.  The sign of $\epsilon$ is left unspecified while $W >0$.

By using inverse scattering techniques, we can construct a potential $V(x)$ with whatever eigenvalues we choose and which also satisfies the evenness condition.  In fact, the HH bifurcation is generic and will occur regardless of the evenness of $V(x)$.  The behavior of the system above the critical amplitude may, however, affect the nonlinear behavior of the system in the supercritical regime.
\subsection{Symmetries}
\label{sec:symmetries}
Let $\fN(\psi)= H \psi - \abs{\psi}^2 \psi$.  Then for any real potential $V(x)$, $\fN$ possesses $O(2)$ symmetry.  More specifically, defining the operators $R_\phi f(x) = e^{i\phi}f(x)$, and $\cZ f(x) = \bar{f}(x)$ corresponding to multiplication by an arbitrary complex phase and complex conjugation, we see that 
\begin{equation}
\fN(R_\phi \psi)=R_\phi \fN(\psi) \text{\ and\ }  \fN(\cZ \psi) = \cZ \fN(\psi).
\label{O2}
\end{equation}
Finally, define the operator $R_-(f(x))=f(-x)$. If, in addition, $V(x)$ is an even function, $\fN$ is also equivariant to the $\ZZ_2$ operation
\begin{equation}
R_- \fN(\psi(x))=\fN ( R_- \psi(x)).
\label{ZZ2}
\end{equation} 
Putting these together gives shows that system~\eqref{NLS} has $O(2)\times \ZZ_2$ symmetry.  Bifurcations in systems with such symmetries generally have codimension greater than or equal to bifurcations in similar systems without such symmetries.  Earlier studies have noted that the results can be generalized to a larger class of nonlinearities for which $\fN(\psi)$ is equivariant under~\eqref{O2} and~\eqref{ZZ2}.  The same is almost certainly true in the present case as well.  We choose to work with the simple cubic nonlinearity described above because a more general nonlinearity would invalidate relation~\eqref{permutation} below and increase even further the number of terms in equation~\eqref{3modes}.  As in Kirr et al., we will show that the reduced ODE system has the same symmetries.

\subsection{An alternate formulation}
\label{sec:alternate}
If we make the change of variables $\psi = \sqrt{\cN} \tilde \psi$ in equation~\eqref{NLS} and $\Psi = \sqrt{\cN} \tilde \Psi$ in~\eqref{stationary_NLS}, we get the modified evolution equation
\begin{equation}
\label{NLS_N}
i \partial_t \tilde\psi = H \tilde\psi -\cN|\tilde\psi|^2 \tilde\psi,
\end{equation}
and stationary equations
\begin{equation}\begin{split} \label{stationary_NLS_N}
\W \tilde\Psi & = H \tilde\Psi -\cN \tilde\Psi^3;\\
\int_{-\infty}^{\infty} \tilde\Psi^2(x) dx &= ||\tilde\Psi||_2^2 = 1.
\end{split}\end{equation}
This formulation presents a natural environment for studying the $\cN \to 0$ limit.  Since this system is well-defined regardless of the sign of $\cN$, we can study all the bifurcations for $\cN\in \RR$, which gives a fuller picture of the dynamics, unifying the focusing and defocusing NLS equations.  In section~\ref{sec:finiteDim}, we derive finite-dimensional models of systems~\eqref{NLS} and~\eqref{stationary_NLS}.  A similar change of variables will allow us to put a small parameter $N$ of either sign in front of the the nonlinear terms in, for example, system~\eqref{3modes} and other equations derived from it.  Also, it should be noted, that in this formulation there will generally be no bifurcation at $\cN=0$: for almost all potentials $V(x)$,  a smooth family of functions will pass right through any solution to system~\eqref{stationary_NLS_N} with $\cN=0$.

\subsection{The language of stability and resonance}
\label{sec:language}
Suppose that $\cN=0$ in the systems~\eqref{NLS_N} and~\eqref{stationary_NLS_N} and that the linear eigenvalue problem has $n$ linearly independent solutions $(\Psi_n,\W_n)$.  Then%
\footnote{In a finite dimensional model, the eigenfunction $\Psi_k(x)$ would be replaced by an eigenvector~$\vec{v}^{(k)}$.}
\begin{equation}
\psi(x,t) = \sum_{j=1}^{n} c_j e^{-i \W_j t} \Psi_j (x)
\label{quasiperiodic}
\end{equation}
solves equation~\eqref{NLS_N}.  In general, this solution is quasiperiodic: each individual component is periodic, but in general, the periods will be irreconcilable, and the solution as a whole is non-periodic.   Topologically, such a solution lies on an $n$-dimensional torus ${\mathbb T}^n$ in the $2n$ dimensional phase space, which can be thought of as the product of $n$ circles or equivalently as an $n$-dimensional hypercube, with opposite (hyper-)faces identified.
 A \emph{resonance relation} is a solution to the equation
\begin{equation}
\sum_{j=1}^{n} k_j \W_j = \innerProd{\vec k}{\vec\W} \text{ with } \vec{k} \in {\mathbb Z}^n\setminus\{0\}.
\label{resonanceGeneral}
\end{equation}
The sum
$$
\nu(\vec k)=\sum_{j=1}^{n} \abs{k_j}
$$
defines the order of a given resonance.  
For example under assumption~\eqref{nearResonanceAssumption} with $\epsilon=0$, the vector $\vec k = (1,-2,1)$ satisfies equation~\eqref{resonanceGeneral} and defines a resonance of order 4.
The number of independent solutions of equation~\eqref{resonanceGeneral} with a given order defines the multiplicity of a that resonance at that order. If the system has no such resonances, then each solution~\eqref{quasiperiodic} is dense on ${\mathbb T}^n$.  The number of linearly independent vectors $\lambda$ that solve equation~\eqref{resonanceGeneral} is the \emph{multiplicity} of the resonance.   If the system is resonant with multiplicity $m$, then the solutions are confined to, and dense on, $n-m$-dimensional subsets of ${\mathbb T}^n$ which are themselves topologically equivalent to ${\mathbb T}^{n-m}$.  To understand this, think of the two-dimensional case.  If there are two non-resonant frequencies, the solution is dense on a two-torus, so its closure is two-dimensional, but if the two frequencies are rationally related, then each one-dimensional orbit will be closed.  These closed orbits must lie on the level set of an additional conservation law.  When there is a near-resonance,
$$
\sum_{j=1}^{n} k_j \W_j\ll 1
$$
but non-zero, and nonlinear terms are nonzero but small, there will be a nearly conserved quantity that allows us to use averaging to decrease the dimension of the system and obtain simpler equations that are valid for a finite time.

Solutions of equation~\eqref{stationary_NLS_N} (but not equation~\eqref{quasiperiodic}!) are known as \emph{relative fixed points}.  Simply put, when viewed in an appropriate reference frame oscillating with frequency $\W$, they are time-invariant.  Similarly, there may exist \emph{relative periodic orbits}, which are themselves quasi-periodic, but appear periodic when viewed in an appropriate reference frame.

\begin{figure}[htb] 
   \centering
   \includegraphics[width=4in]{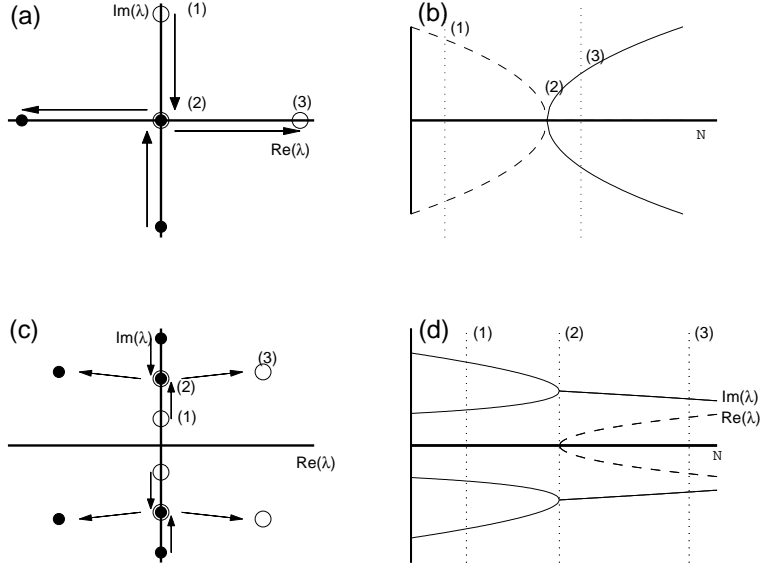}
      \caption{(a) The path of the eigenvalues as an parameter is varied in the Hamiltonian pitchfork bifurcation. (b) The real and imaginary parts of the eigenvalues.  (c) The path of the eigenvalues as an parameter is varied in the HH bifurcation. (d) The real and imaginary parts of the eigenvalues. After Luzzatto-Fegiz and Williamson~\cite{Luzzatto:2010}.}
   \label{fig:schematicBifurcations}
\end{figure}

The linear stability of a given solution to some general system will be determined by the eigenvalues $\lambda_j$ of a certain matrix $M$.  The imaginary parts of the eigenvalues determine the frequency with which small perturbations oscillate about the solution, and the real parts will determine the growth ($\Re \lambda_j>0$) or decay rate ($\Re\lambda_j<0$) of perturbations.  The solution is therefore unstable if there exist any eigenvalues $\lambda_j$ with $j>0$.   Points in parameter space where the stability changes are called bifurcation points, and there are different types.  The manifestation of the symmetry-breaking (or Hamiltonian pitchfork) bifurcation for NLS/GP is discussed in great detail in~\cite{KirKevShl:08,Marzuola:2010}.  We will describe the HH bifurcation in greater detail in section~\ref{sec:HH}.  In Hamiltonian systems, it is well-known that if $\lambda$ is an eigenvalue, then so are $-\lambda$, $\bar\lambda$, and $-{\bar\lambda}$.  This implies that the eigenvalues can occur in four types of groupings, up to multiplicity: complex quadruplets $\{\lambda,\bar \lambda, -\lambda,-\bar \lambda\}$ with nonzero real and imaginary parts, real-valued pairs $\{\lambda,-\lambda\}$, purely imaginary pairs $\{i\mu,-i\mu\}$ and zero eigenvalues of even algebraic multiplicity.  The symmetry-breaking, or Hamiltonian pitchfork occurs when, as a parameter is varied, a purely imaginary pair of eigenvalues collide at the origin, producing a purely real pair.  Here, small perturbations to the origin will initially grow monotonically due to the real positive eigenvalue.  See figure~\ref{fig:schematicBifurcations}(a) and~(b). The Hamiltonian Hopf bifurcation occurs when two pairs of pure imaginary eigenvalues collide at a nonzero point on the imaginary axis, and the four eigenvalues recombine to form a quartet of fully complex eigenvalues.  The dynamics in the near-linear regime is oscillatory due to the imaginary parts of the eigenvalues; see figure~\ref{fig:schematicBifurcations}(c) and~(d).  The major goal of this paper is to investigate the behavior of solutions in the nonlinear regime.

\section{Construction of the linear potential $V(x)$}
\label{sec:construction}

By explicit construction following Harrell~\cite{Har:80}, Kirr et al.\  demonstrate that, given a potential $V(X) $ with exactly $n$ discrete eigenmodes, the dual well potential $V^{(2)}_L(X)$ given by~\eqref{2well} will have exactly $2n$ eigenmodes, and that the eigenvalues come in pairs, each pair exponentially close to each other and to the corresponding eigenvalue of the single-well potential.\   Kapitula et al.\  discuss this same idea for a three-well potential given by~\eqref{3well}. As $L \to \infty$, the three eigenvalues of this system all converge to a multiplicity-three eigenvalue---a highly degenerate situation, for which a complete analysis is rather more complicated, with many bifurcations occurring quite near to each other.  

Another way to proceed is to specify the eigenvalues $\W_j = - \k_j^2, \, j= 1\ldots n$ and to use inverse scattering methods to construct a reflectionless potential with exactly these eigenvalues~\cite{Yukon:1980}.  This will be unique except for $n$ integrating factors $\xi_j$ (corresponding to the positions of the solitons) that arise from solving the associated Gel'fand-Levitan-Mar\v{c}enko equations.   The solution will exactly be a two-soliton solution of the Korteweg-de Vries equation~\cite{Drazin:1993}.  B\"{a}cklund transformations and Darboux transformations can be used to more easily find this solution.  The Darboux transformation is very similar to the B\"{a}cklund transformation, but it yields not only the potential, but its eigenvectors, which will be useful in what follows~\cite{Ablowitz:2004,MatSal:91}. There is a unique way to choose the constants $\xi_j$ such that $V(x)$ is an even function (corresponding to the situation where all $n$ solitons collide at the origin at $t=0$).

When $n=2$, the general formula for this 2-soliton is 
\begin{equation}
\label{2soliton}
V(x)=\frac
{4(\k_2^2-\k_1^2)(\k_2^2\cosh{2\k_1 x}+\k_1^2\cosh{2\k_0 x})}
{\left( (\k_1-\k_2)\cosh{(\k_2+\k_1)x} + (\k_1+\k_2)\cosh{(\k_1-\k_2)x} \right)^2}
\end{equation}
with $\k_1>\k_2>0$.  This has (un-normalized) ground state and excited states
\begin{align*}
\Psi_1 &= \frac{\cosh{\k_2 x}}{(\k_1-\k_2)\cosh{(\k_1+\k_2)x} + (\k_1+\k_2)\cosh{(\k_1-\k_2)x} } \\
\text{and\ } 
\Psi_2 &= \frac{\sinh{\k_1 x}}{(\k_1-\k_2)\cosh{(\k_2+\k_1)x} + (\k_1+\k_2)\cosh{(\k_1-\k_2)x} }
\end{align*}
and frequencies $\W_j = -\k_j^2$.
When $\k_1=2$ and $\k_2=1$, this potential reduces to the familiar initial condition for the KdV two-soliton 
$$
V(x)=-6 \sech^2{x}
$$
with frequencies $\W_1=-4$ and $\W_2=-1$.  If we choose $\k_1=\sqrt{1+\epsilon}$ and $\k_2=\sqrt{1-\epsilon}$, then for $0<\epsilon\ll 1$, the potential~\eqref{2soliton} takes the form of dual-well potential, very similar to that studied by Kirr et al.\    However the eigenvalues and eigenfunctions are now known exactly.

To compute the Hopf bifurcation, we may construct a potential $V(x)$ as a three-soliton solution to KdV.  The three-soliton has a very similar form to the two-soliton in equation~\eqref{2soliton}, but with ten terms in the numerator and four in the denominator.  It is also computed via the Darboux transformation.  This solution is given in appendix~\ref{sec:appendix}.  If the parameters are chosen such that the three eigenvalues are spaced very closely together (close to a triply-degenerate eigenvalue), then this potential takes the form of three nearly identical potentials spaced equidistantly apart at a large distance, as was studied by Kapitula et al.\   In a similar, but much more complex, vein, Hirsh et al.\ have used inverse scattering in order to design potentials that support modes of a user-prescribed shape~\cite{Hirsh:2009}.

\begin{figure}
\begin{center}
\includegraphics[width=3.5in]{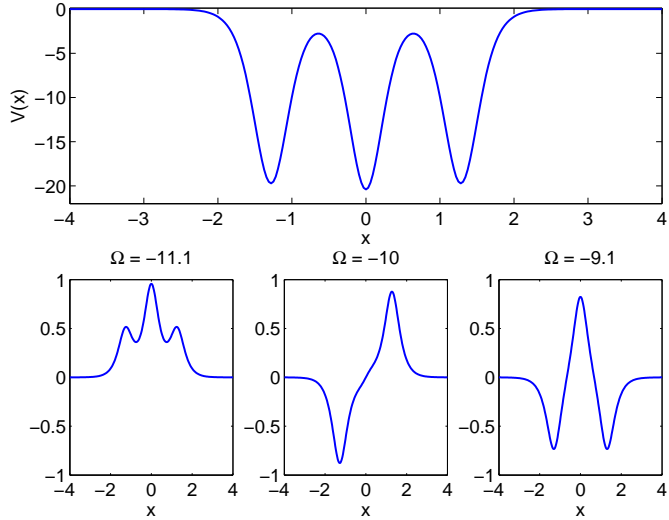}
\caption{(Top) The potential with $\W_2=-10$, $W=1$ and $\epsilon=-0.1$ in assumption~\eqref{nearResonanceAssumption}. (Bottom) Its corresponding eigenfunctions.}
\label{potentialModes_9_10_11}
\end{center}
\end{figure}

An example that displays the HH bifurcation, and which we will use in our subsequent numerical studies, is shown in figure~\ref{potentialModes_9_10_11}.  Here the potential is chosen with $\W = (-11.1,-10,-9.1)$. In fact, for the rest of this paper, this potential will be used, except where otherwise noted.  It is the mode corresponding to $\W_2-10$ that undergoes the HH bifurcation. 

\section{The finite-dimensional model}
\label{sec:finiteDim}

\subsection{Derivation of the model}
\label{sec:derivation}
We decompose the solution to equation~\eqref{NLS_N} as the following time-dependent linear combination:
\begin{equation}
\psi = c_1(t) \Psi_1(x) + c_2(t) \Psi_2(x) + c_3 \Psi_3(x) + \eta(x;t)
\label{decomposition}
\end{equation}
where the eigenvectors are orthonormal and, for each $t$ and $j$, $\eta(x;t)$ is in the orthogonal complement to the discrete eigenspace, i.e.\  
$$
\innerProd{\Psi_i}{\Psi_j}=\delta_{i,j} \text{ and }
\innerProd{\eta(\cdot,t)}{\Psi_j} = 0, \text{ for } i,j=1,2,3.
$$
We define the projection operators on to the discrete eigenmodes
\begin{equation}
\Pi_j \zeta= \innerProd{\Psi_j}{\zeta} \Psi_j, \text{ for } j=1,2,3
\label{proj_j}
\end{equation}
and onto the continuous spectrum
\begin{equation}
\Pi_{\rm cont} \zeta = \zeta - (\Pi_1 + \Pi_2 + \Pi_3) \zeta. 
\label{proj_cont}
\end{equation}

Following the methodology of Marzuola and Weinstein, we substitute the decomposition~\eqref{decomposition} into the PDE~\eqref{NLS} and apply to it the four projection operators defined above, giving evolution equations for the components of the decomposition.  The following system of equations is equivalent to the PDE~\eqref{NLS_N} under the assumption that $V(x)$ is even and supports exactly three modes.  
\begin{subequations}
\label{3modesPlus}
\begin{align}
 \nonumber    i \frac{dc_1}{dt} -\W_1 c_1 + \cN \bsl
    a_{1111} \abs{c_1}^2c_1 +
    a_{1113}(c_1^2 \bar c_3 + 2 \abs{c_1}^2 c_3) +
    a_{1122}(2c_1\abs{c_2}^2 + \bar c_1 c_2^2) \qquad &\\
     +   a_{1133}(2c_1\abs{c_3}^2 + \bar c_1  c_3^2) + 
    a_{1223}(c_2^2 \bar c_3 + 2\abs{c_2}^2 c_3) +
    a_{1333}\abs{c_3}^2 c_3 \bsr
    & =R_1(c_1,c_2,c_3,\eta) \label{c1} \\
 \nonumber    i \frac{dc_2}{dt} -\W_2 c_2 + \cN \bsl
    a_{1122}(c_1^2 \bar c_2 + 2\abs{c_1}^2 c_2) +
    2 a_{1223}(c_1 c_2 \bar c_3 + c_1 \bar c_2 c_3 + \bar c_1 c_2 c_3)  \qquad &\\+
    a_{2222}\abs{c_2}^2 c_2 
   + a_{2233}(2c_2\abs{c_3}^2+ \bar c_2 c_3^2)\bsr
    & =R_2(c_1,c_2,c_3,\eta) \label{c2}\\
 \nonumber    i \frac{dc_3}{dt} -\W_3 c_3 + \cN \bsl
    a_{1113} \abs{c_1}^2c_1 +
    a_{1133}(c_1^2 \bar c_3 + 2 \abs{c_1}^2 c_3) +
    a_{1223}(2c_1\abs{c_2}^2 + \bar c_1 c_2^2) \qquad &\\
     +   a_{1333}(2c_1\abs{c_3}^2 + \bar c_1c_3^2) + 
    a_{2233}(c_2^2 \bar c_3 + 2\abs{c_2}^2 c_3) +
    a_{3333}\abs{c_3}^2 c_3 \bsr
    & =R_3(c_1,c_2,c_3,\eta)  \label{c3}\\  
    i \partial_t \eta - H \eta  + \cN \abs{\eta}^2 \eta& = R_{\rm cont}(c_1,c_2,c_3,\eta); \label{eta_eqn}
\end{align}
\end{subequations}
where
$$
a_{jklm}= \innerProd{\psi_j}{\psi_k \psi_l \psi_m}
$$
where we have used that if $\{\pi_j,\pi_k,\pi_l,\pi_m\}$ is any permutation of $\{j,k,l,m\}$, then 
\begin{equation}
a_{\pi_j \pi_k \pi_l \pi_m}=a_{jklm}
\label{permutation}
\end{equation}
and that $a_{jklm}= 0$ if $j+k+l+m \equiv 1\mod{2}$. These parameters will be calculated numerically as needed for the simulations presented below. The $R_j$ and $R_{\rm cont}$ terms are the projections onto the appropriate eigenspaces of remaining nonlinear terms of~\eqref{NLS_N} and are presented in full in appendix~\ref{sec:projection_appendix}.

Ignoring the contributions of $\eta(x;t)$ to the solution, we derive a finite-dimensional approximation to~\eqref{3modesPlus}
\begin{subequations}
\label{3modes}
\begin{align}
 \nonumber    i \frac{dc_1}{dt} -\W_1 c_1 + N \bsl
    a_{1111} \abs{c_1}^2c_1 +
    a_{1113}(c_1^2 \bar c_3 + 2 \abs{c_1}^2 c_3) +
    a_{1122}(2c_1\abs{c_2}^2 + \bar c_1 c_2^2) \qquad &\\
     +   a_{1133}(2c_1\abs{c_3}^2 + \bar c_1  c_3^2) + 
    a_{1223}(c_2^2 \bar c_3 + 2\abs{c_2}^2 c_3) +
    a_{1333}\abs{c_3}^2 c_3 \bsr & =0  \\
 \nonumber    i \frac{dc_2}{dt} -\W_2 c_2 + N \bsl
    a_{1122}(c_1^2 \bar c_2 + 2\abs{c_1}^2 c_2) +
    2 a_{1223}(c_1 c_2 \bar c_3 + c_1 \bar c_2 c_3 + \bar c_1 c_2 c_3)  \qquad &\\+
    a_{2222}\abs{c_2}^2 c_2 
   + a_{2233}(2c_2\abs{c_3}^2+ \bar c_2 c_3^2) \bsr & =0 \\
 \nonumber    i \frac{dc_3}{dt} -\W_3 c_3 + N \bsl
    a_{1113} \abs{c_1}^2c_1 +
    a_{1133}(c_1^2 \bar c_3 + 2 \abs{c_1}^2 c_3) +
    a_{1223}(2c_1\abs{c_2}^2 + \bar c_1 c_2^2) \qquad &\\
     +   a_{1333}(2c_1\abs{c_3}^2 + \bar c_1c_3^2) + 
    a_{2233}(c_2^2 \bar c_3 + 2\abs{c_2}^2 c_3) +
    a_{3333}\abs{c_3}^2 c_3 \bsr & =0;
\end{align}
\end{subequations}
We have the following slight change of notation in this equation.  System~\eqref{3modesPlus}, being equivalent to equation~\eqref{NLS_N} conserves the $L^2$ norm
$$
\abs{c_1}^2 + \abs{c_2}^2 + \abs{c_3}^2 + \norm{\eta}{2}^2 = 1.
$$
This implies that 
$$
\abs{c_1}^2 + \abs{c_2}^2 + \abs{c_3}^2  \le 1.
$$
System~\eqref{3modes} possesses a finite-dimensional conserved quantity
$$
\abs{c_1}^2 + \abs{c_2}^2 + \abs{c_3}^2  = 1.
$$
The conserved quantities in systems~\eqref{NLS_N} and~\eqref{3modes} are not equivalent, since the contribution of $\eta(x,t)$ is ignored in the latter.  Recall that $\cN$ represents the total intensity and the sign of the nonlinearity in equation~\eqref{NLS}.  Since the meaning of $\cN$ is slightly changed from equation~\eqref{NLS_N} to system~\eqref{3modes}, we introduce the new constant $N$.

Note that system~\eqref{3modes} possess the same $O(2) \times \ZZ_2$ symmetry as the PDE~\eqref{NLS} from which they are derived.  The $O(2)$ symmetry is defined in the obvious way, analogous to that used in equation~\eqref{O2}.  The $\ZZ_2$ symmetry is due to the equivariance of equation~\eqref{3modes} generated by the operation of 
$$R_{\rm flip}(c_1,c_2,c_3) = (c_1,-c_2,c_3).$$
Note that composing this with the operator $R_\phi$ with $\phi=\pi$ flips the signs on $c_1$ and $c_3$, leaving $c_2$ unchanged.

\subsection{Stationary Solutions}
\label{sec:stationary}
We first look for stationary solutions of system~\eqref{3modes} of the form 
$$
\begin{pmatrix} c_1(t) \\ c_2(t) \\ c_3(t) \end{pmatrix} =
\begin{pmatrix}x \\ y \\ z \end{pmatrix} e^{-i \W t}
$$
This calculation is well-covered by Kapitula et al.~\cite{KapKevChe:06} in the case where the $a_{ijkl}$ coefficients satisfy some properties that significantly simplify all the equations and the resulting analysis.  We will repeat the parts we will need in what follows.
It is simple to show that the solution to equation~\eqref{stationary_NLS} with real potential $V(x)$ is, up to a constant phase factor, a real-valued function. This allows us, without loss of generality to assume $x, y, z \in \RR$ and that the stationary solution is of the form%
\footnote{See Kirr et al.\ for a more thorough justification~\cite{KirKevShl:08}.}
\begin{subequations}
\label{3modesReal}
\begin{align}
  (\W-\W_1) x + N(
    a_{1111} x^3 +
    3a_{1113}x^2  z  +
    3a_{1122}x y^2  +
    3a_{1133}x z^2 + 
    3a_{1223}y^2 z+
    a_{1333}z^3)& = 0  \\
(\W-\W_2) y +N(
   3 a_{1122}x^2  +
    6a_{1223}x z +
    a_{2222}y^2+
     3a_{2233}z^2)y & = 0 \label{yEqn}\\
 (\W-\W_3) z + N(
    a_{1113} x^3 +
    3a_{1133}x^2 z +
    3a_{1223}x y^2 +
    3a_{1333}x  z^2 + 
    3 a_{2233}y^2  z +
    a_{3333} z^3 )& =0  \\  
 x^2 +y^2 +z^2 - 1 &=0 \label{L2first}
\end{align}
\end{subequations}

\subsubsection*{Odd solutions}
Note that $\psi_1$ and $\psi_3$ lie in the invariant subspace of even functions, and thus that this system has solutions with $y=0$ and $x$ and $z$ nonzero.  Similarly, since only $\psi_2$ lies in the odd invariant subspace, there are solutions with only $y$ nonzero.  It is these whose stability we will investigate.  This family of solutions,
\begin{equation}
\label{oddSolution}
(x_{\rm odd},y_{\rm odd},z_{\rm odd},\W_{\rm odd})=
(0,1, 0, \W_2-a_{2222}N),
\end{equation}
may lose stability in a HH bifurcation.

\subsubsection*{Even solutions}
If $y=0$ then equation~\eqref{L2first} allows us to write $x=\cos\th$ and $z=\sin\th$.  Eliminating $\W$ from the remaining equations yields a single equation
\begin{multline}
N\Big(-a_{1333} \sin^4{\th} +
(a_{3333}-3 a_{1133}) \sin^3{\th} \cos{\th}+ 
3( a_{1333}- a_{1113}) \sin^2{\th} \cos^2{\th} \\
+(3 a_{1133}-a_{1111}) \sin {\th} \cos^3{\th}
+a_{1113} \cos^4{\th} \Big) 
+(\W_1-\W_3) \sin{\th} \cos{\th}=0
\label{evenSolution}
\end{multline}
which has period $\pi$, allowing us to consider the domain $0\le\th<\pi$.
For $N\ll 1$, this has two solutions: one near $\th=0$ and one near $\th=\pi/2$ which correspond to the fixed points $(x,y,z,\W)=(1,0,0,\W_1)$ and $(x,y,z,\W)=(0,0,1,\W_3)$ of the linear problem.  Both these solutions are found numerically to be stable for all $N$, which is consistent with the Krein signatures of their linearizations.  Additional even solutions may appear due to saddle-node bifurcations as discussed in~\cite{KapKevChe:06} and are shown in figure~\ref{fig:stationary}.

\subsubsection*{General Solutions}
Finally, there are solutions with all three components nonzero.  Cancelling out a factor of $y$ in equation~\eqref{yEqn}, then system~\eqref{3modesReal} depends on $y$ only through $y^2$.  We therefore define polar coordinates
$$ x = r \cos{\th}, \, z= r \sin{\th}, \, 0 \le r \le 1, $$
 and use equation~\eqref{L2first} to get $y^2=1-r^2$.  Note that the cases $r=0$ and $r=1$ correspond to even and odd solutions discussed above.  We use equation~\eqref{yEqn} to solve for $\W$ in terms of $r$ and $\th$.  After plugging this value into the $x$- and $z$-equations, we have a system in $r$ and $\th$ alone.  Eliminating $r$ from this system produces a single equation for $\th$:
\begin{equation}0=\beta_4\sin^4{\th} 
  +\beta_3 \sin^3{\th} \cos\th 
  +\beta_2 \sin^2{\th} \cos^2{\th} 
  +\beta_1 \sin\th \cos^3{\th} 
  +\beta_0 \cos^4{\th} 
\label{generalSolution}
 \end{equation}
where
\begin{align*}
\beta_4&=N(-a_{1333} a_{2222} +9 a_{1223} a_{2233} +3 a_{1333} a_{2233} -3 a_{1223}
   a_{3333}) -(3 a_{1223}-a_{1333}) (\W_2-\W_3);\\
 \beta_3&=N(18 a_{1223}^2 -9 a_{2233}^2 -6 a_{1223}
   a_{1333} -3 a_{1133} a_{2222} +9 a_{1122} a_{2233} +9 a_{1133} a_{2233} -3_{1122}
   a_{3333} +a_{2222} a_{3333}) +\\
   &\phantom{=} (a_{2222}  -6 a_{2233} +a_{3333})  \W_1+
  (-3 a_{1122} +3 a_{1133} +3 a_{2233}  -a_{3333})  \W_2 +
  (3 a_{1122} -3 a_{1133} -a_{2222} +3 a_{2233})  \W_3;\\
 \beta_2 &= N(27 a_{1122} a_{1223} -9 a_{1122} a_{1333} -3 a_{1113} a_{2222}   
   +3 a_{1333} a_{2222} +9 a_{1113} a_{2233} -27 a_{1223} a_{2233}) +\\
   &\phantom{=}(-9 a_{1223} +3a_{1333}) \W_1+(3 a_{1113} -3 a_{1333}) \W_2
   +(-3 a_{1113} +9 a_{1223}) \W_3; \\
   \beta_1 &=N(9 a_{1122}^2-18 a_{1223}^2 -9 a_{1122}
   a_{1133} +6 a_{1113} a_{1223} -a_{1111} a_{2222} +3 a_{1133} a_{2222} +3 a_{1111}  
   a_{2233}-9 a_{1122} a_{2233}) +\\
   &\phantom{=}(-3 a_{1122} +3 a_{1133} 1+a_{2222} -3 a_{2233})  \W_1
   +(a_{1111}  -3 a_{1122} -3 a_{1133}  +3 a_{2233}) \W_2
   +(-a_{1111}  +6 a_{1122}  -a_{2222})  \W_3;\\
   \beta_0 &=N(-3 a_{1113} a_{1122} +3 a_{1111} a_{1223} -9 a_{1122} a_{1223}
   +a_{1113} a_{2222} )  + (a_{1113}-3 a_{1223})(\W_1-\W_2).
   \end{align*}
   
\begin{figure}[htbp] 
   \centering
   \includegraphics[width=3in]{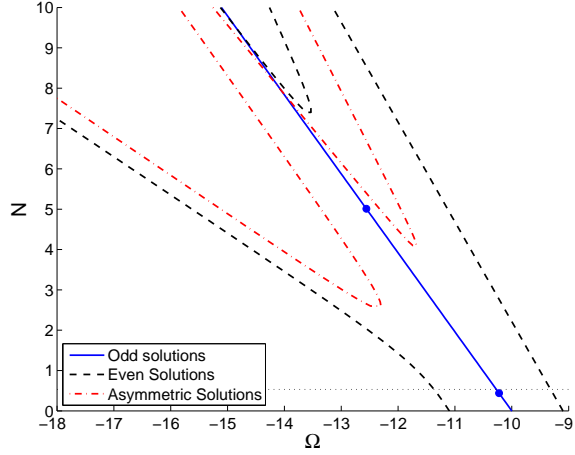} 
   \caption{The complete set of stationary solutions to system~\eqref{3modesReal}, with the symmetry of the corresponding PDE solutions indicated by line style.  The amplitudes of the Hopf bifurcations discussed in sections~\ref{sec:stability} and~\ref{sec:numerics} are indicated by points on the curve of odd solutions.  The bifurcation creating heteroclinic orbits of the even subspace as given by equation~\eqref{nuF} is shown as a horizontal dotted line.}
   \label{fig:stationary}
\end{figure}

It is straightforward, though messy to find the saddle-node bifurcation values of $N$ where new solutions to equations~\eqref{evenSolution} and~\eqref{generalSolution} arise (although this becomes much neater if the coefficients $a_{ijkl}$ are assumed to satisfy condition (3.6) of~\cite{KapKevChe:06}.  This calculation shows that in both cases, the bifurcations happen for $N=\Or{(W)}$ which we assume to be $\Or{(1)}$ as $\epsilon \to 0$.  A complete bifurcation diagram of solutions to system~\eqref{3modesReal} is shown in figure~\ref{fig:stationary}.  This figure does not show more complicated solutions to system~\eqref{3modes} such as quasiperiodic orbits whose existence we demonstrate in later sections.  

\subsection{Model Reduction via Symmetries}
\label{sec:reduction}
The HH bifurcation does not lead to any new fixed points.  Further, as the $c_j$ may evolve, it is not sufficient to assume that each component is real.   We therefore must work directly with equation~\eqref{3modes} rather than with the simpler equation~\eqref{3modesReal}.  

System~\eqref{3modes} may be written in Hamiltonian form as
\begin{equation}
\begin{split}
H=&\W_1 \abs{c_1}^2 + \W_2 \abs{c_2}^2 + \W_3 \abs{c_3}^2 - N {\bigg [}
   \thalf a_{1111} \abs{c_1}^4  
  + a_{1113}\abs{c_1}^2(c_1 \bar c_3 + \bar c_1 c_3) \\
  &+ a_{1122}\left(\thalf c_1^2 \bar c_2^2 + 2 \abs{c_1}^2 \abs{c_2}^2 + \thalf \bar c_1^2 c_2^2 \right)
  + a_{1133}\left(\thalf c_1^2 \bar c_3^2 + 2 \abs{c_1}^2 \abs{c_3}^2 + \thalf \bar c_1^2 c_3^2 \right)\\
  &+ a_{1223} \left(2\abs{c_2}^2(c_1 \bar c_3 + \bar c_1 c_3) + c_1 \bar c_2^2 c_3 + \bar c_1 c_2^2 \bar c_3  \right)
  + a_{1333} \abs{c_3}^2 (c_1 \bar c_3 + \bar c_1 c_3)\\
  &+ \thalf a_{2222} \abs{c_2}^4
  + a_{2233} \left( \thalf c_2^2 \bar c_3^2 + 2 \abs{c_2}^2 \abs{c_3}^2 + \thalf \bar c_2^2 c_3^2 \right)
  + \thalf a_{3333} \abs{c_3}^4 {\bigg ]}
\end{split}
\label{Hc}
\end{equation}
with evolution equations 
$$ i \dot{c}_j = \frac{\partial H}{\partial {\bar c}_j}.$$
Because $H$ is equivariant under the group action
$$
(c_1,c_2,c_3) \to e^{i\th}(c_1,c_2,c_3),
$$
the dynamics also conserve the (squared) $l^2$-norm:
\begin{equation}
\label{l2}
\abs{c_1}^2+\abs{c_2}^2+\abs{c_3}^2 =1
\end{equation}
as a consequence of the equivariance of system~\eqref{3modes} to the operator $R_\phi$ given in section~\ref{sec:symmetries} (see also equation~\eqref{L2first}).  Physically, this simply states that the number of particles or photons is conserved. 
We may use this property to reduce the problem from three degrees of freedom to two.  Taking advantage of the phase-invariance of $H$, we define new evolution variables
\begin{equation}
\label{complexChange}
c_1(t) = \s_1(t) e^{i \th(t)};\;
c_2(t) = \r(t) e^{i \th(t)}; \;
c_3(t) =\s_3(t) e^{i \th(t)}.
\end{equation}
where $\r(t),\th(t) \in \RR$.  The Hamiltonian~\eqref{Hc} is independent of $\th$, and conservation law~\eqref{l2} tells us that $\r(t) = (1 -\abs{\s_1(t)}^2-\abs{\s_3(t)}^2)^{1/2}$. Using this we may write down the reduced Hamiltonian dependent on just $\s_1$, $\s_3$, and their complex conjugates:
\begin{equation}
\begin{split}
H=&(\W_1-\W_2) \abs{\s_1}^2 +  (\W_3-\W_2) \abs{\s_3}^2 
 - N \bsl
   \thalf a_{1111} \abs{\s_1}^4  
  + a_{1113}\abs{\s_1}^2(\s_1 \bar \s_3 + \bar \s_1 \s_3) \\
  &+ a_{1122} (1- \abs{\s_1}^2 - \abs{\s_3}^2)(\abs{\s_1}^2+ 2 (\Re{\s_1})^2)
    + a_{1133}(\thalf \s_1^2 \bar \s_3^2 + 2 \abs{\s_1}^2 \abs{\s_3}^2 + \thalf \bar \s_1^2 \s_3^2 )\\
  &+ a_{1223} (1- \abs{\s_1}^2 - \abs{\s_3}^2)(\s_1 \s_3 + 2 \s_1 \sbar_3 + 2 \sbar_1 \s_3 + \sbar_1 \sbar_3)
  + a_{1333} \abs{\s_3}^2 (\s_1 \bar \s_3 + \bar \s_1 \s_3)\\
  &+ \thalf a_{2222} (1  - \abs{\s_1}^2 - \abs{\s_3}^2)^2
  + a_{2233} (1- \abs{\s_1}^2 - \abs{\s_3}^2)(\abs{\s_3}^2+ 2 (\Re{\s_3})^2)
    + \thalf a_{3333} \abs{\s_3}^4 \bsr. \nonumber
\end{split}
\end{equation}
This gives evolution equations 
\begin{subequations}
\label{complexform}
\begin{equation}
\begin{split}
i \dot \s_1 = &(\W_1-\W_2) \s_1 + N \bsl
-a_{1111} \abs{\s_1}^2 \s_1
-a_{1113} (2\abs{\s_1}^2 \s_3 + \s_1^2  \sbar_3) \\&
+ a_{1122} \left((2\abs{\s_1}^2+\abs{\s_3}^2-1)(2\s_1 + \sbar_1) + \thalf(\s_1^2-\sbar_1^2)\s_1
\right)
-a_{1133}  (2 \abs{\s_3}^2 \s_1+\sbar_1 \s_3^2) \\&
+a_{1223} \left((2\abs{\s_1}^2+\abs{\s_3}^2-1)(2 \s_3 +\sbar_3) + \s_1^2(\s_3+2\sbar_3) \right) \\&
-a_{1333} \abs{\s_3}^2 \s_3
-a_{2222} (\abs{\s_1}^2 +\abs{\s_3}^2-1)\s_1 
+\thalf a_{2233}  (4 \abs{\s_3}^2+\sbar_3^2+\s_3^2) \s_1 \bsr
\end{split}
\label{complex1}
\end{equation}
and
\begin{equation}
\begin{split}
i \dot \s_3 = &
(\W_3-\W_2)\s_3 + N \bsl
-a_{1113} \abs{\s_1}^2 \s_1
+\thalf a_{1122} (4\abs{\s_1}^2+\sbar_1^2+\s_1^2)\s_3 
-a_{1133} (2 \abs{\s_1}^2 \s_3+ \s_1^2 \sbar_3)\\&
+a_{1223} \left((\abs{\s_1}^2+2\abs{\s_3}^2-1)(2 \s_1 +\sbar_1) + \s_3^2(\s_1+2\sbar_1) \right) \\&
-a_{1333} (2\abs{\s_3}^2 \s_1+\sbar_1 \s_3^2)
-a_{2222} (\abs{\s_1}^2 +\abs{\s_3}^2-1)\s_3  \\&
+a_{2233} \left((\abs{\s_1}^2+2\abs{\s_3}^2-1)(2\s_3 + \sbar_3) + \thalf(\s_3^2-\sbar_3^2)\s_3
\right)
-a_{3333} \abs{\s_3}^2 \s_3 \bsr.
\end{split}
\label{complex2}
\end{equation}
\end{subequations}
This reduction involves fixing a reference phase $\th(t)$ and thus leads to equations that are not equivariant with respect to operator $R_\phi$.

The full solution may be recovered using the conservation law 
$$
\r^2  = 1 - \abs{\s_1}^2 - \abs{\s_3}^2,
$$
and the auxiliary equation for $\th(t)$:
\begin{multline*}
\dot \th(t)=  -\W_2+ N \bsl a_{2222} (1-\abs{\s_1}^2 - \abs{\s_3}^2)
   + \thalf a_{1122}(\abs{\s_1}^2 +2\Re(\s_1^2))\\
   +a_{1223} (2 \sbar_3 \s_1+2 \sbar_1 \s_3+\sbar_1 \sbar_3+\s_1\s_3)
   +\thalf a_{2233} (\abs{\s_3}^2 +2\Re(\s_3^2) ) \bsr.
\end{multline*}

\section{Linear stability}
\label{sec:stability}

We are particularly interested in the stability of the antisymmetric mode $\Psi_2(x,N)$, the nonlinear continuation of the linear eigenmode $\Psi_2$ of equation~\eqref{linearEig}.  

\subsection{Linearization of PDE solutions}
Letting $(\Psi,\W)$ be a solution of system~\eqref{stationary_NLS_N} and consider small time-dependent perturbations of the form 
$$\psi(x,t) = \Psi(x) + (u(x,t)+i v(x,t)) e^{-i \W t}.$$
Then, linearizing and making the standard assumption that
$$ u=U(x) e^{\lambda t}, \, v=V(x) e^{\lambda t}$$
one finds the eigenvalue problem
$$
\lambda \binom{U}{V} = 
\begin{pmatrix}
0 & -(\W + \partial_x^2 - V(x) + \cN) \\
\W + \partial_x^2 -V(x) + 3 \cN
\end{pmatrix}
 \binom{U}{V}.
$$

\subsection{Linearization of ODE}

First, we determine the linear stability of the solution~\eqref{oddSolution}, which corresponds to $\s_1=\s_3=0$ in system~\eqref{complexform}. 
By inserting the form 
$$
\begin{pmatrix}
\s_1(t)  \\ \s_3(t)
\end{pmatrix} = 
\begin{pmatrix}
u_1(t) + i v_1(t)  \\
u_3(t) + i v_3(t)
\end{pmatrix}
$$
into system~\eqref{complexform}, the linearized equations become
\begin{equation}
\label{Linearized}
\frac{d}{dt}
\begin{pmatrix} u_1 \\  u_3 \\ v_1 \\ v_3 \end{pmatrix} = 
\begin{pmatrix} 0_2 & M_1 \\ M_2 & 0_2 \end{pmatrix}
\begin{pmatrix} u_1 \\  u_3 \\ v_1 \\ v_3 \end{pmatrix} 
\end{equation}
where $0_2$ is a $2\times2$ matrix of zeros, 
$$
M_1=\begin{pmatrix}
\W_1-\W_2 + (a_{2222}-a_{1122})N &  -a_{1223} N \\
-a_{1223}N &  \W_3-\W_2 + (a_{2222}-a_{2233})N
\end{pmatrix}
$$
and
$$
M_2=\begin{pmatrix}
-(\W_1-\W_2) + (3a_{1122}-a_{2222})N  & 3a_{1223} N  \\
3a_{1223} N  & -(\W_3-\W_2) + (3a_{2233}-a_{2222})N 
\end{pmatrix}.
$$
Note in the limit $N \to 0$, the matrix $M=\left(\begin{smallmatrix} 0_2 & M_1 \\ M_2 & 0_2 \end{smallmatrix}\right)$ has eigenvalues $\pm i(\W_2-\W_1)$ and $\pm i(\W_3-\W_2)$ on the imaginary axis, and that for all $N$, the matrix $M$ is symplectic, i.e.\  $M=J K$ where $K$ is symmetric and 
$$
J= \begin{pmatrix} 0_2 & -I \\ I & 0_2 \end{pmatrix}
\text{ so that }
K = \begin{pmatrix} M_2 & 0_2 \\ 0_2 & -M_1 \end{pmatrix}.
$$

\subsection{Analytical criterion for ODE bifurcation}
\label{sec:HH}
The HH bifurcation is the result of a \emph{Krein collsion}, where two (pairs of complex-conjugate) eigenvalues collide on the imaginary axis, as is shown schematically in figure~\ref{fig:schematicBifurcations}(c) and~(d)~\cite{MacKay:1987}.  If $\xi$ is any vector in the eigenspace belonging to the eigenvalues $\pm i \w$ on the imaginary axis.  Then the \emph{Krein signature} 
$$
\cK(\xi) = \sgn{\left(\half \xi^{T} J K \xi\right)}
$$
is constant on the entire eigenspace, and is thus a property of the eigenspace rather than of any particular nonzero vector in that space.  If two eigenvalues have opposite Krein signature, then, upon collision, they will generically split into a Krein quartet, indicating that the origin has become unstable, with oscillatory dynamics due to their nonzero imaginary parts.

In the present ODE, in the $N \to 0$ limit, the eigenvalues, i.e.\ the frequencies in the linear system, are $\pm i(\W_2-\W_1)$ and $\pm i(\W_2-\W_3$.  The Krein signatures are $\cK(\pm i(\W_2-\W_1))=\sgn{(\W_1-\W_2)}$ and $\cK(\pm i(\W_3-\W_2))=\sgn{(\W_3-\W_2)}$, implying by assumption~\eqref{A1}, that their Krein signatures are different. Thus, their collision will lead to instability. In fact, the Krein signature can be interpreted as the direction of phase rotation, and since $\W_2$ lies between $\W_1$ and $\W_3$, the Krein signatures can be determined without performing this calculation.  We expect, and find numerically below, that the frequency at which the collision takes place is near $\pm\abs{\W_2-\W_3} \approx \pm\abs{\W_2-\W_3} \approx W$.

To detect the HH bifurcation, we construct $P(\lambda;N)$, the characteristic polynomial of $M$, which, as is generic for Hamiltonian systems, is a quadratic polynomial in $\lambda^2$.  Letting $q=\lambda^2$, we define the simpler quadratic polynomial $p(q;N)$.  There will be a double eigenvalue at the value of $N$ where the discriminant of $p(q)$ is zero. We further make assumption~\eqref{nearResonanceAssumption}.  The discriminant is a quartic polynomial $\Pi(N)$.  Defining $\nu = N/\epsilon$, and factoring out a common factor of $\epsilon^2$, we find

\begin{equation}
\Pi(\nu) = d_4(\epsilon) \nu^4 + d_3(\epsilon) \nu^3 + d_2 (\epsilon)\nu^2 + d_1 (\epsilon)\nu + d_0(\epsilon) = 0
   \label{discriminant}
 \end{equation}
 where
 \begin{align*}
d_4 = & (3 a_{1122}-4 a_{2222}+3 a_{2233})^2 (a_{1122}^2-2 a_{2233} a_{1122}+4 a_{1223}^2+a_{2233}^2) \epsilon^2 \\
d_3 = & 8 (a_{1122}-a_{2233}) (a_{1122}-a_{2222}+a_{2233}) (3 a_{1122}-4 a_{2222}+3 a_{2233}) W \epsilon\\
&-8 (3 a_{1122}-4 a_{2222}+3 a_{2233}) (a_{1122}^2-2 a_{2233} a_{1122}+4 a_{1223}^2+a_{2233}^2) \epsilon^2\\
d_2 = & 16 (a_{1122}-a_{1223}-a_{2222}+a_{2233}) (a_{1122}+a_{1223}-a_{2222}+a_{2233}) W^2 \\ 
& -8 (a_{1122}-a_{2233}) (7 a_{1122}-8 a_{2222}+7 a_{2233}) W \epsilon \\
&+16 (a_{1122}^2-2 a_{2233} a_{1122}+4 a_{1223}^2+a_{2233}^2) \epsilon^2\\
d_1 =& 32 (a_{1122}-a_{2233}) W \epsilon -32 (a_{1122}-a_{2222}+a_{2233}) W^2\\
d_0 =& 16 W^2
\end{align*}
We may solve this numerically or by a perturbation expansion of the form:
$$
\nu = n_1 + \Or{(\epsilon})
$$
and find that there are double eigenvalues at
\begin{equation}
N_{{\rm HH},\pm} = \frac{\epsilon}{-a_{1122}\pm a_{1223}+a_{2222}-a_{2233}} 
+Or{(\epsilon^2)}. 
\label{Ncritical}
\end{equation}
Thus, there will be a HH bifurcations for small values of $\epsilon$.  The $\Or{(\epsilon^2)}$ term (not shown) contains a factor of $W^{-1}$, showing that if $W\ll 1$, the divergence of the bifurcation value from a simple linear function of $\epsilon$ is greater.  This is exactly the case of a near-triply degenerate eigenvalue, the case studied by Kapitula et al.

Before proceeding to simulate and analyze system~\eqref{complexform}, we make some observations: 
\begin{itemize}
\item The first two terms correspond to $|c_1|^2$ and $|c_3|^2$ and are unchanged. Similarly any term multiplying a coefficient $a_{ijkl}$ where each term in the subscript is 1 or 3 is unchanged.  Only the terms that had a contribution from $c_2$ (those with 2's in the subscripts) are altered.  
\item Except for a scaling factor, the real parts of $\s_1$ and $\s_3$ can be interpreted as position variables, and their imaginary parts as canonical momentum variables.
\end{itemize}

\section{Numerical Simulations}
\label{sec:numerics}
\subsection*{Bifurcation study: spectrum of linearization}
\begin{figure}[htbp] 
   \centering
   \includegraphics[width=2.5in]{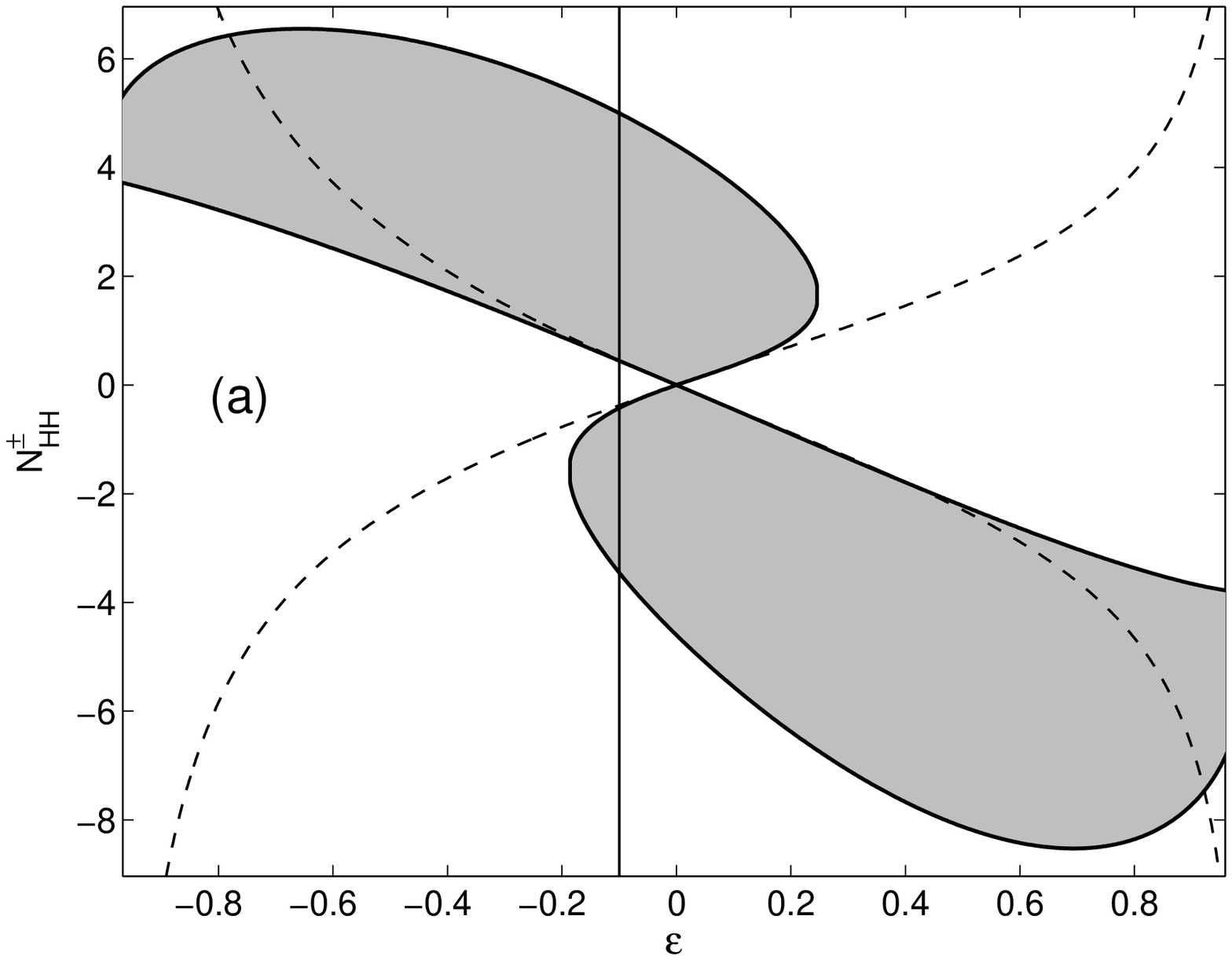} 
   \includegraphics[width=2.5in]{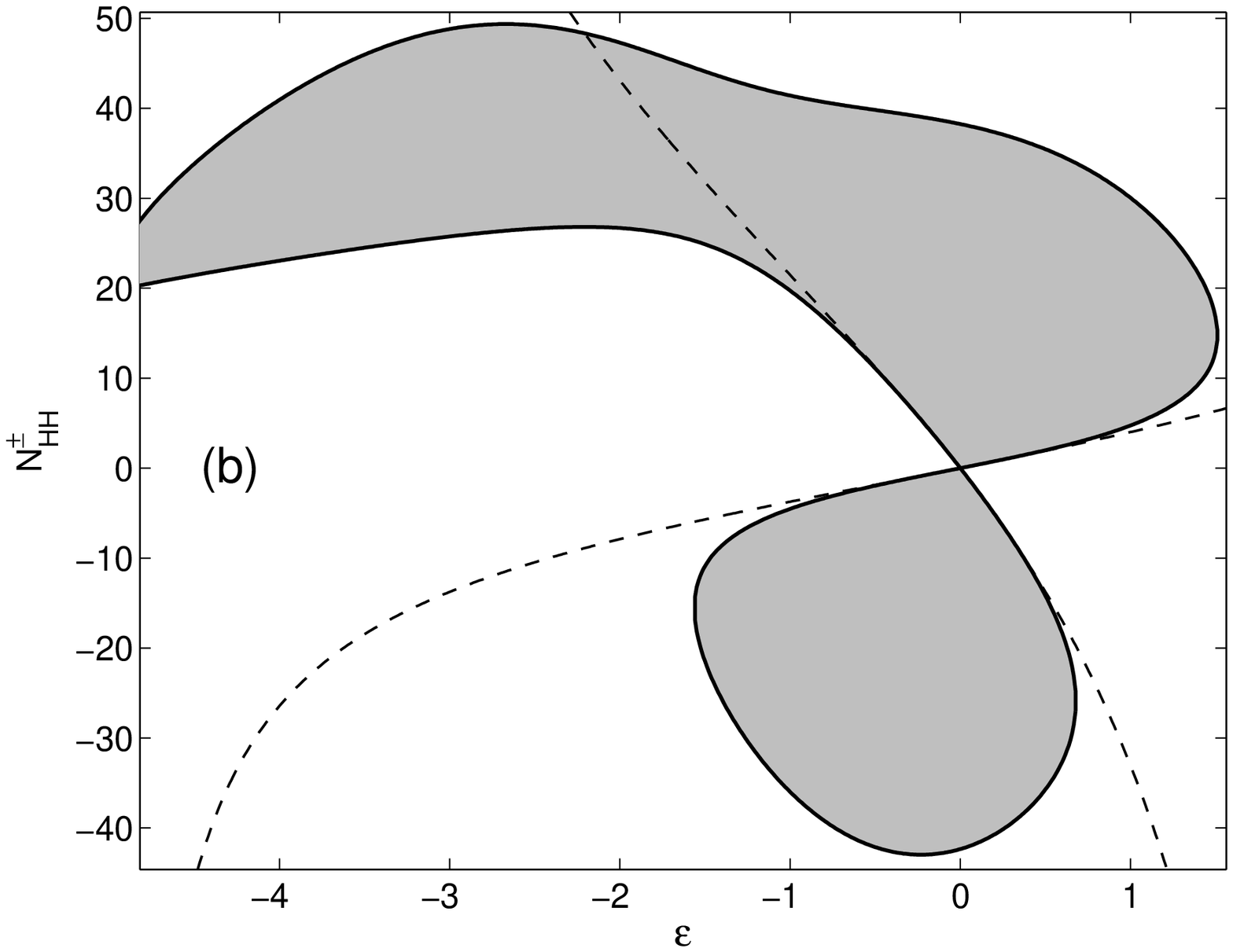} 
   \caption{Stability diagrams.  The trivial solution to system~\eqref{Linearized} is linearly stable inside the shaded regions, and loses stability to HH bifurcations along the thick black lines.  The dashed lines show the approximate values~\eqref{Ncritical} of $\NHH$, computed to $\Or{(\epsilon})$. This corresponds to potentials with frequencies \textbf{(a)} $(\W_1,\W_2,\W_3) = (-11+\epsilon,-10,-9+\epsilon)$ and \textbf{(b)} $(\W_1,\W_2,\W_3) = (-15+\epsilon,-10,-5+\epsilon)$. The vertical line in (a) gives the values of the parameters used in all other figures in this paper.}       
    \label{fig:nCritical}
\end{figure}

We first consider whether equation~\eqref{Ncritical} provides a good approximation to the critical value $\NHH$ when the eigenvalues are of the form given by equation~\eqref{nearResonanceAssumption}.   We show two examples. In both cases we choose $\W_2=-10$, while  $\epsilon$ in equation~\eqref{nearResonanceAssumption} is allowed to vary.  The first subfigure shows $W=1$ and in the second, $W=5$.   The potential pictured in figure~\ref{potentialModes_9_10_11} corresponds to choosing $\epsilon=0.1$ in the first subfigure, and its shape does not change much as $\epsilon$ is varied.  The result is shown in figure~\ref{fig:nCritical}, and demonstrates the large effect $W$ has on higher-order terms in this approximation.  We also see from this figure, that for small values of $\epsilon$, the system undergoes HH bifurcations at both positive and negative values of $N$, but that for larger values of $\epsilon$, one of these bifurcations may cease to exist.  This change in character occurs for values of $\epsilon$ where two roots of the discriminant~\eqref{discriminant} collide and annihilate each other.  This will happen at values of $\epsilon$ where the discriminant of $\Pi(N)$ vanishes.  This is formally a sixth-degree polynomial in~$\epsilon$, but the coefficients $a_{ijkl}$ are themselves functions of~$\epsilon$, so the equation is in fact transcendental.

Next we investigate how well the bifurcation structure of ODE system~\eqref{complexform} compares with that of standing waves of system~\eqref{stationary_NLS} when the potential $V(x)$ is given in Appendix~\ref{sec:appendix}.  We find numerically approximate solutions to~\eqref{stationary_NLS} by first  replacing the derivatives with their pseudospectral approximations, and solving the resulting (finite dimensional) equations using Matlab's \texttt{fsolve} command.  We then form the linearization about the standing wave, again using the pseudospectral approximation for the derivatives.  This approximation is implemented as a function and the spectrum is calculated using the Matlab \texttt{eigs} command.  The discrete spectrum of this problem is compared with the numerically calculated spectrum of the matrix in equation~\eqref{Linearized}.

\begin{figure}[htbp] 
   \centering
   \includegraphics[width=2.75in]{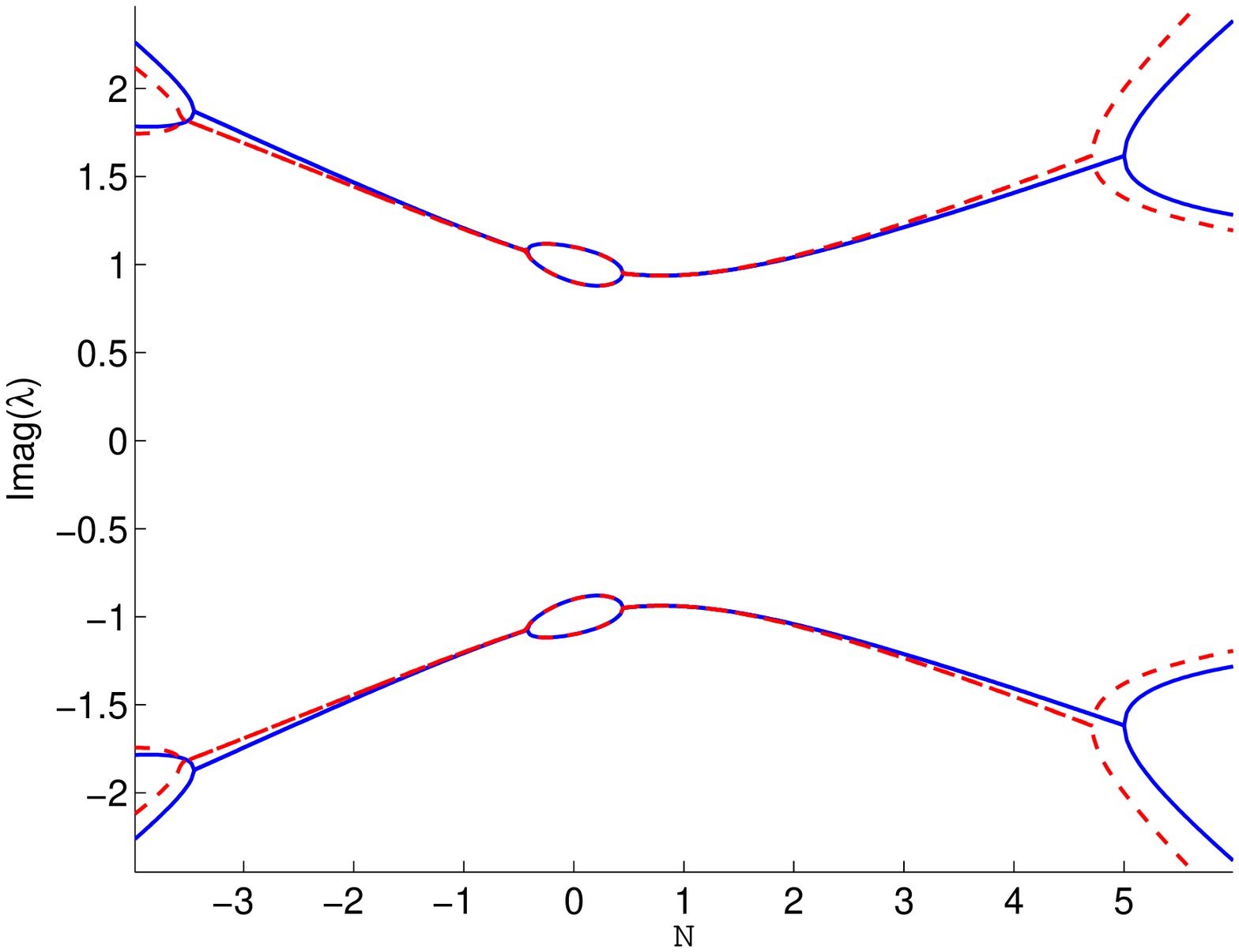} 
   \includegraphics[width=2.75in]{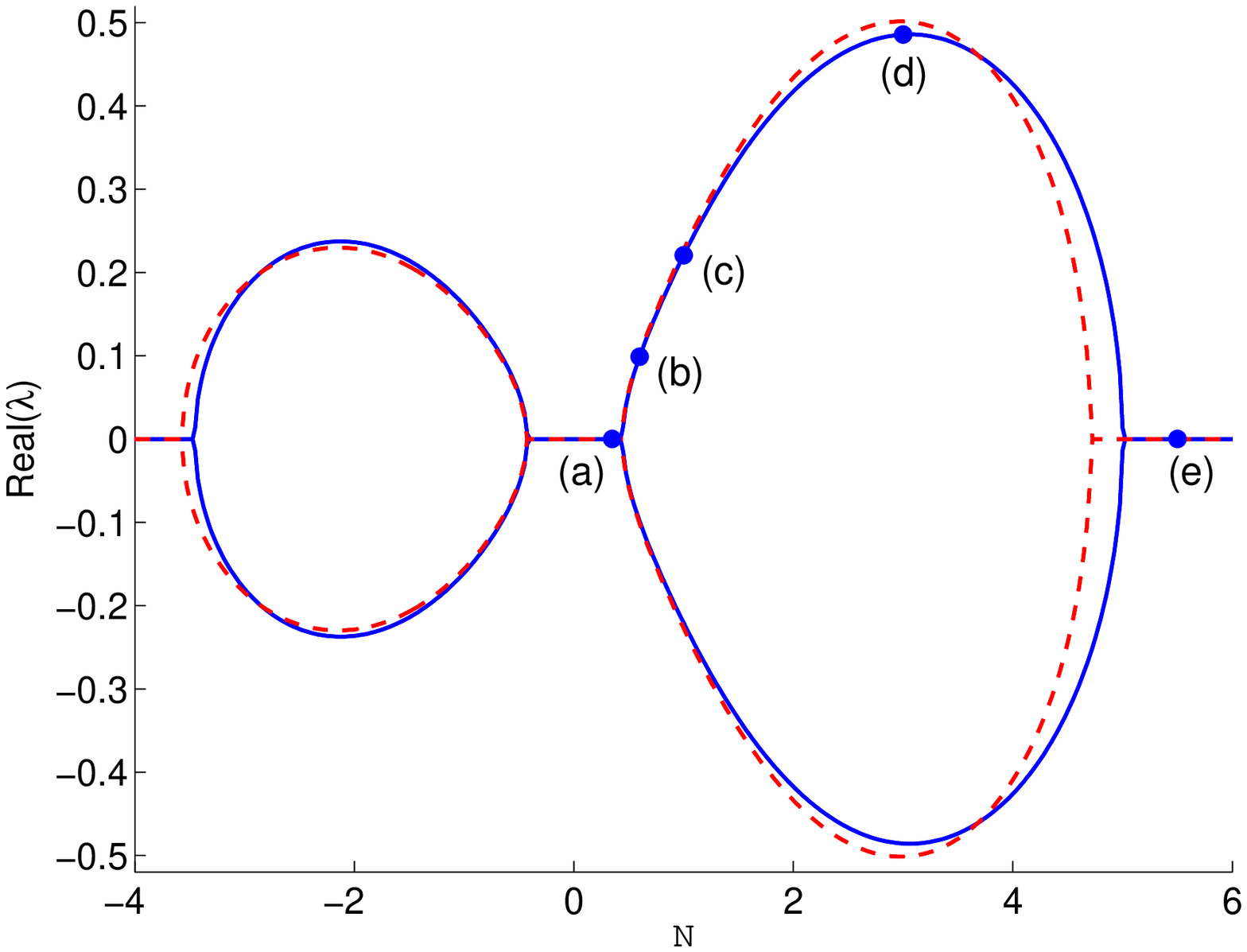} 
   \caption{The imaginary and real parts of the discrete eigenvalues of the reduced ODE (solid) PDE standing wave (dashed).  Both show four HH bifurcations.  Both the PDE and ODE solutions lose stability near $N=\pm 0.44$ the solutions regain stability $N=4.71$ and $N=-3.58$(PDE) and $N=-3.46$ and $N=5.01$ (ODE), where it regains stability. The parameter values are $\W = (-11.1,-10,-9.1)$. }
   \label{fig:ODE_PDE_Hopf}
\end{figure}

When $\epsilon=0$, the discrete and continuous spectrum of the linearized operator lie entirely on the imaginary axis.  We can define the Hopf bifurcation point as the value of $N$ for which the eigenvalues on the imaginary axis, in pairs, collide and move off into the complex plane as a quartet.  In figure~\ref{fig:ODE_PDE_Hopf}, we see that the discrete spectrum of the ODE closely resembles that of the PDE system.  The PDE system also has a double eigenvalue at the origin, which is unaffected by the bifurcation and which is not shown.

\subsection*{ODE dynamics}
\begin{figure}[htbp] 
   \centering
\includegraphics[width=\mywidth]{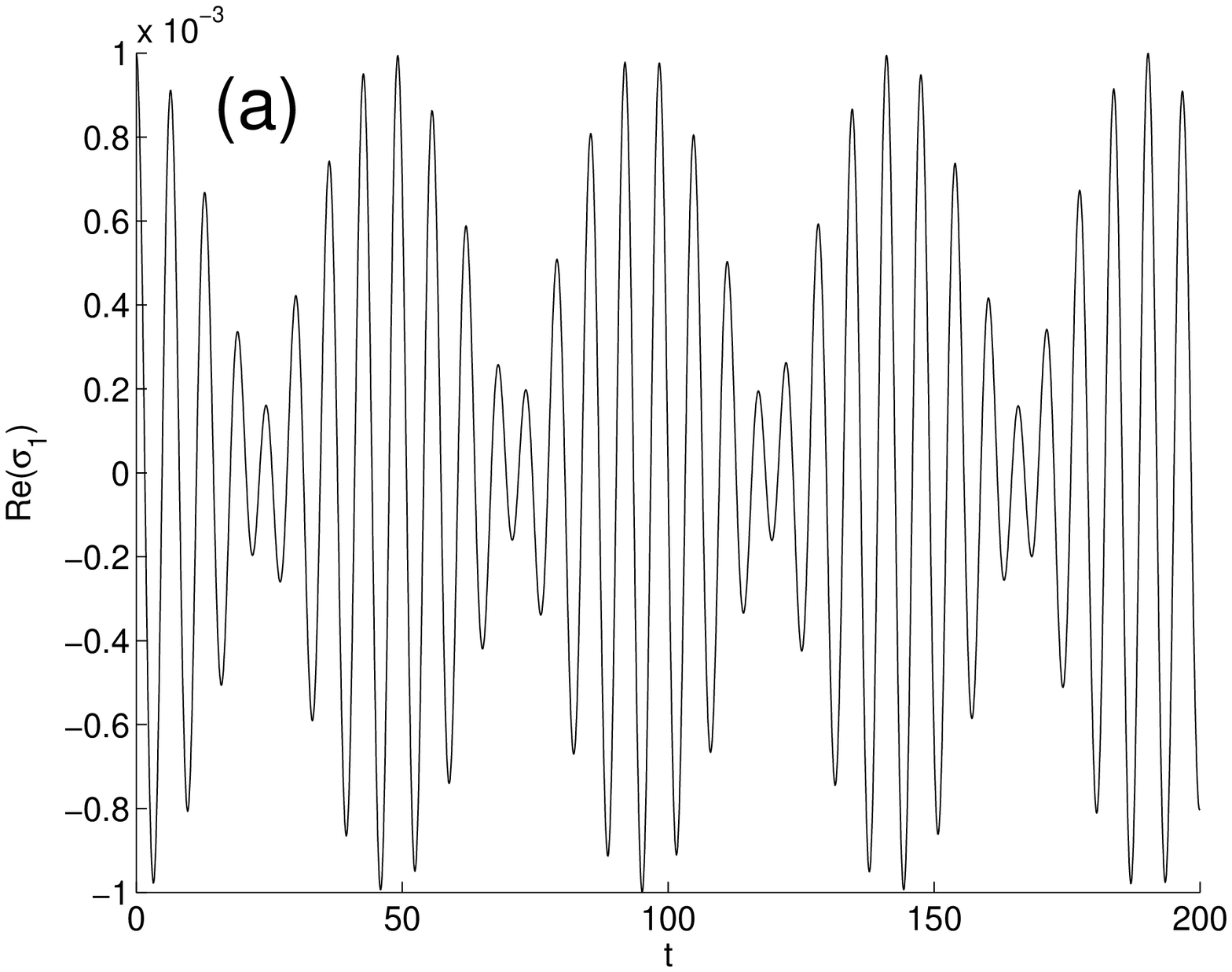}
\includegraphics[width=\mywidth]{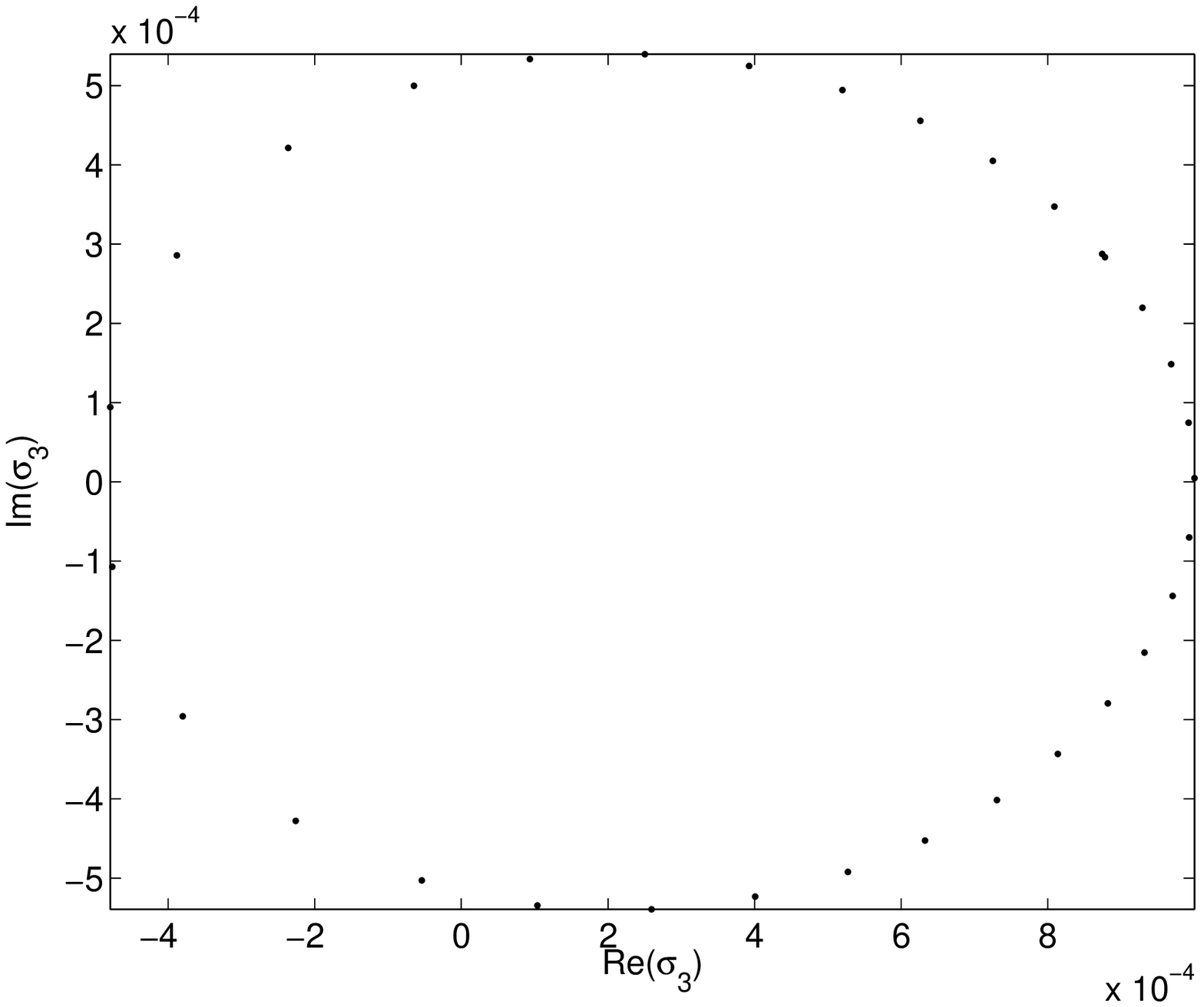}
\includegraphics[width=\mywidth]{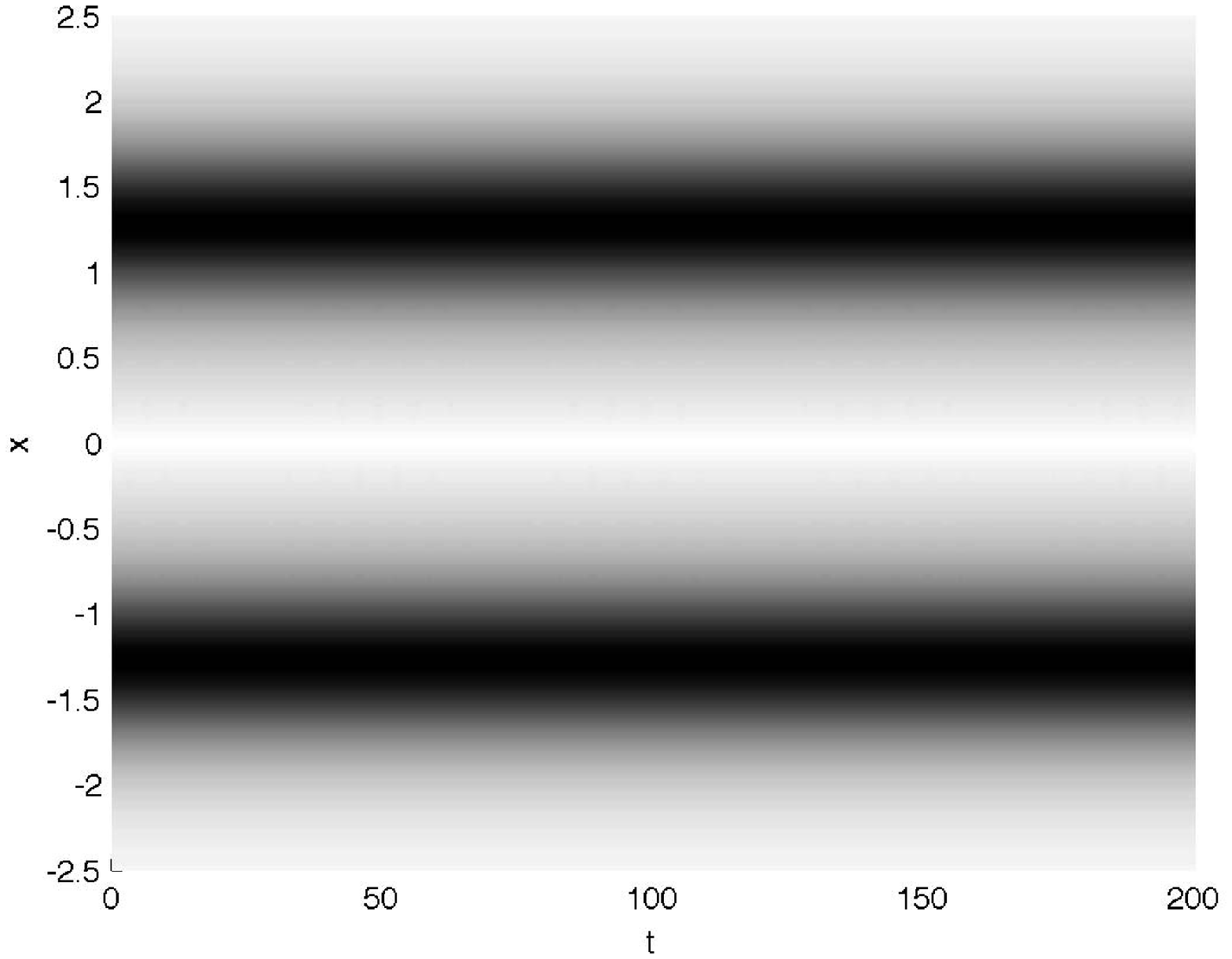}
\includegraphics[width=\mywidth]{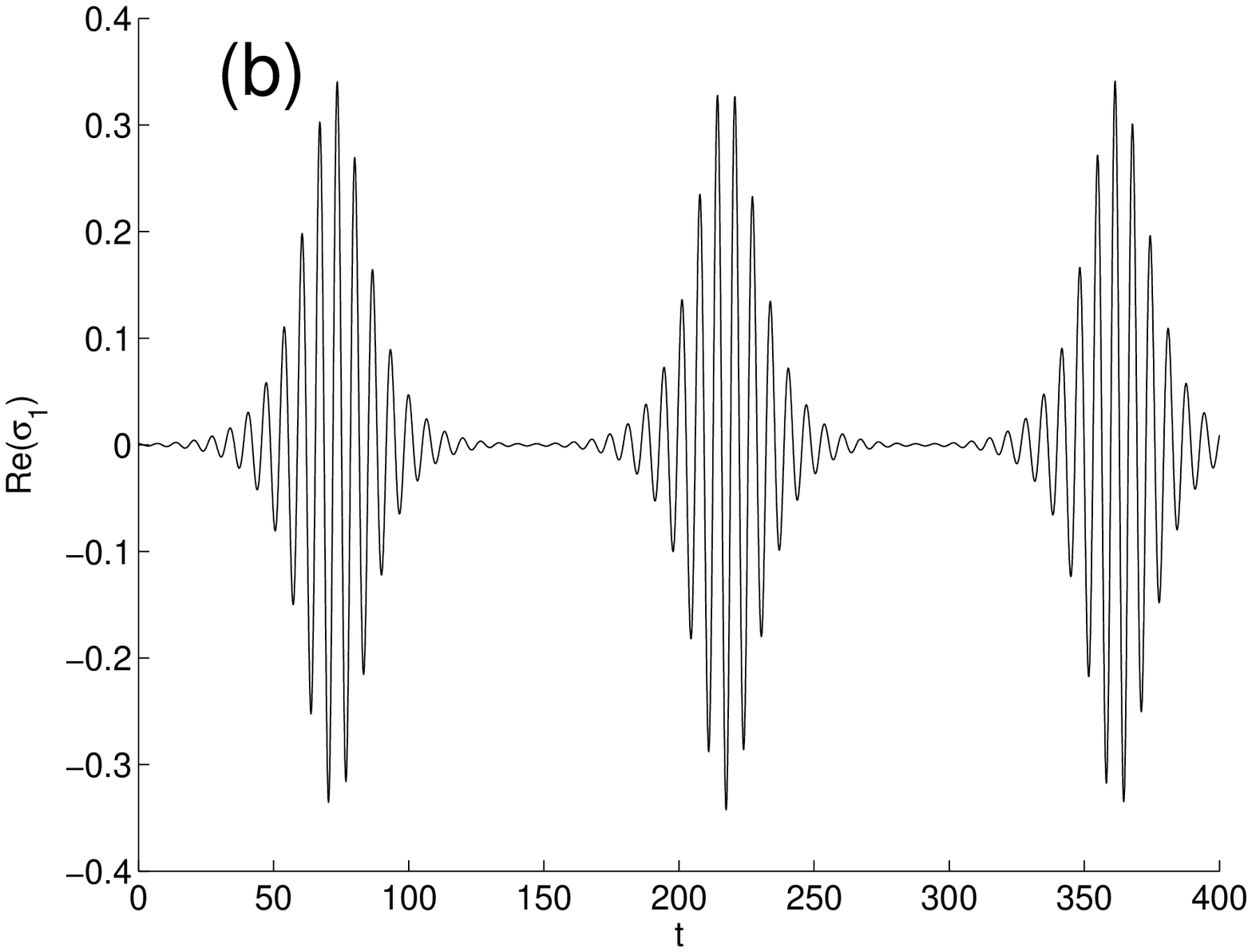}
\includegraphics[width=\mywidth]{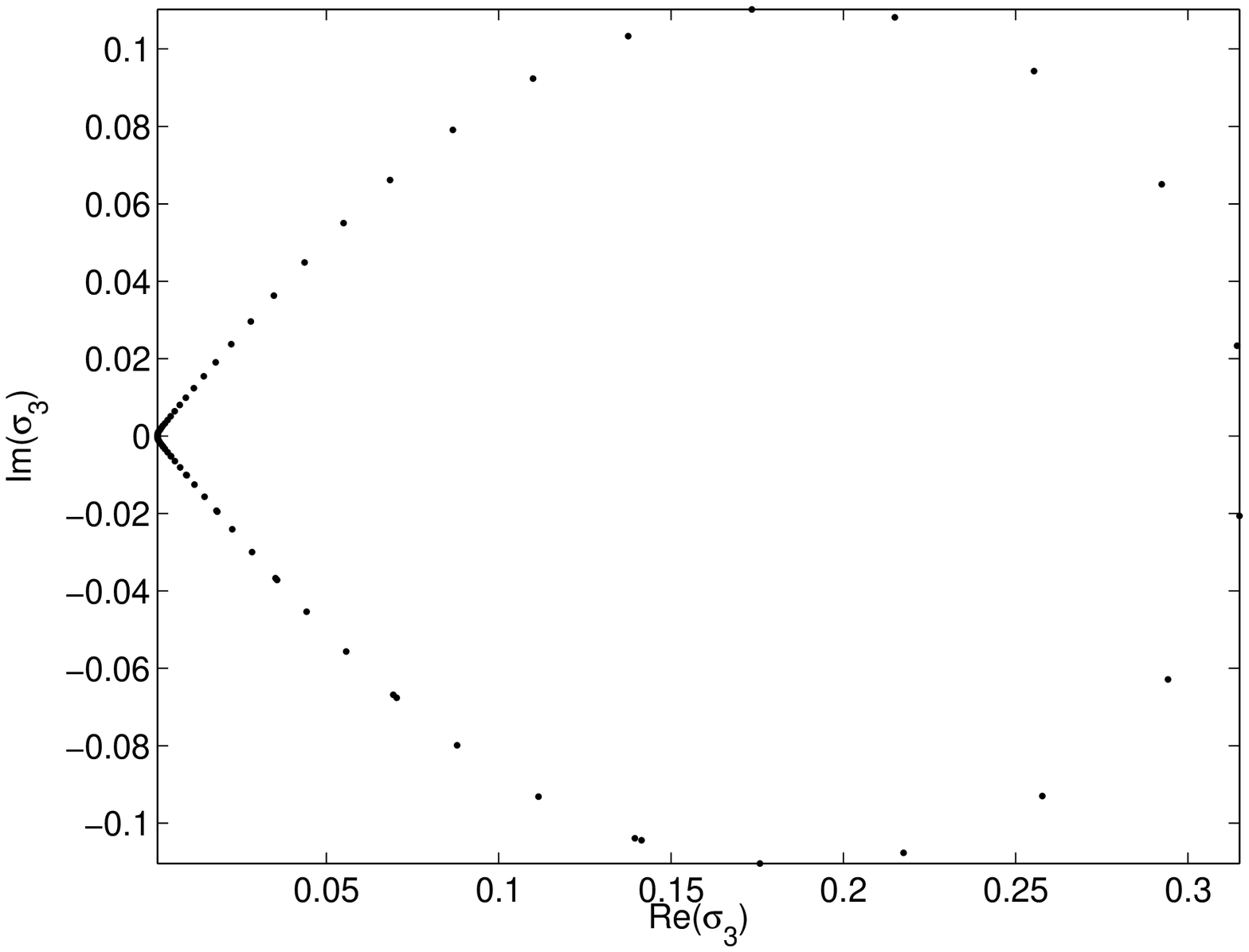}
\includegraphics[width=\mywidth]{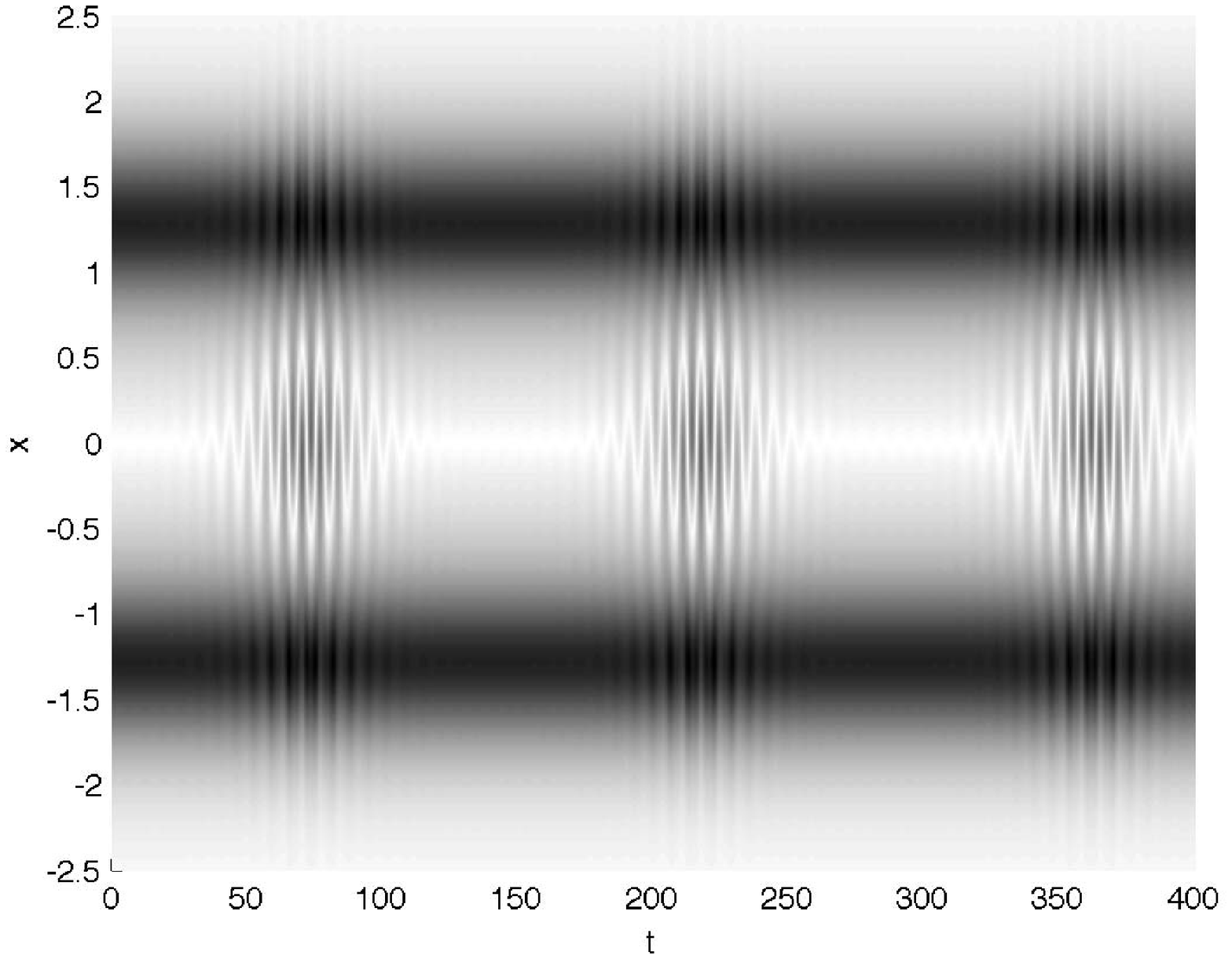}
\includegraphics[width=\mywidth]{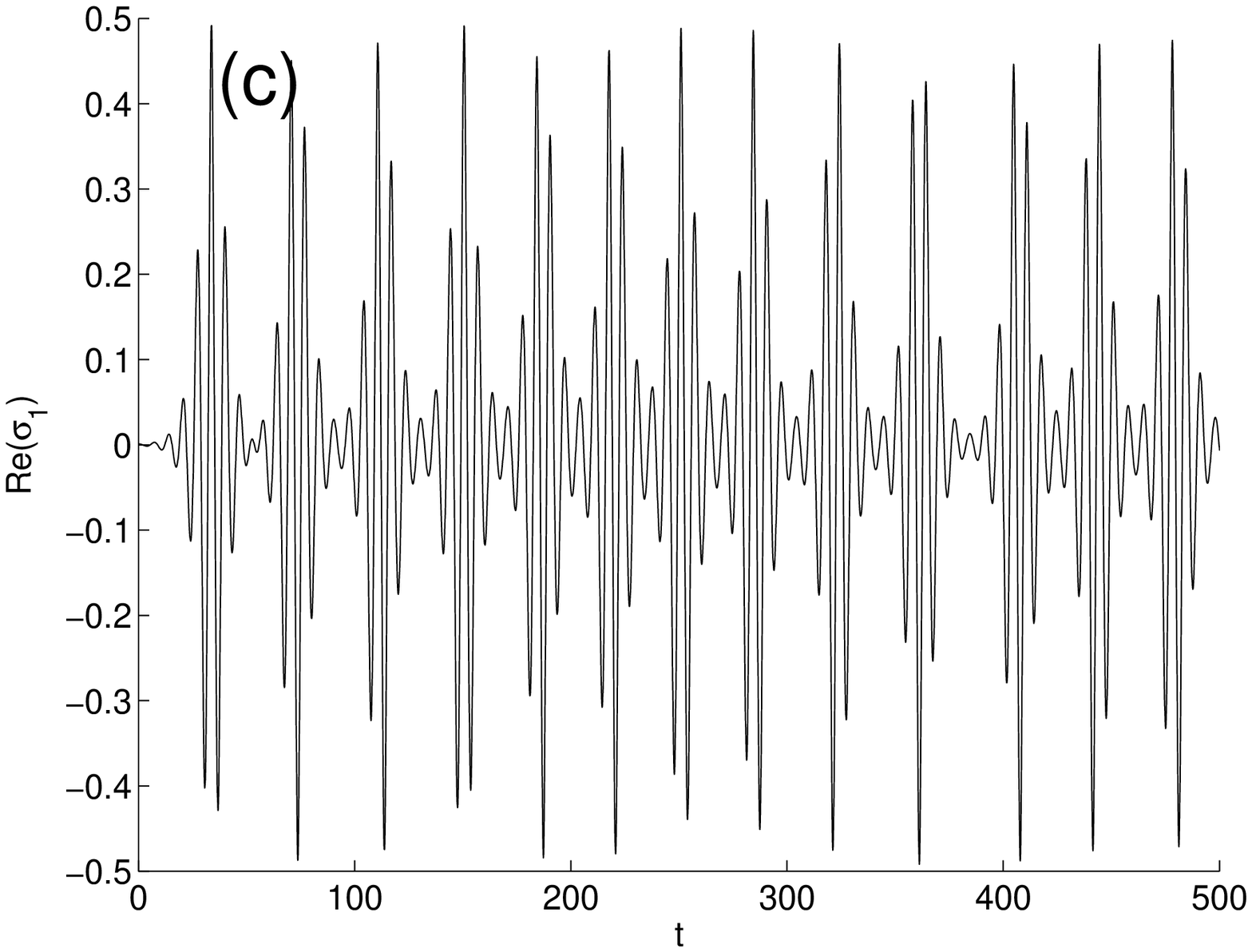}
\includegraphics[width=\mywidth]{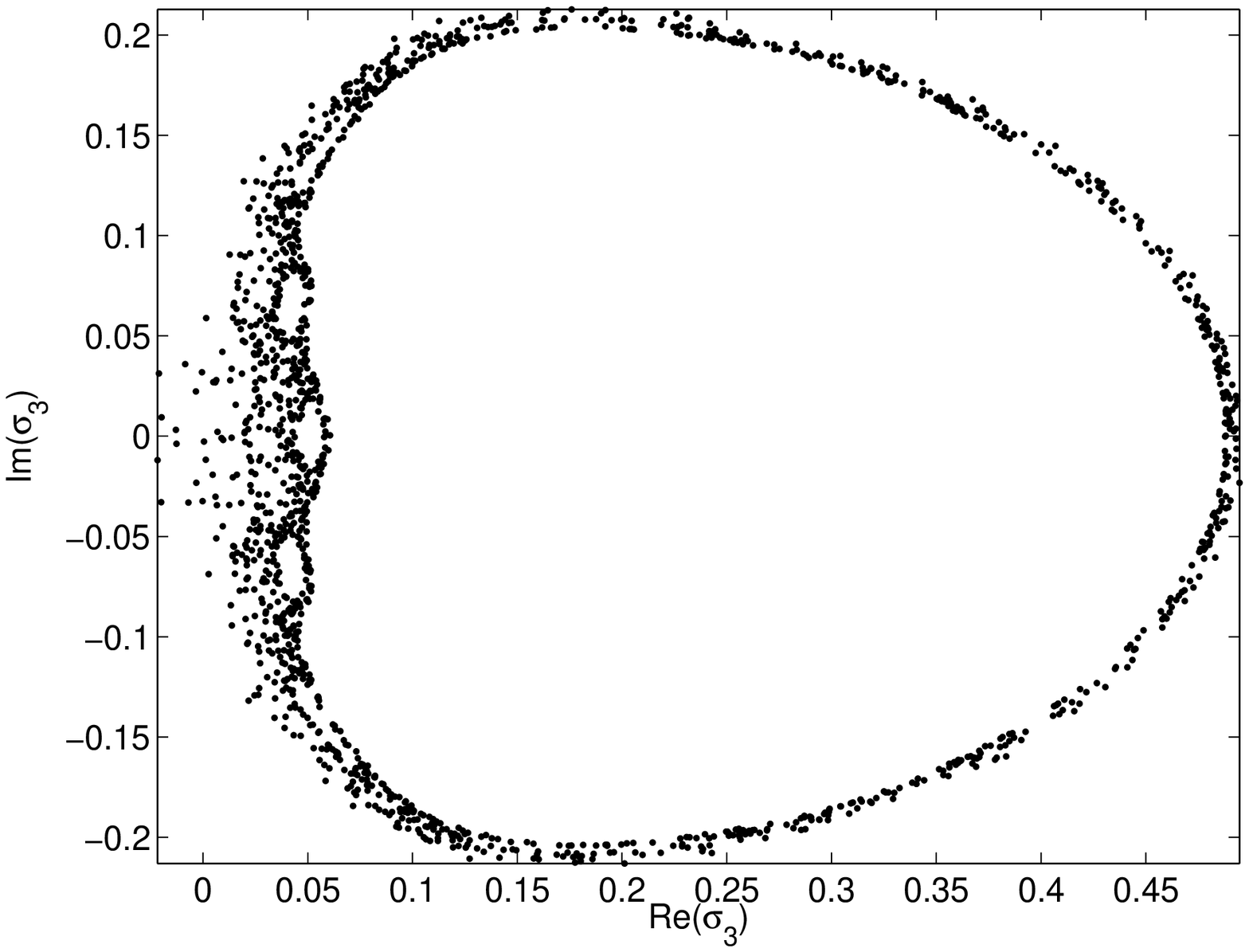}
\includegraphics[width=\mywidth]{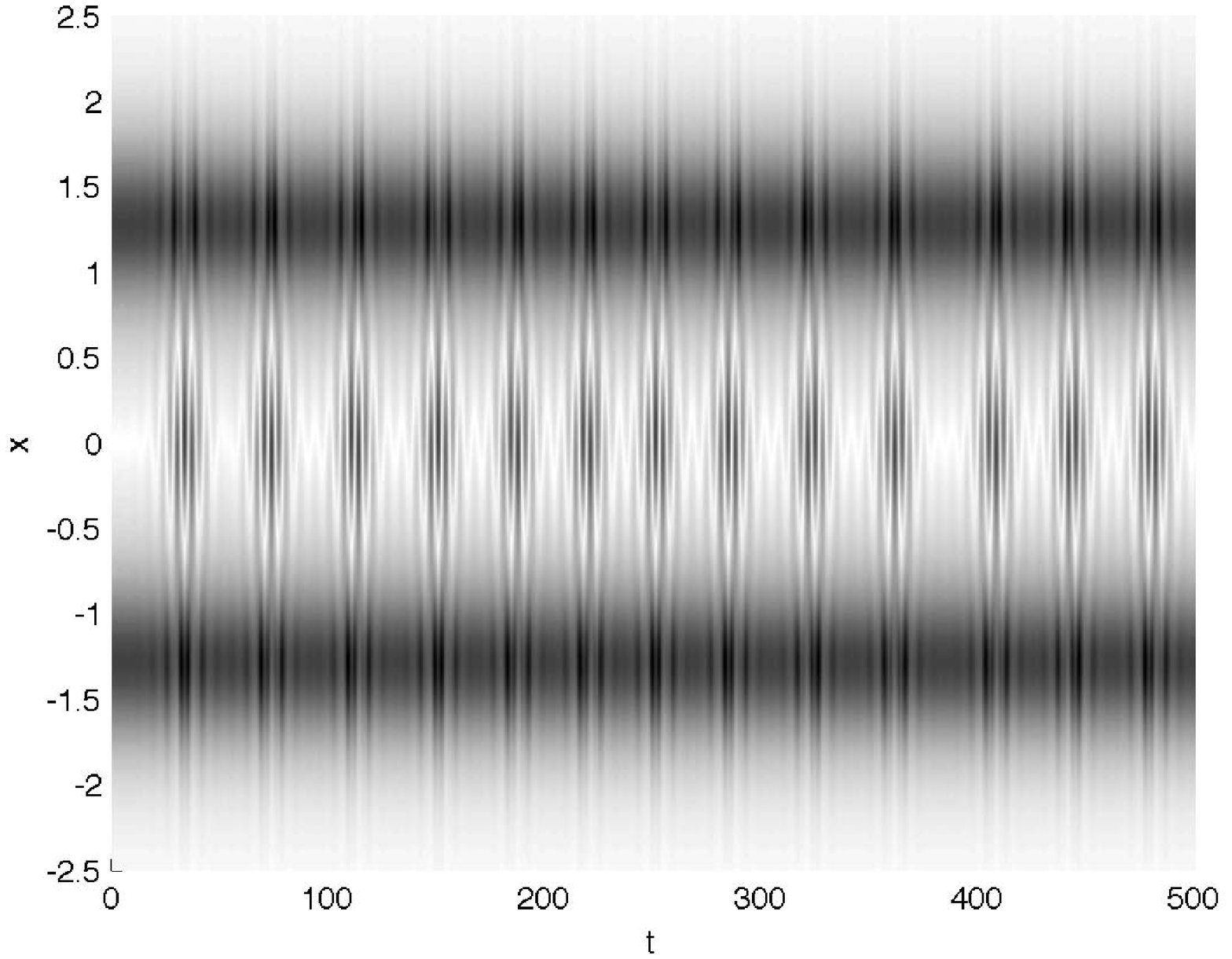}
\includegraphics[width=\mywidth]{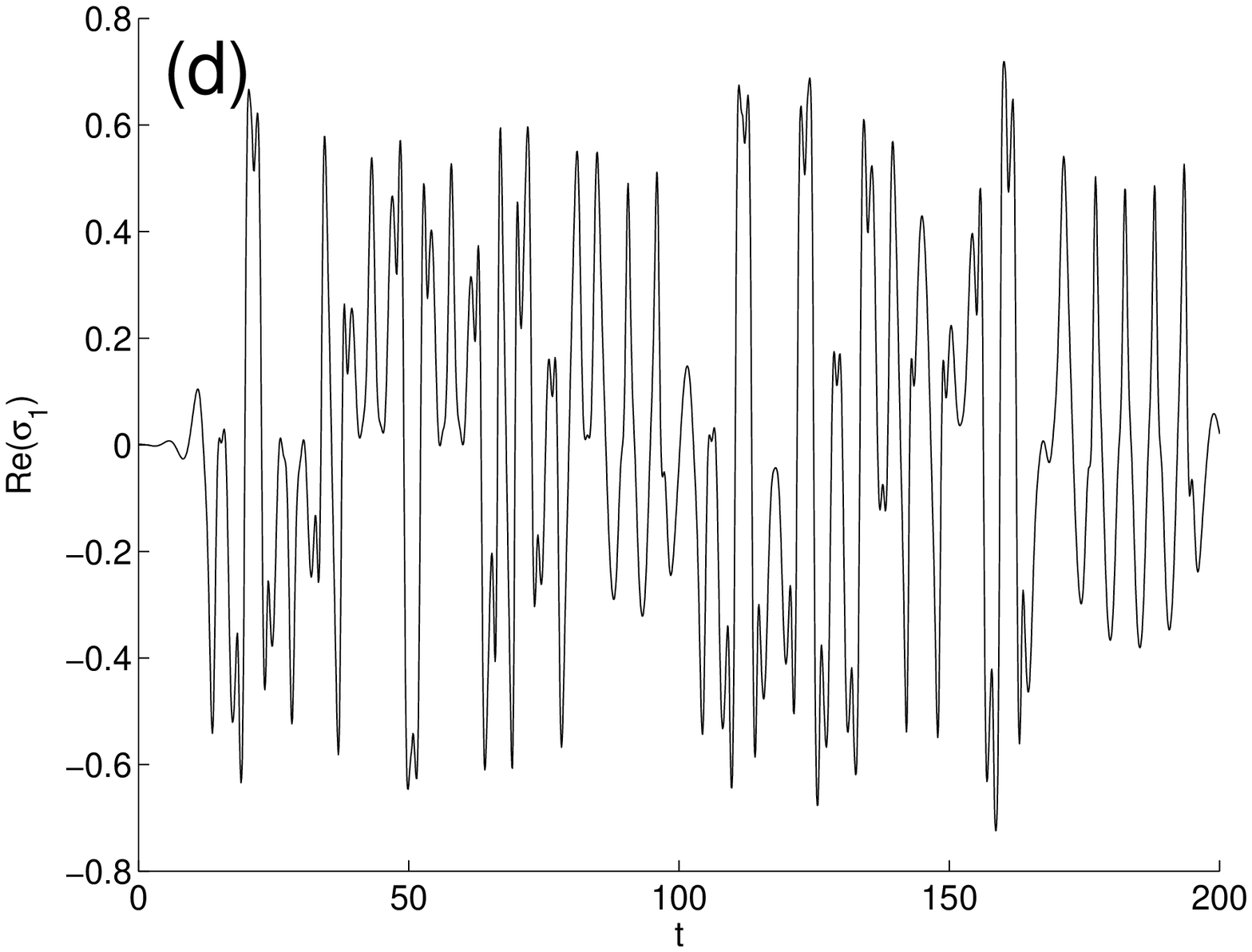}
\includegraphics[width=\mywidth]{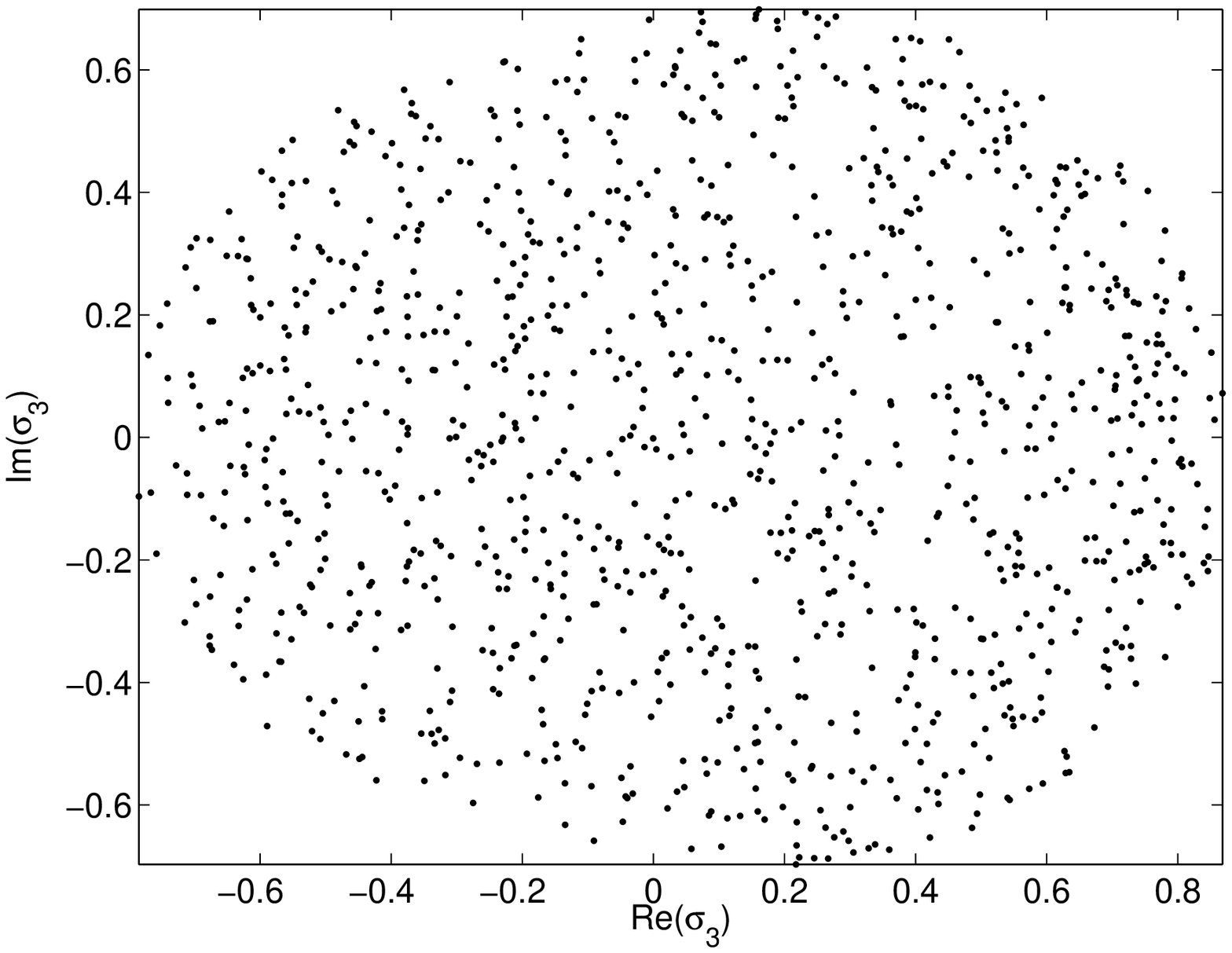}
\includegraphics[width=\mywidth]{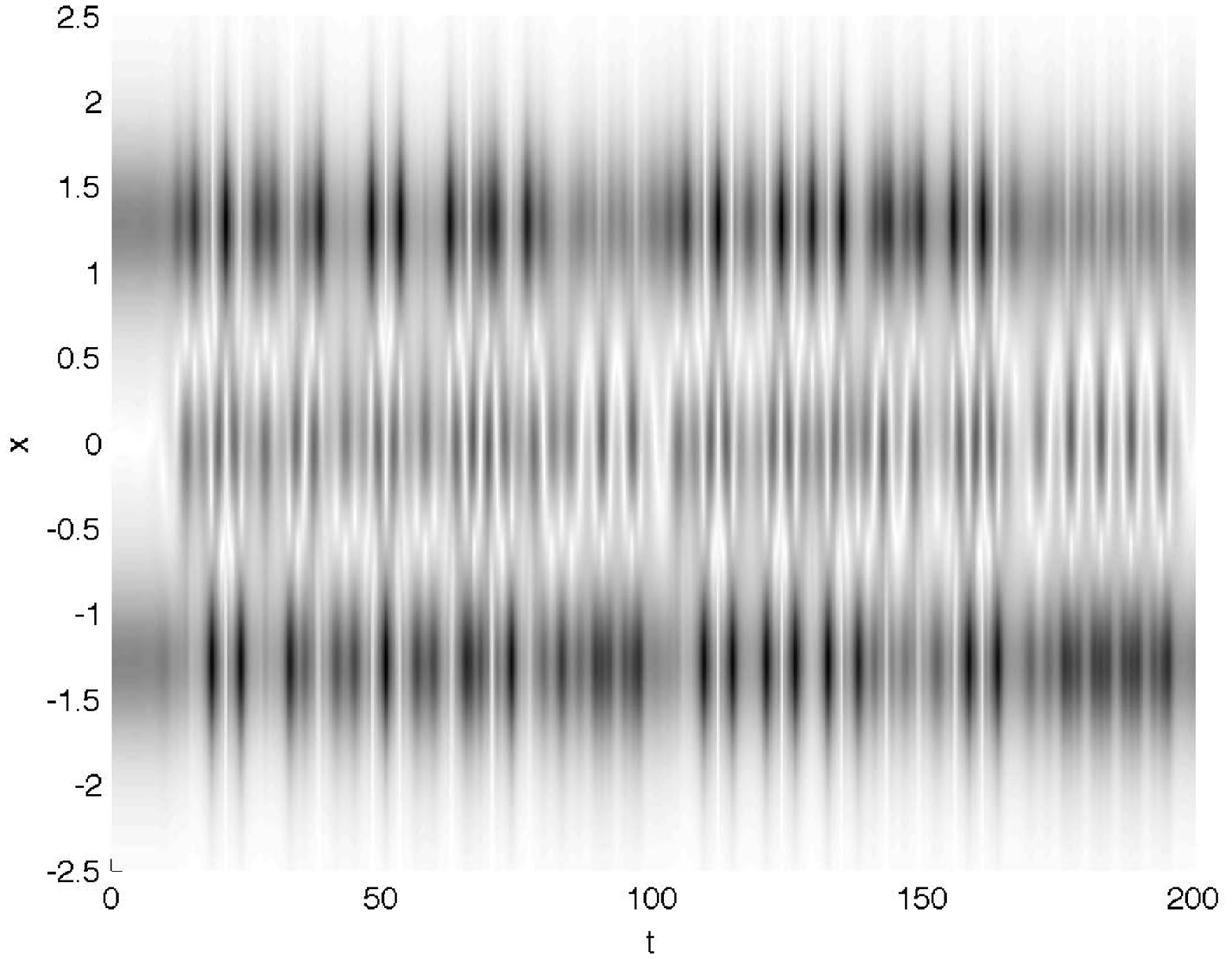}
\includegraphics[width=\mywidth]{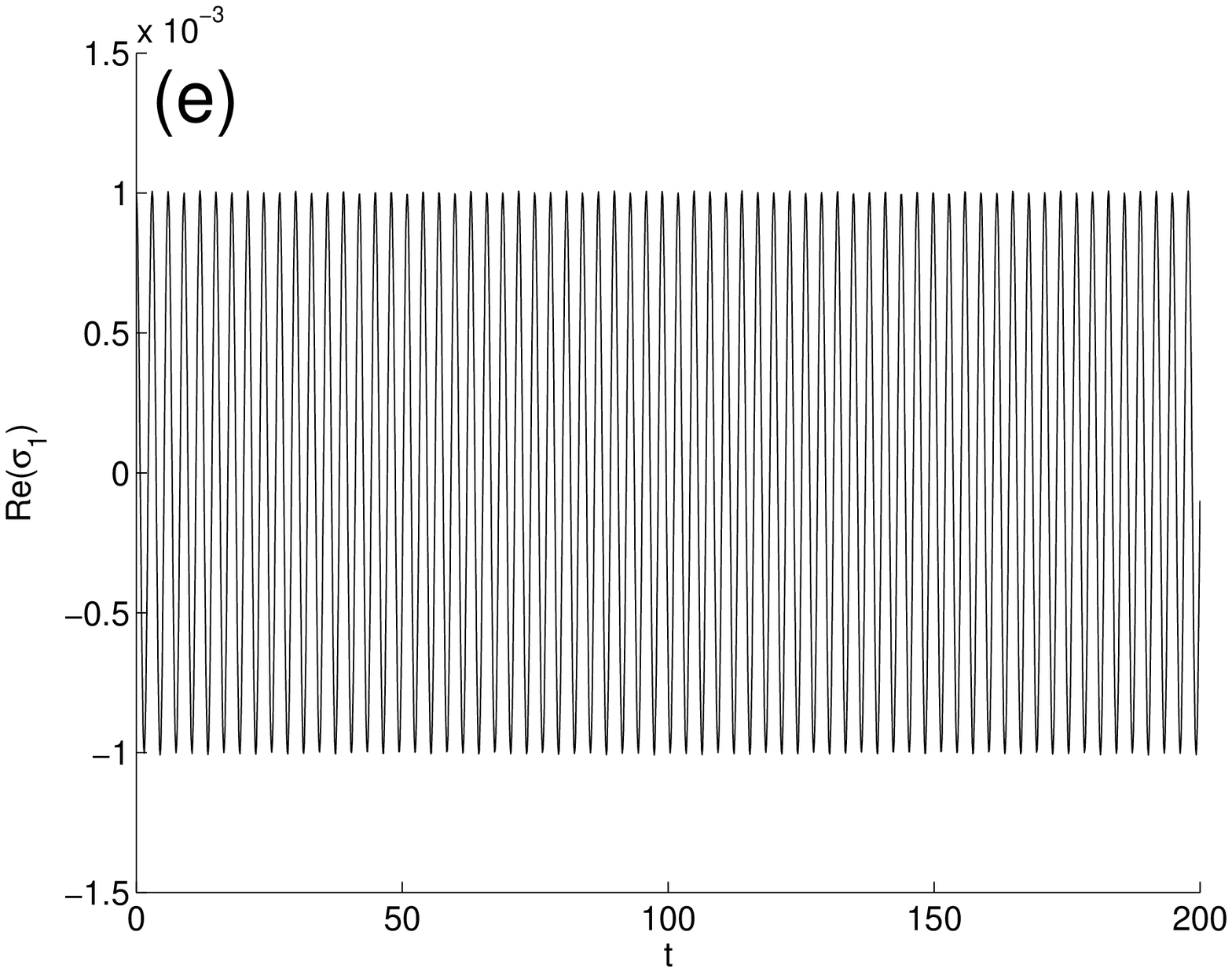}
\includegraphics[width=\mywidth]{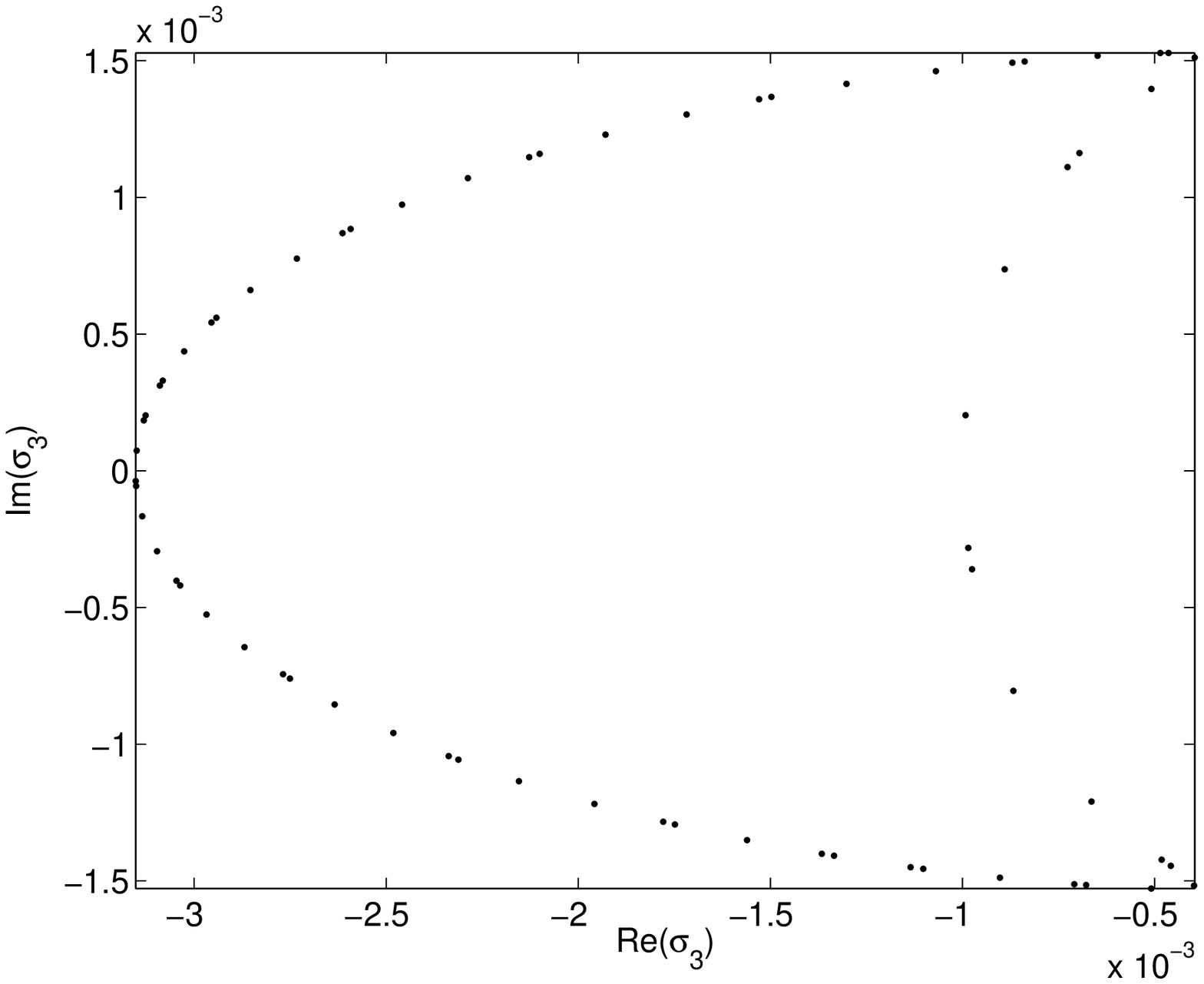}
\includegraphics[width=\mywidth]{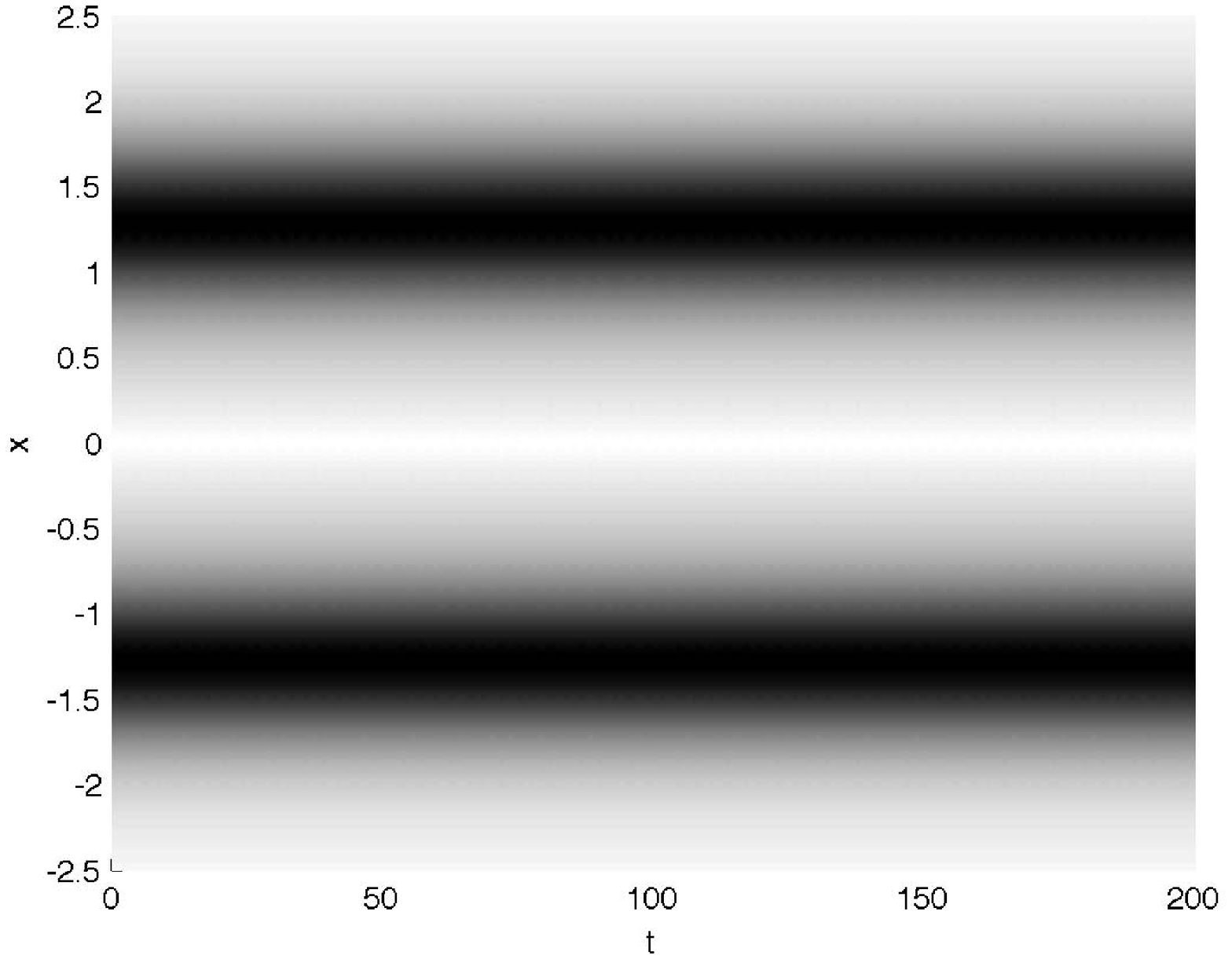}
   \caption{Simulations of ODE system~\eqref{complexform}.  The rows, labeled (a)-(e) correspond to the values of $N$ indicated on figure~\ref{fig:ODE_PDE_Hopf}.  The first column shows $\Re{\s_1}(t)$.  The second shows the intersection of the solution with Poincar\'{e} section $\Sigma_{N,H}$.  The third column shows a reconstruction of $\abs{u(t)}$ using ansatz~\eqref{complexChange} (darker areas indicated larger values).  This shows, as is predicted by figure~\ref{fig:ODE_PDE_Hopf}, that cases (a) and (e) are stable, and that instabilities, and even chaos, exist in the other three cases (chaos, in fact exists in the fifth case as well).  Note in rows (b) and (c) that the Poincar\'{e} map $\cM$ was run to $t=10000$  and $t=5000$ respectively. Also note the elliptical region toward the left in this figure, into which no points of the trajectory enter. For all simulations, the initial condition is $\sigma_1 = \sigma_3 = 10^{-3}$. The pictures for nearby initial conditions are qualitatively the same.}
   \label{fig:ODEsolutions}
\end{figure}

To simulate system~\eqref{complexform}, we use a Hamiltonian Boundary Value Method (HBVM) of Brugnano et al.\ \cite{Brugnano:2009a,Brugnano:2009,Brugnano:2009b}, which exactly conserves the energy in polynomial Hamiltonian systems. 
We consider the potential shown in figure~\ref{potentialModes_9_10_11}, which has eigenvalues $\{-11.1,-10,-9.1\}$.  The numerically calculated value of $\NHH = 0.381$.  We run several numerical experiments, each with initial conditions $\s_1=\s_3=10^{-5}$ and $\s_2 = \r = (N -\s_1^2-\s_3^2)$, where $N \in \{0.35, 0.6, 1, 3, 5.5\}$ corresponding to the five points labeled in figure~\ref{fig:ODE_PDE_Hopf}. These simulations are shown in the five rows of figure~\ref{fig:ODEsolutions}.  The first column contains a time series of $\Re \s_1$.  The second contains a Poincar\'{e} section, defined on level sets of the amplitude $N$ and the Hamiltonian $H$, by
$$
\Sigma_{N,H}= \{(\s_1,\s_3) | \Im(\s_1) = 0\ \&\ \frac{d}{dt} \Im(\s_1) >0 \}.
$$

Because the values of $N$, $H$, and $\s_3$ on this section uniquely determine $\s_1 \in \RR$, $\Sigma_{N,H}$ is parameterized by the value of $\s_3$.  We denote the mapping from one crossing of $\Sigma_{N,H}$ to the next as $\cM$.
The third column contains a reconstruction of the field amplitude $\abs{\psi(x)}$ computed from~\eqref{decomposition} and~\eqref{complexChange}

In case (a), $N = 0.35$, the solution oscillates quasiperiodically in a neighborhood of the initial condition, with the amplitude of oscillation depending on, and remaining close to, the initial condition.  A Poincar\'{e} section, as defined above, shows these solutions appear to be quasiperiodic; see column two of the figure. In case (b), $N=0.6$, the solution makes large excursions from the initial condition, lying close to an apparent homoclinic orbit.  In the reconstructed field, column 3, we see that these bursts consist of an oscillatory growth of the field in the middle of the potential.   We will discuss the exact nature of this solution in section~\ref{sec:numerics}. For case (c), $N=1.0$, the solution appears ``weakly chaotic,'' in that it  still takes the form of large heteroclinic bursts, only now the time between bursts is irregular.  The chaos is evident from the Poincar\'{e} section.  In case (d), $N=2.25$, the chaos is fully developed, and the trajectory in  Poincar\'{e} section appears to cover a large open set.  This open set avoids, however, an elliptical region toward the left half of the figure, on which the map has a fixed point of elliptical type.  Finally, in case (e) with $N=5.5$, we see that for $N$ sufficiently large, the trivial solution is again stable. 

\begin{figure}[htbp] 
   \centering
   \includegraphics[height=2in]{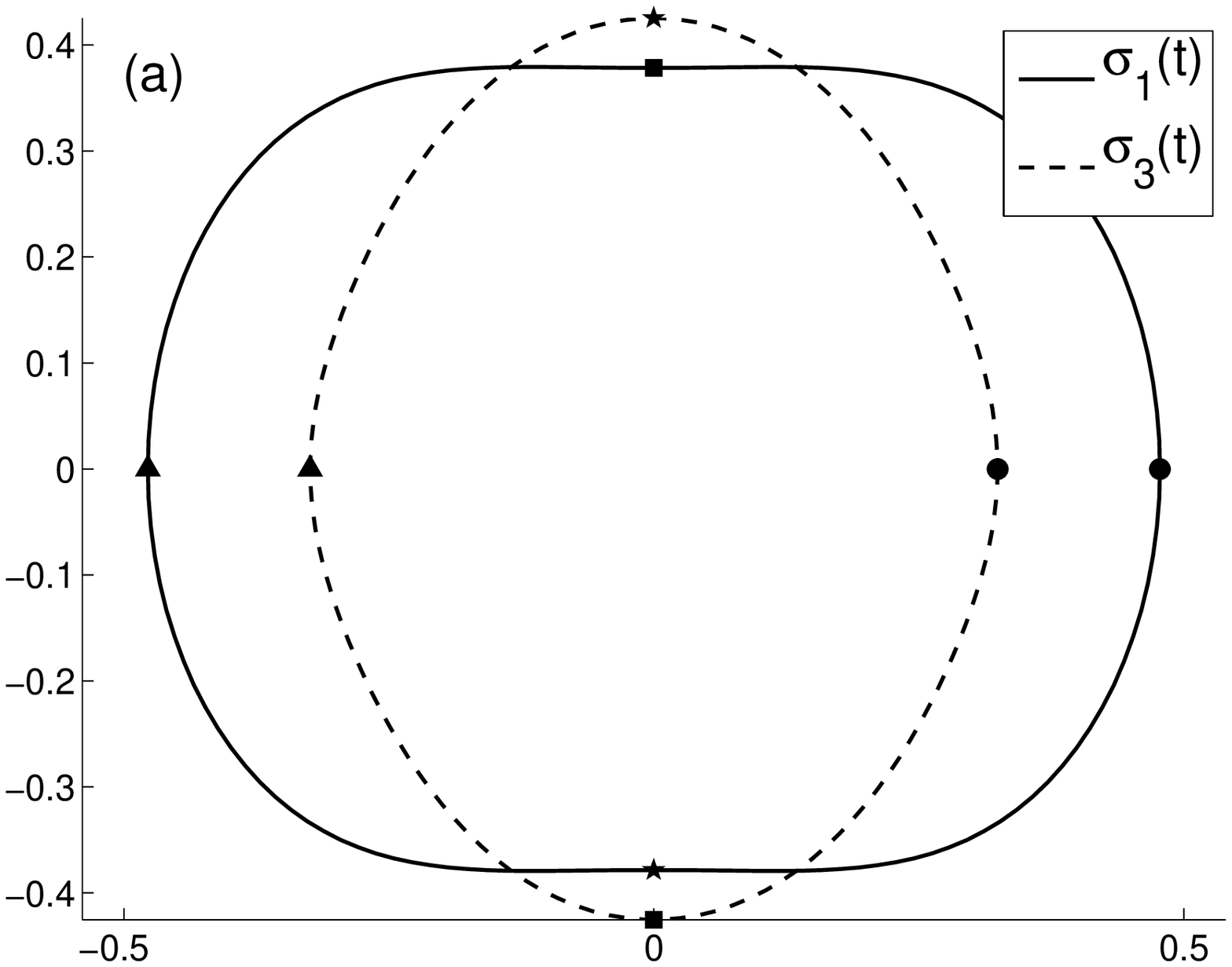} 
   \includegraphics[height=2in]{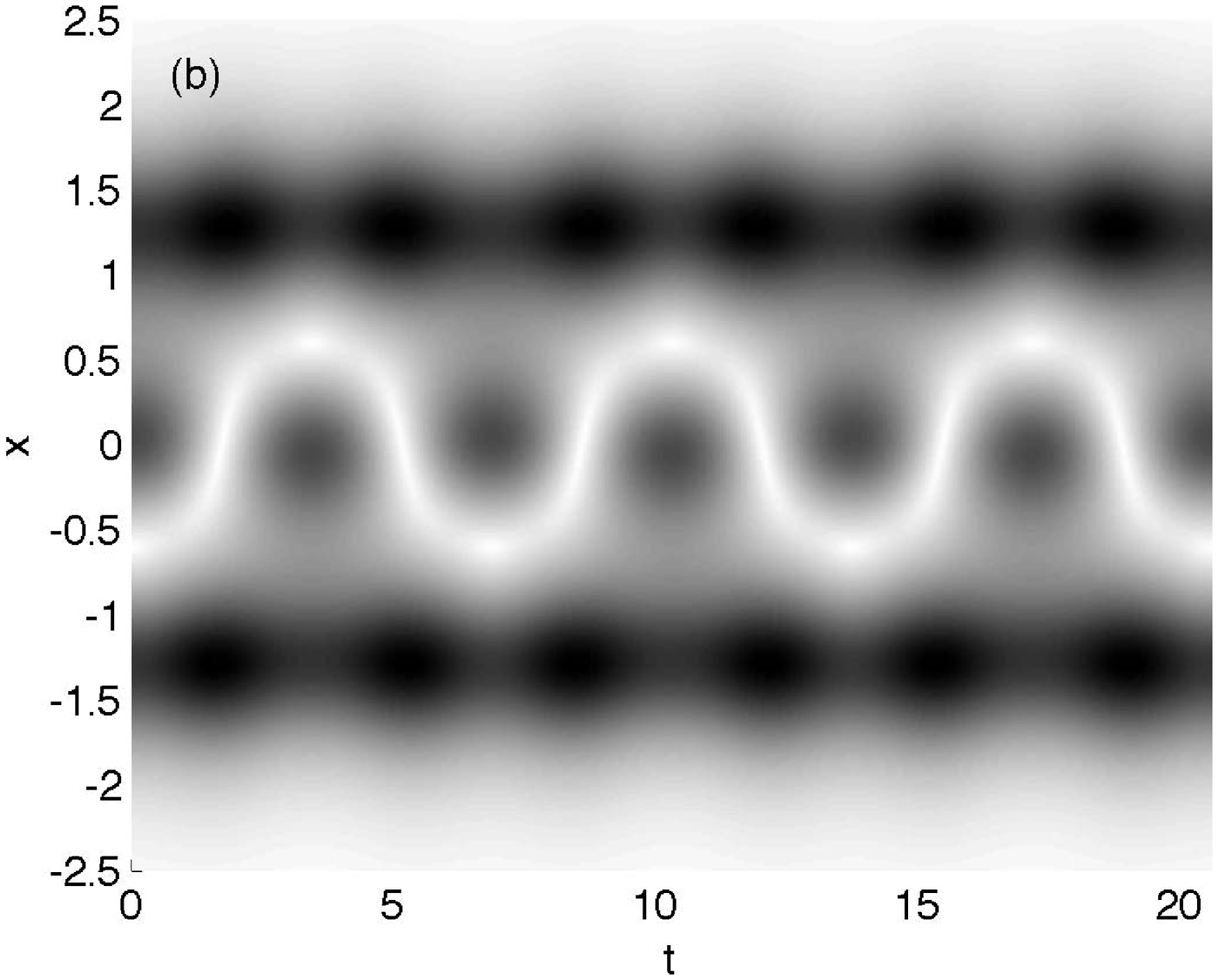} 
   \includegraphics[height=2in]{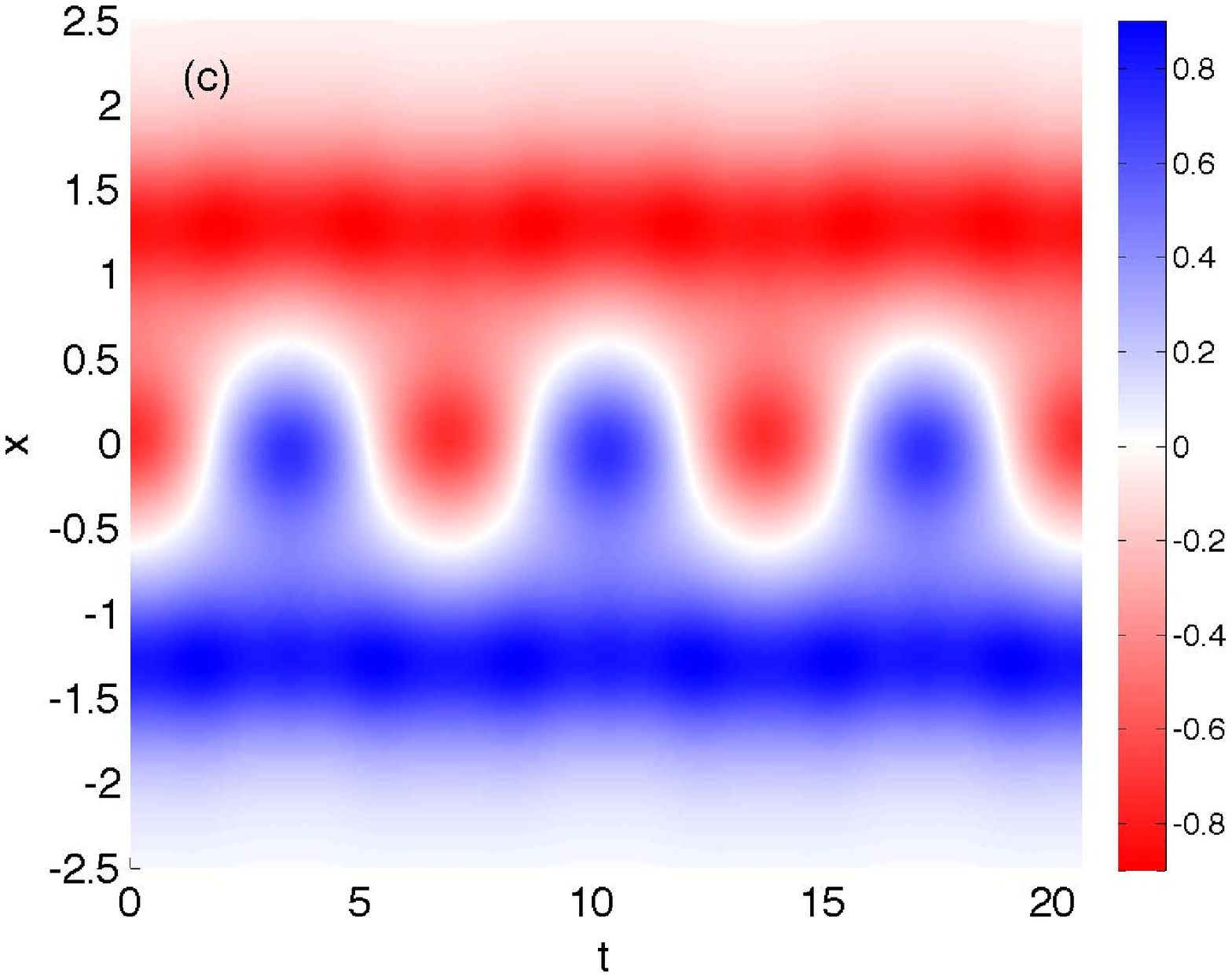} 
   \includegraphics[height=2in]{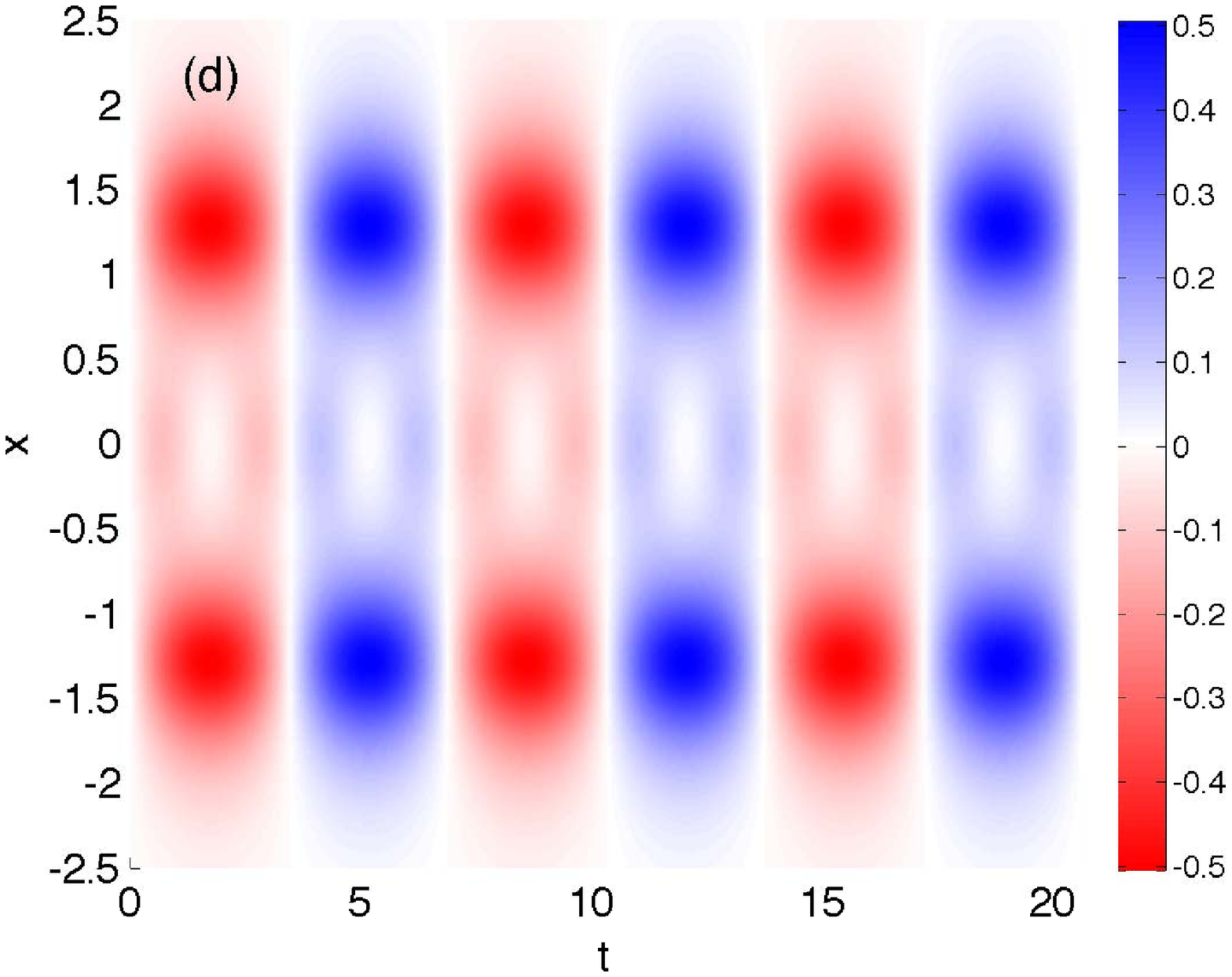} 
   \caption{The periodic orbit of the averaged system with parameters as in figure~\ref{fig:nCritical} with $N=2>\NHH$.  (a) $\s_1(t)$ and $\s_3(t)$.  At $t=0$, the orbits start at the points marked $\bullet$, and proceed at quarter-periods through the points marked $\blacksquare$, $\blacktriangle$, and $\bigstar$.  A reconstruction of the PDE field over three periods of oscillation: (b) absolute value,(c) real part (mod $e^{i\th(t)}$), (d) imaginary part (mod $e^{i\th(t)}$).}
   \label{fig:reconstructedPeriodic}
\end{figure}

At the bifurcation amplitude $\NHH$, a new fixed point  $\s_{\rm p}$ of the Poincar\'{e} map $\cM$ appears at a distance $\Or{(\sqrt{N-\NHH})}$ from zero, corresponding to a new periodic orbit of the two-degree-of-freedom system~\eqref{complexform}, i.e.\  a new relative periodic orbit of  system of the full three-degree-of-freedom system defined by the Hamiltonian~\eqref{Hc}.  
The fixed point $\s_{\rm p}$ appears on the symmetry axis $\s_3 \in \RR$.
A complete periodic solution to reduced system~\eqref{complexform} with $N=2$, and a reconstruction of a PDE solution from this ODE solution are shown in figure~\ref{fig:reconstructedPeriodic}.  In subfigure (a), we see that when $\sigma_1$ and $\sigma_3$ are purely real, they are in phase.  From figure~\ref{potentialModes_9_10_11}, we see that in this case, the modes $\Psi_1$ and $\Psi_3$ add constructively in the middle well and destructively on the two outer wells.  When the 
$\sigma_1$ and $\sigma_3$ are purely imaginary, they are $180^\circ$ out of phase, so that $\Psi_1$ and $\Psi_3$ add destructively in the middle well and constructively on the two outer wells.  

As the system reaches the second HH bifurcation at $N\approx 4.71$ where the trivial solution regains stability, the new periodic orbit does not disappear, nor does the chaotic motion shown in row (d) of figure~\ref{fig:ODEsolutions}.
Instead, a small region (in fact, a KAM island) around the origin appears at this amplitude, on which the solution is regular (quasiperiodic and confined to topological ellipses), and this region grows as $N$ is further increased.  This is confirmed by numerical simulation.
Johansson finds similar Hamiltonian chaos when the parameters in his NLS trimer are in the unstable domain, as well as KAM islands~\cite{Johansson:2004}

\subsection*{PDE dynamics}
For comparison, we compute time-dependent solutions of the PDE system.   For this we use a Matlab code written by T. Dohnal.  It uses fourth-order centered differences to compute spatial derivatives, and an implicit-explicit additive Runge-Kutta method for time stepping~\cite{Kennedy:2003} and most importantly for long-term simulation, uses perfectly matched layers (PML) to handle the outgoing radiation~\cite{Dohnal:2007}.

\begin{figure}[htbp] 
   \centering
   \includegraphics[width=\mywidth]{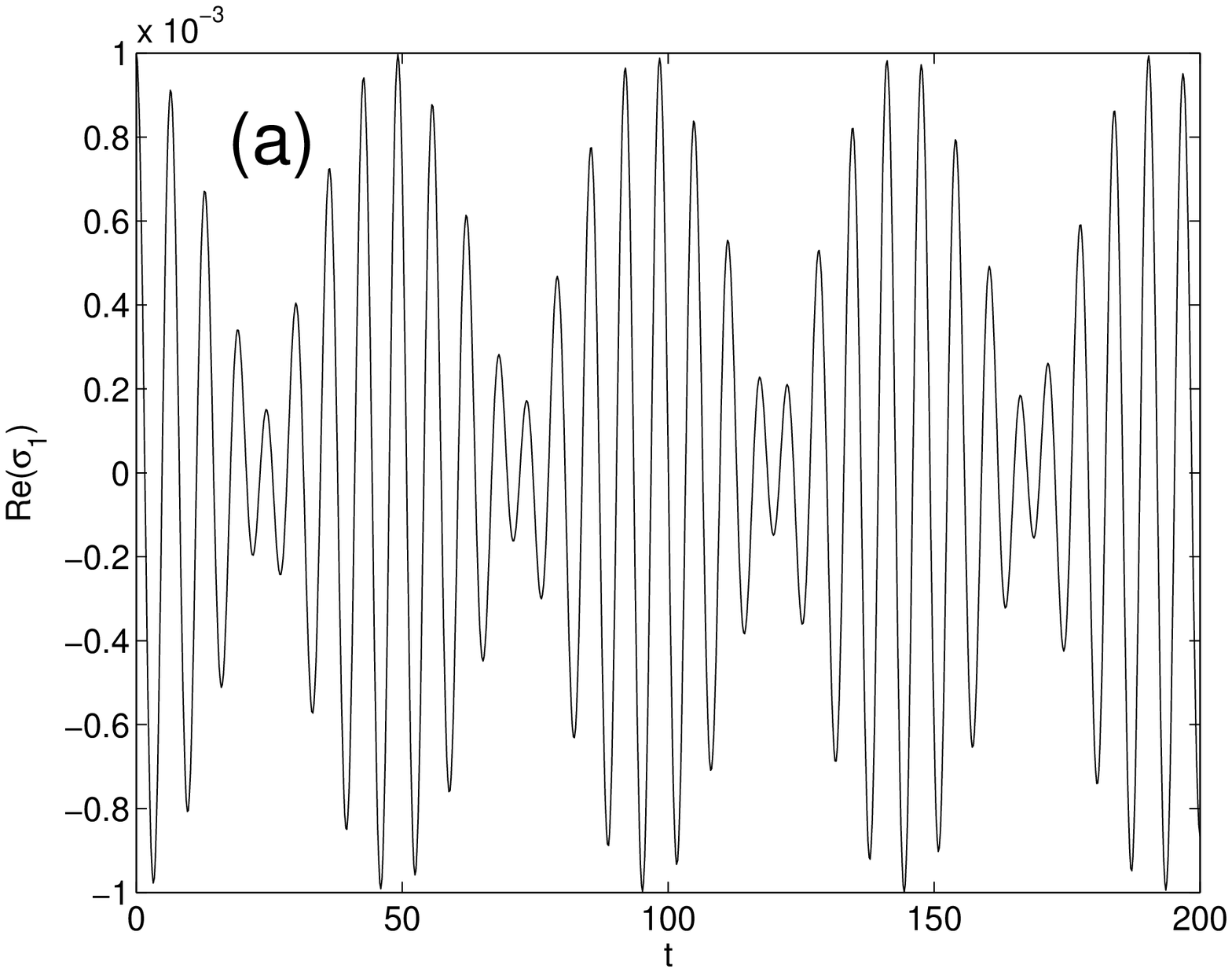}
\includegraphics[width=\mywidth]{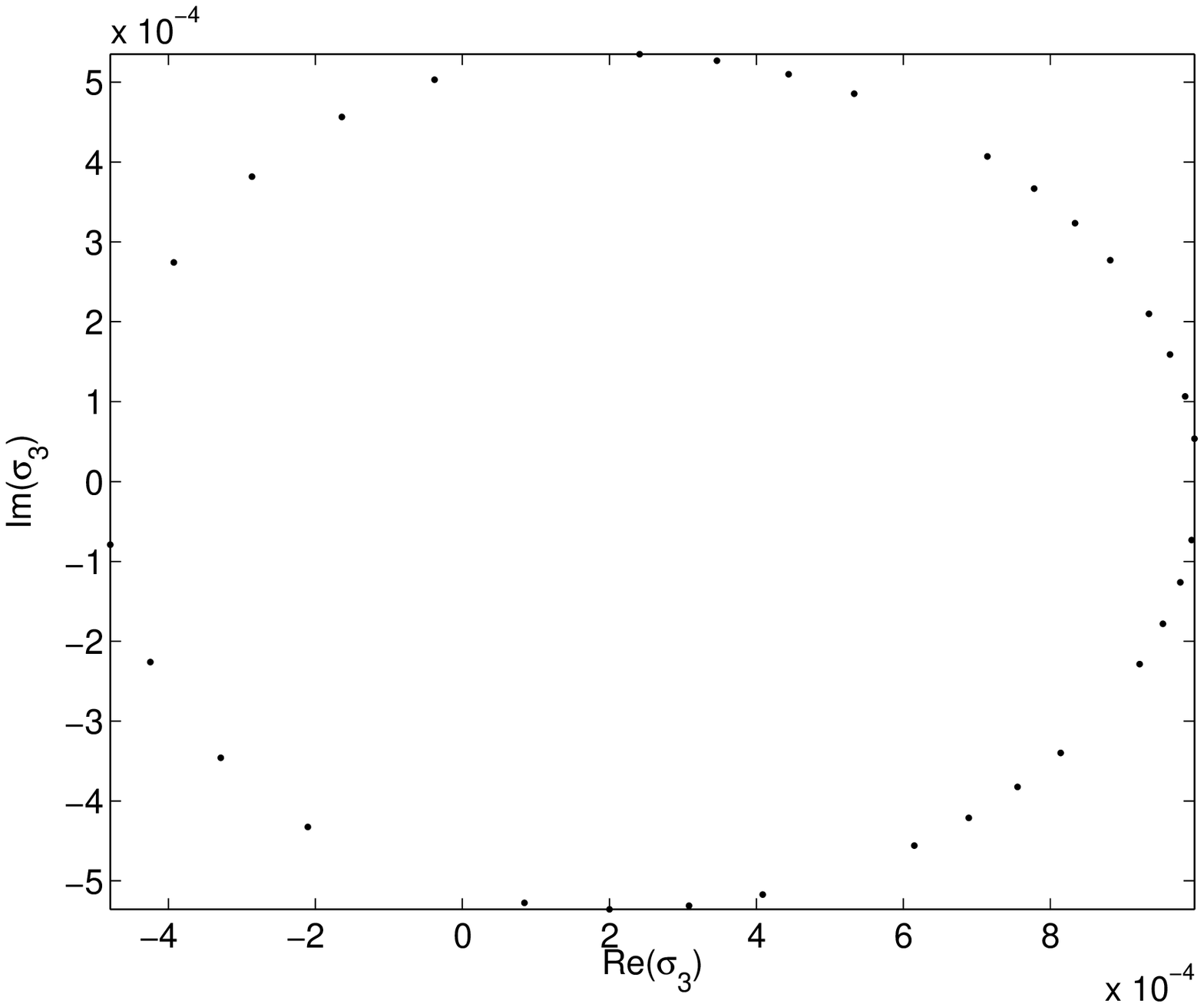}
\includegraphics[width=\mywidth]{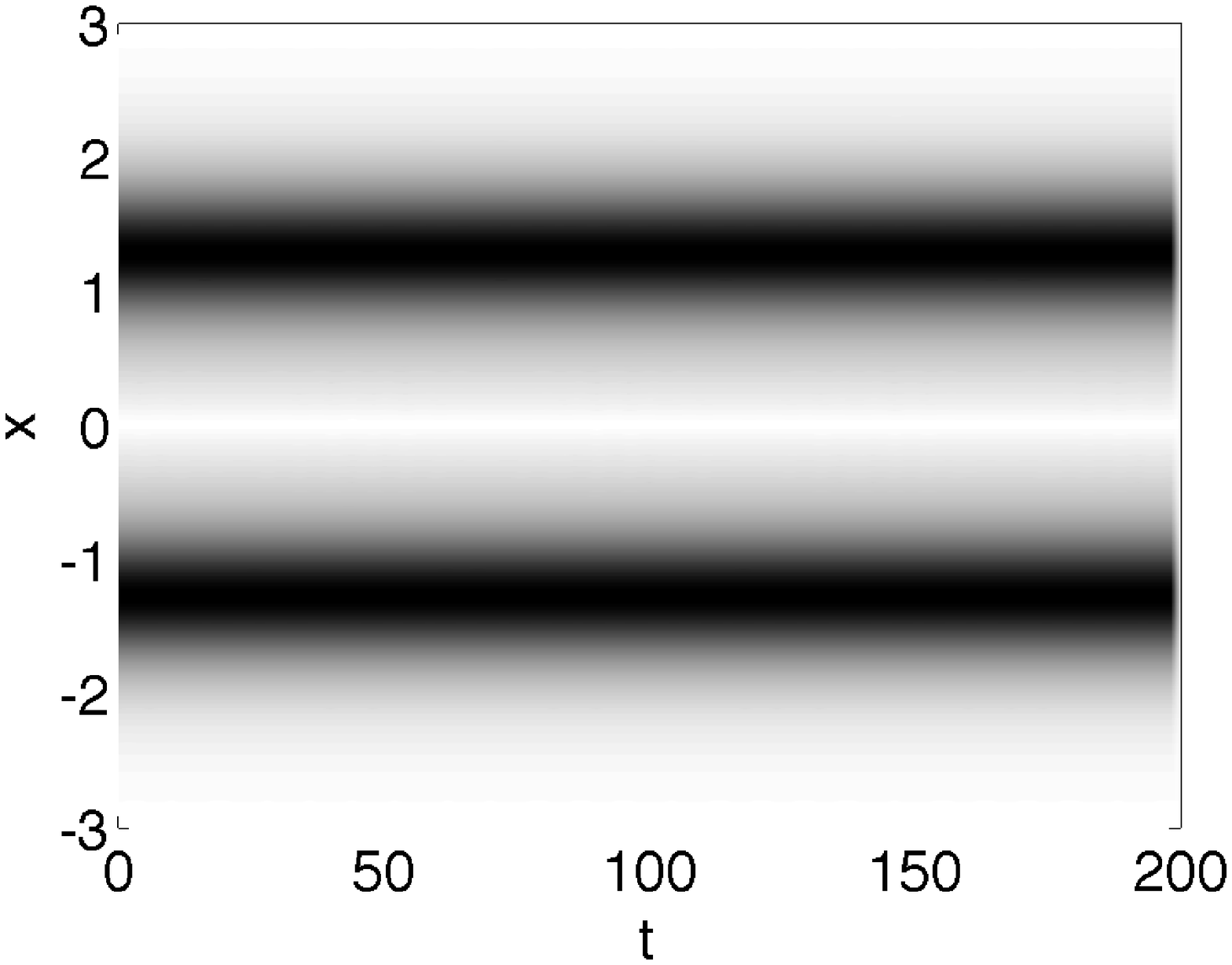}
\includegraphics[width=\mywidth]{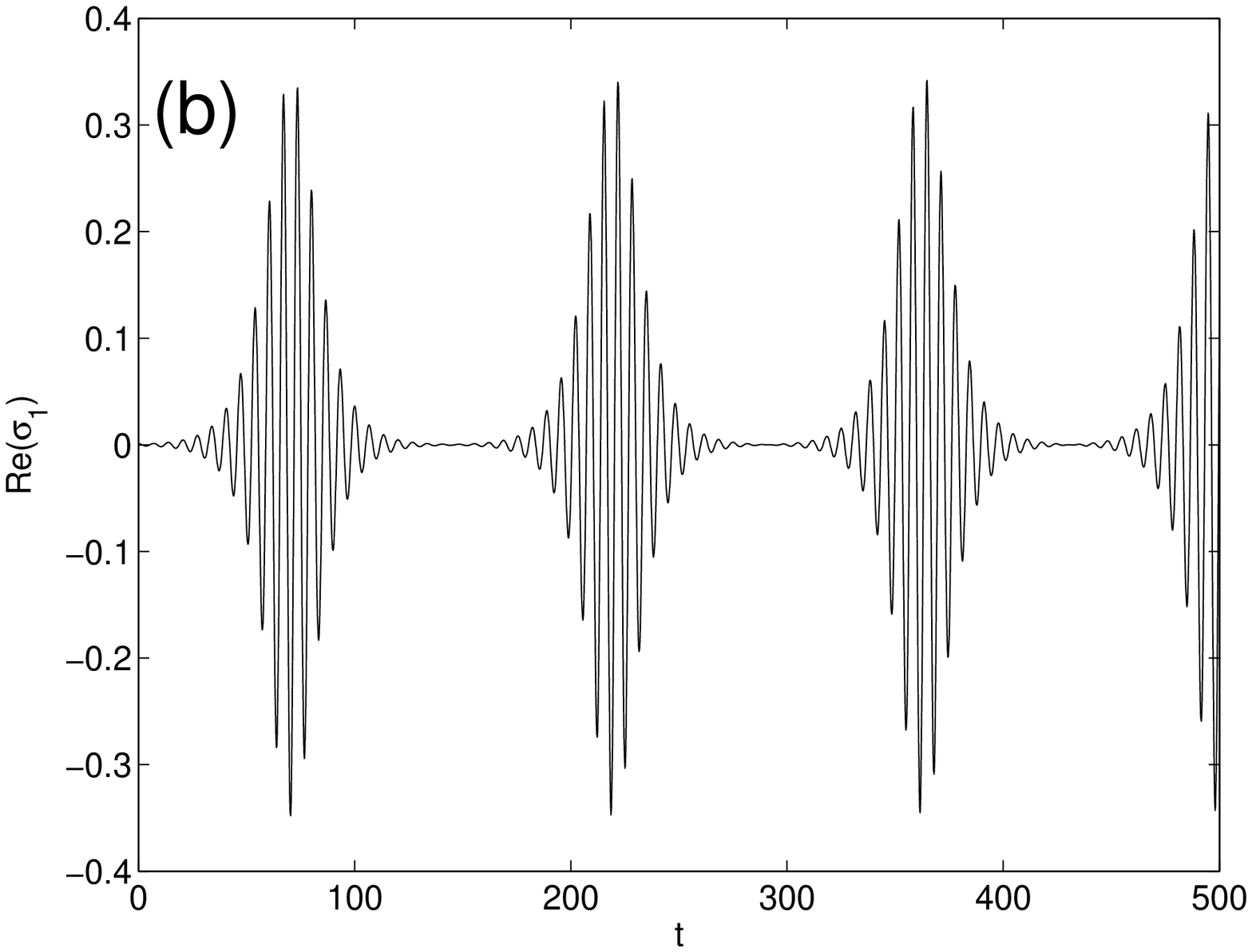}
\includegraphics[width=\mywidth]{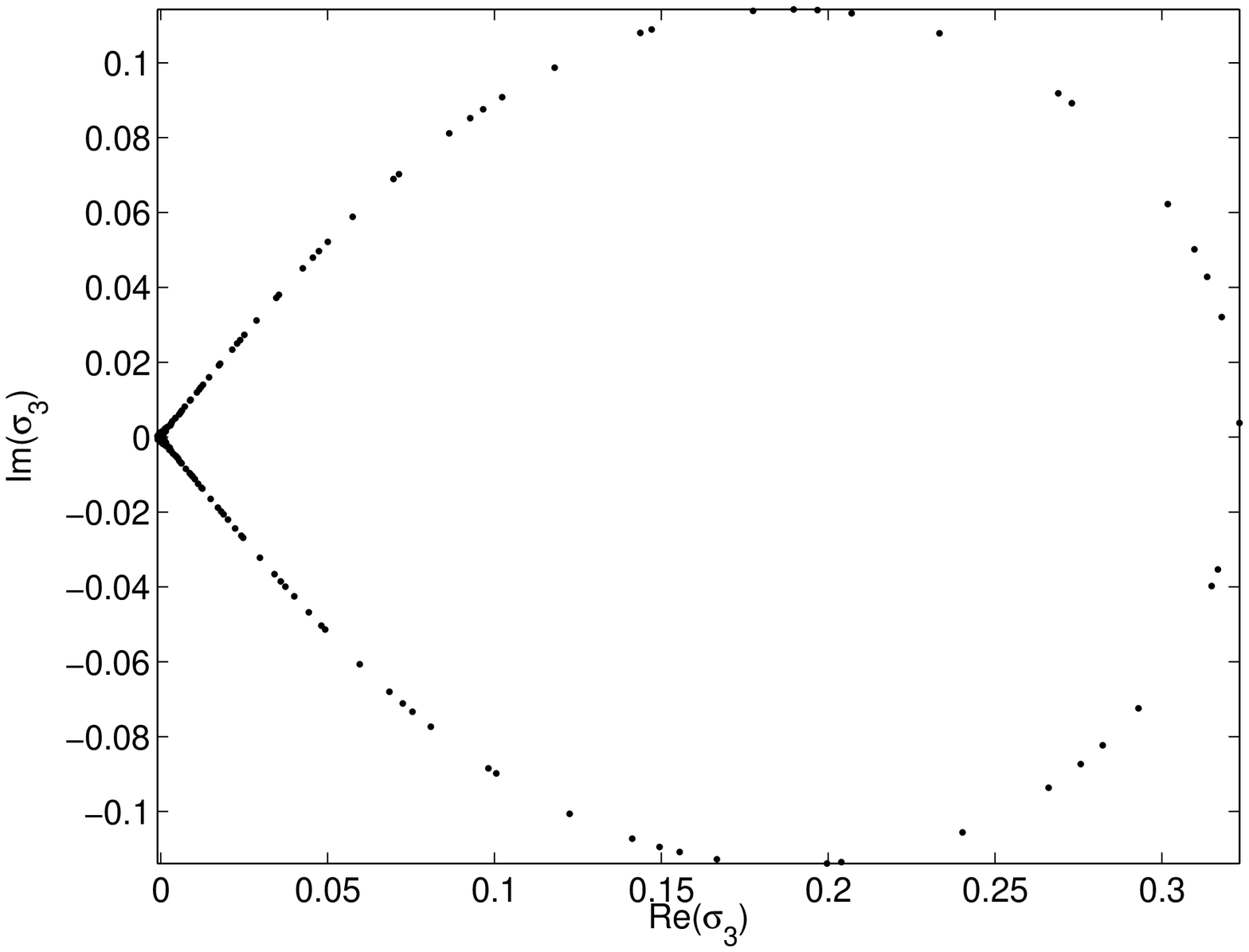}
\includegraphics[width=\mywidth]{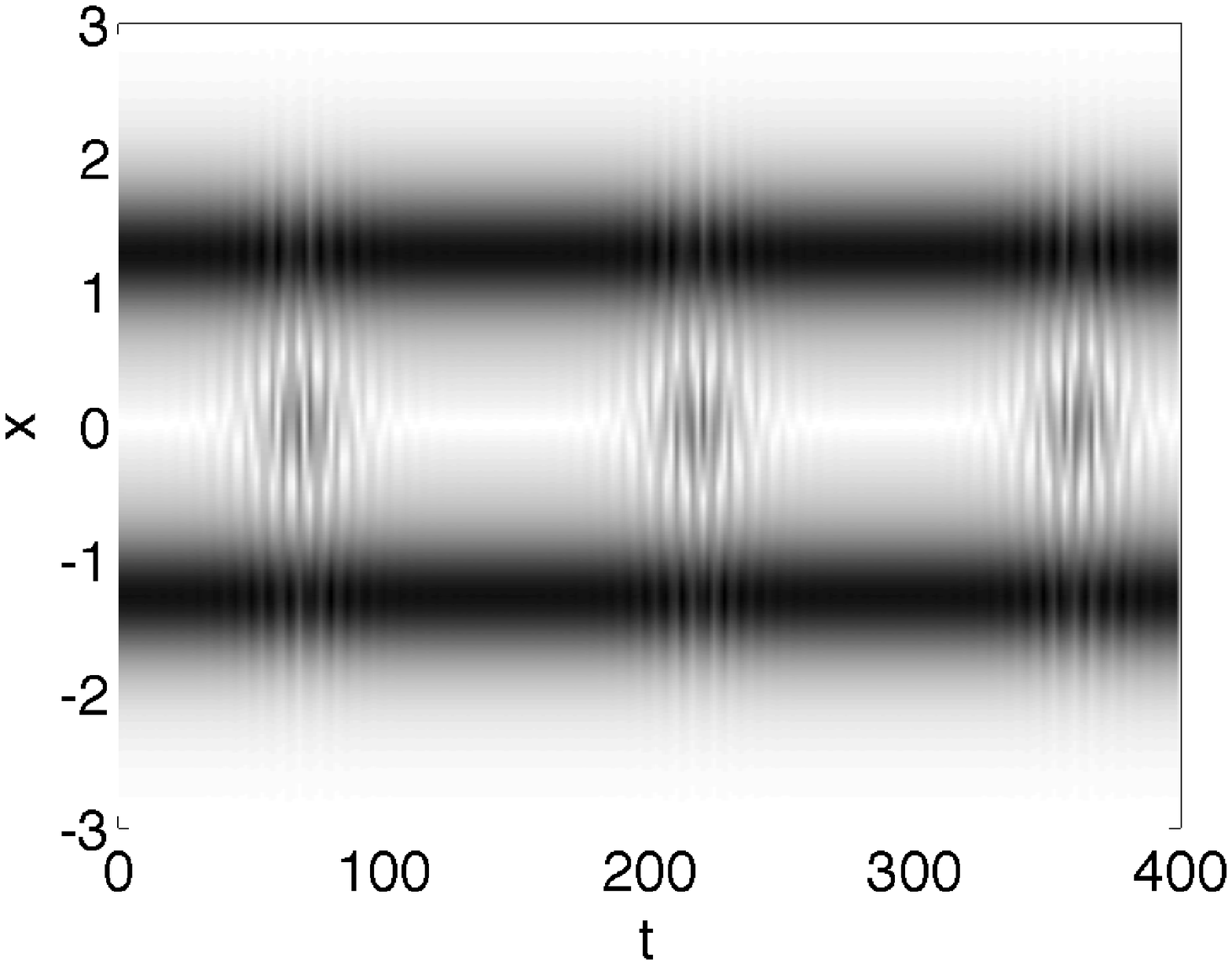}
\includegraphics[width=\mywidth]{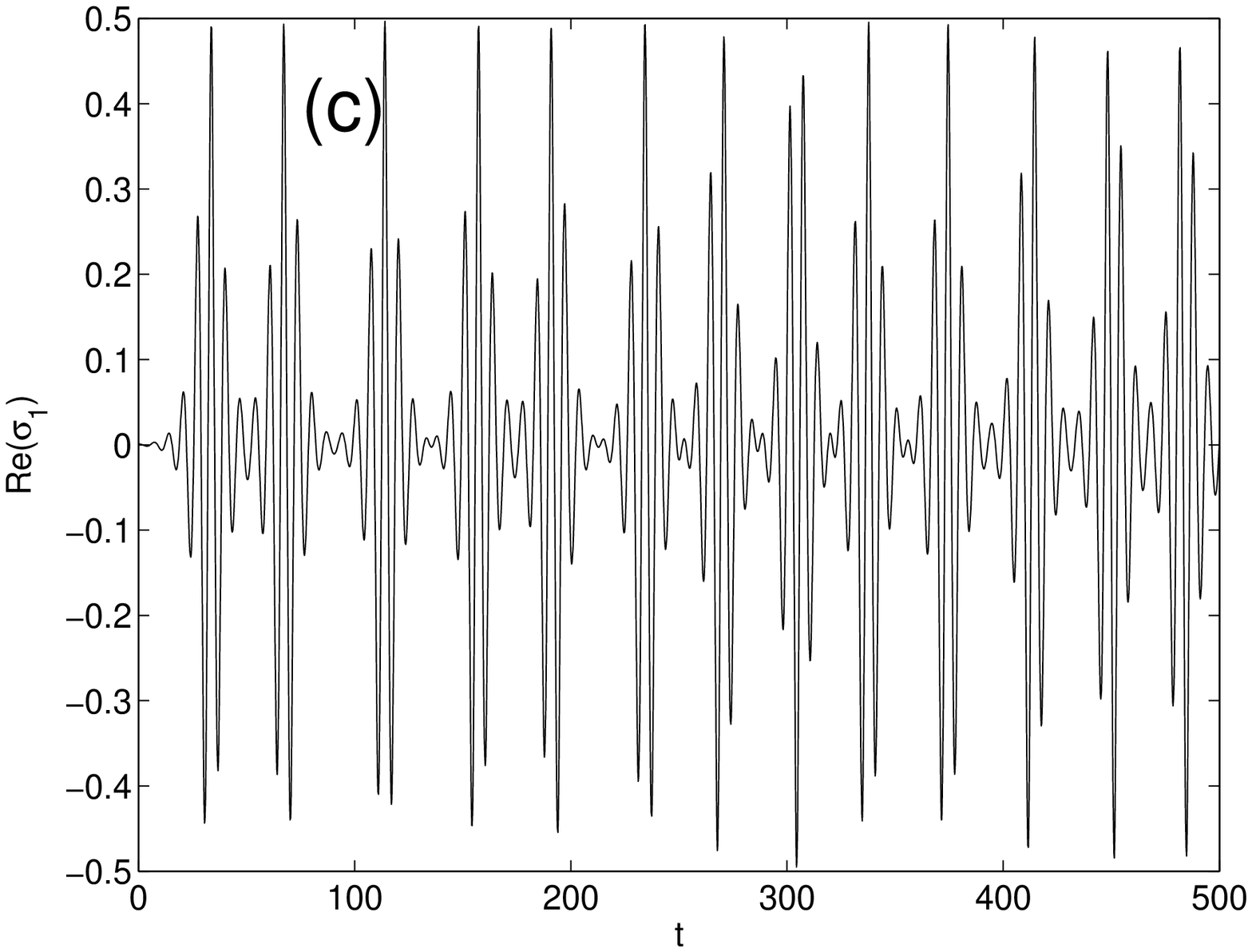}
\includegraphics[width=\mywidth]{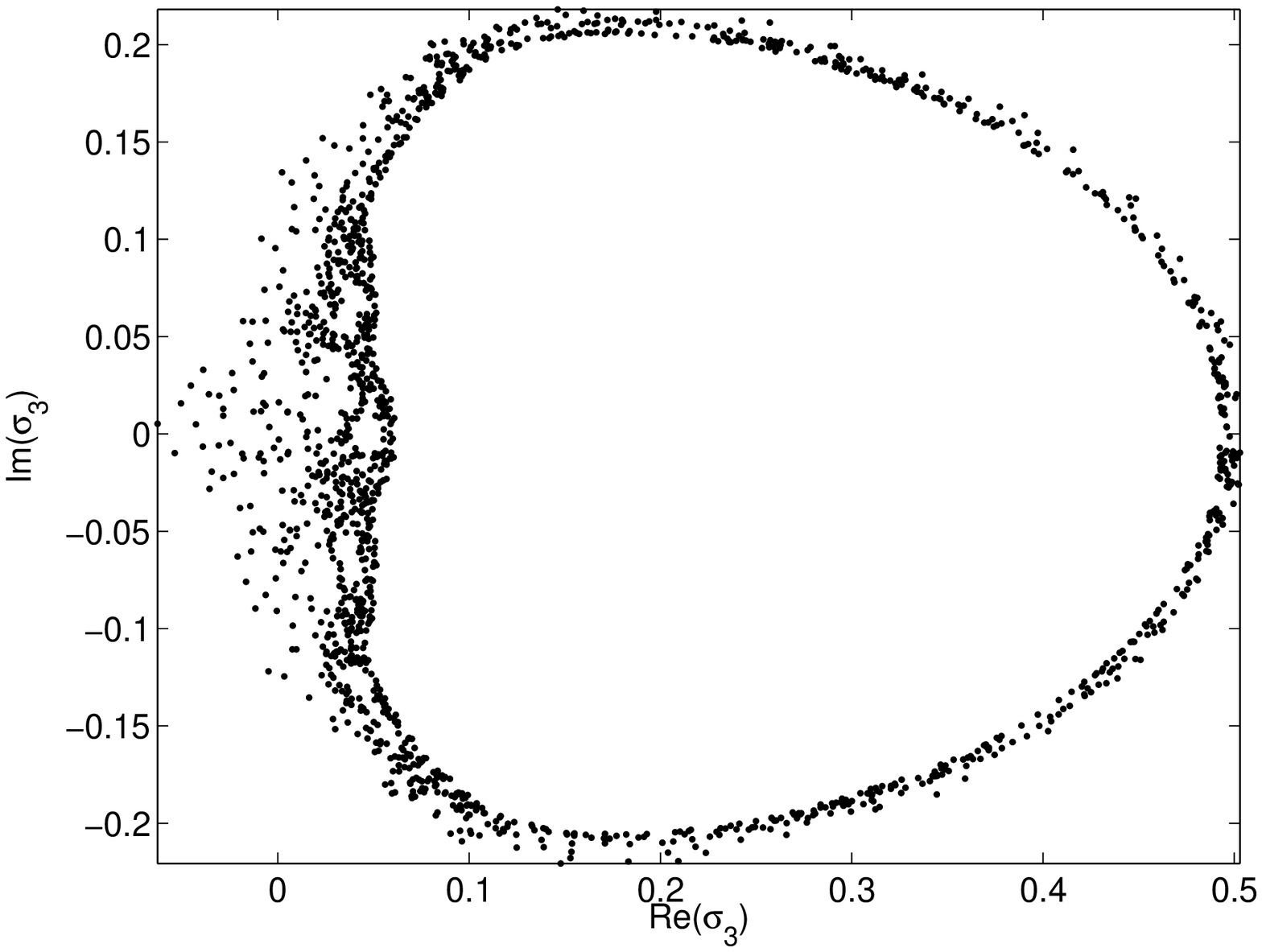}
\includegraphics[width=\mywidth]{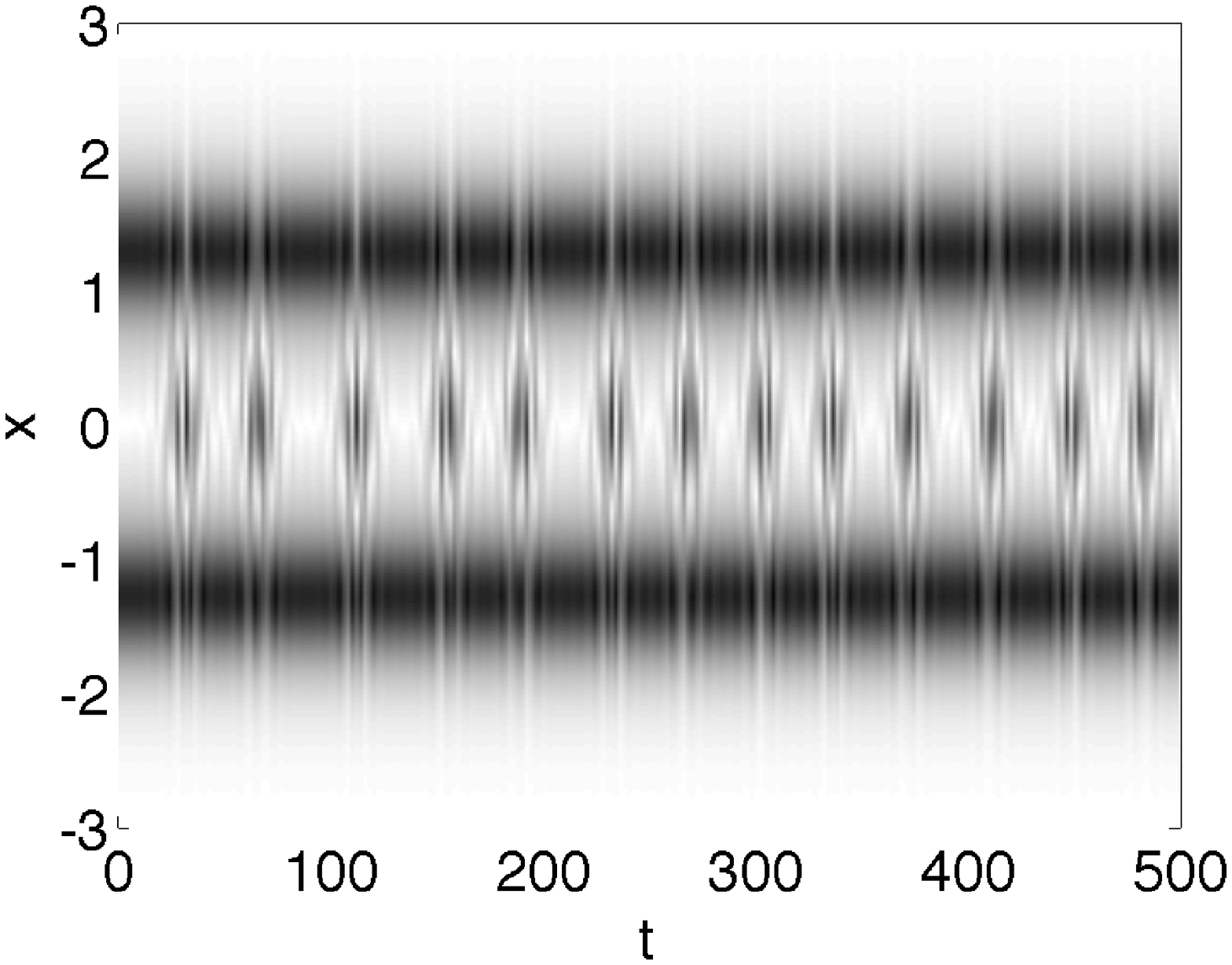}
\includegraphics[width=\mywidth]{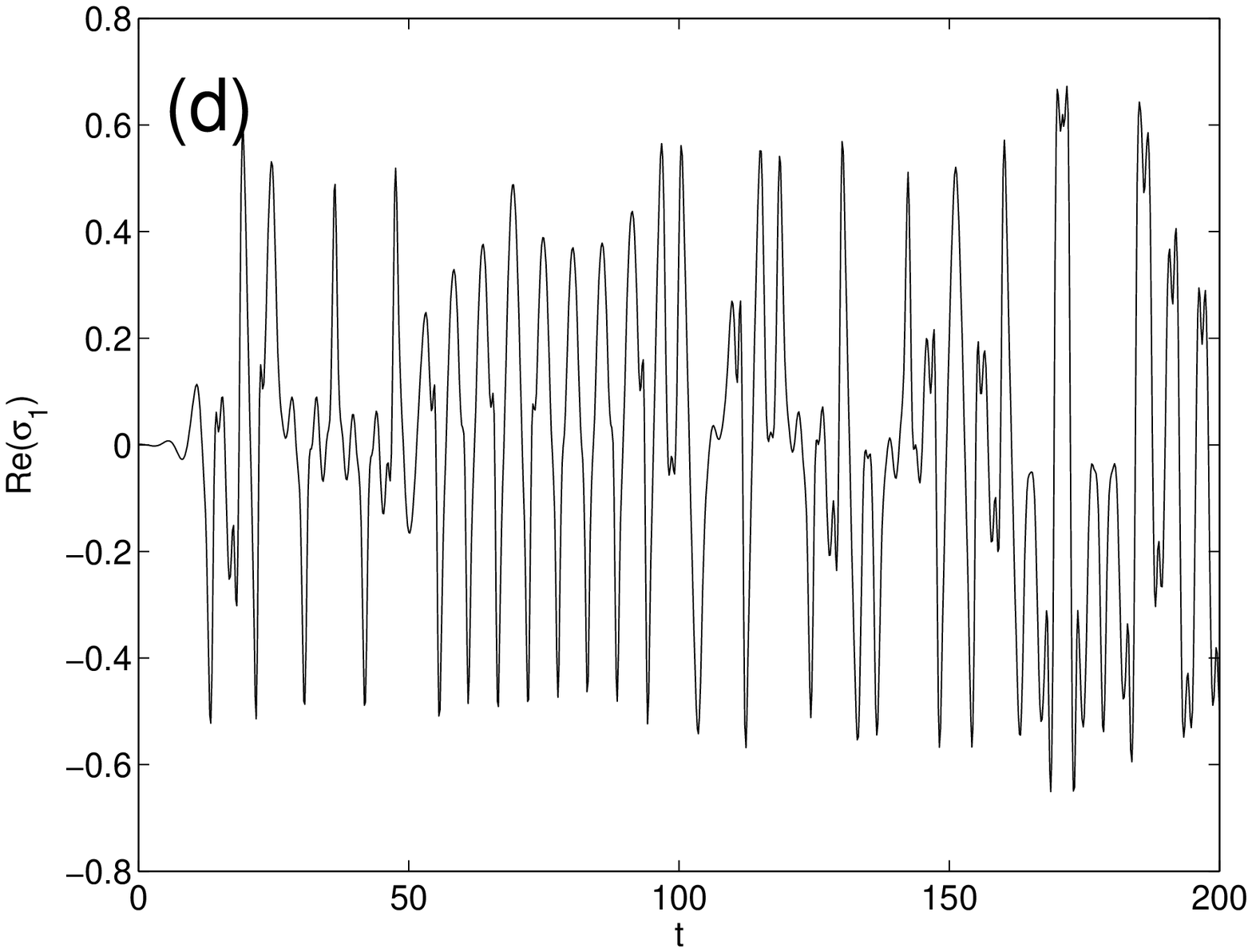}
\includegraphics[width=\mywidth]{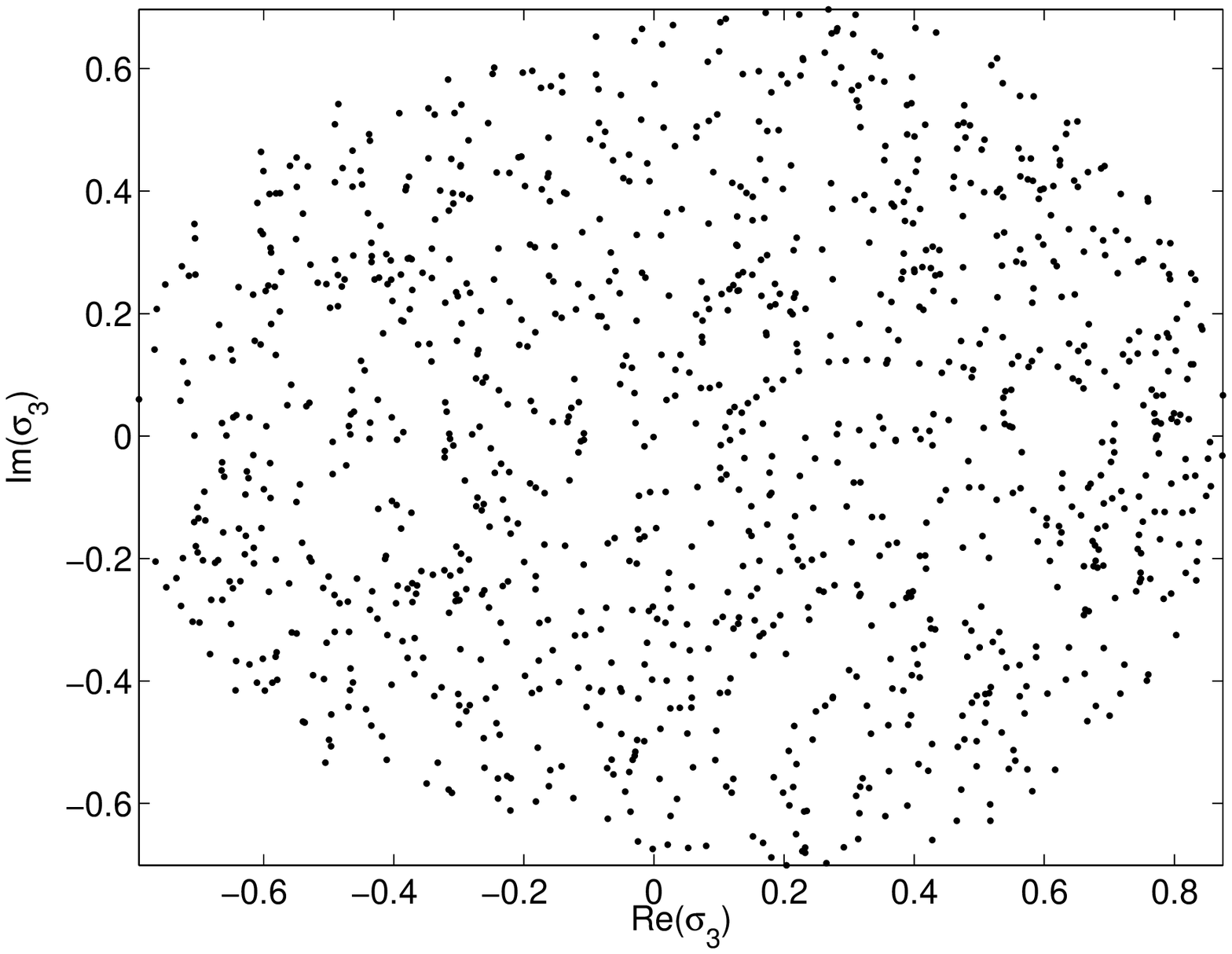}
\includegraphics[width=\mywidth]{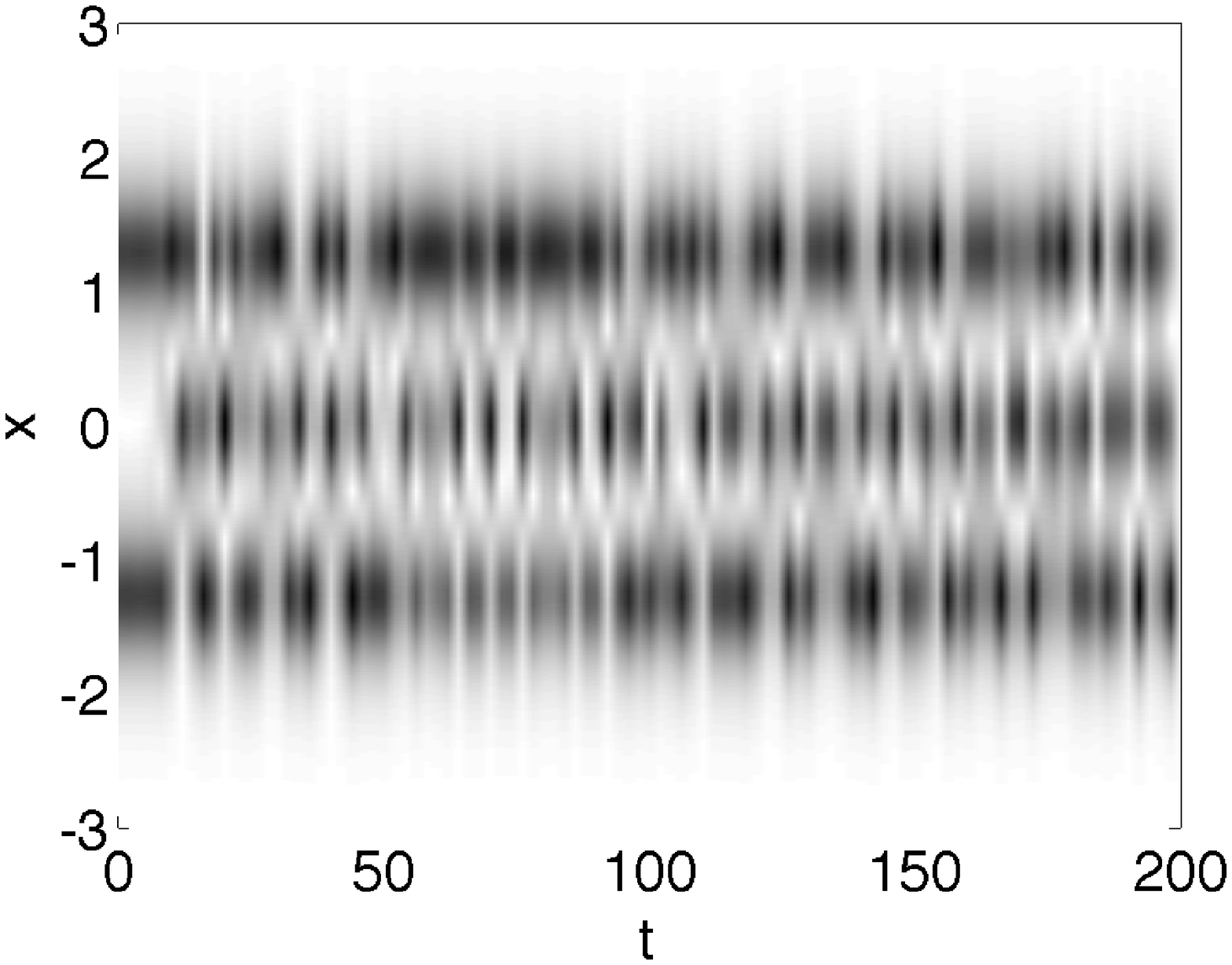}
\includegraphics[width=\mywidth]{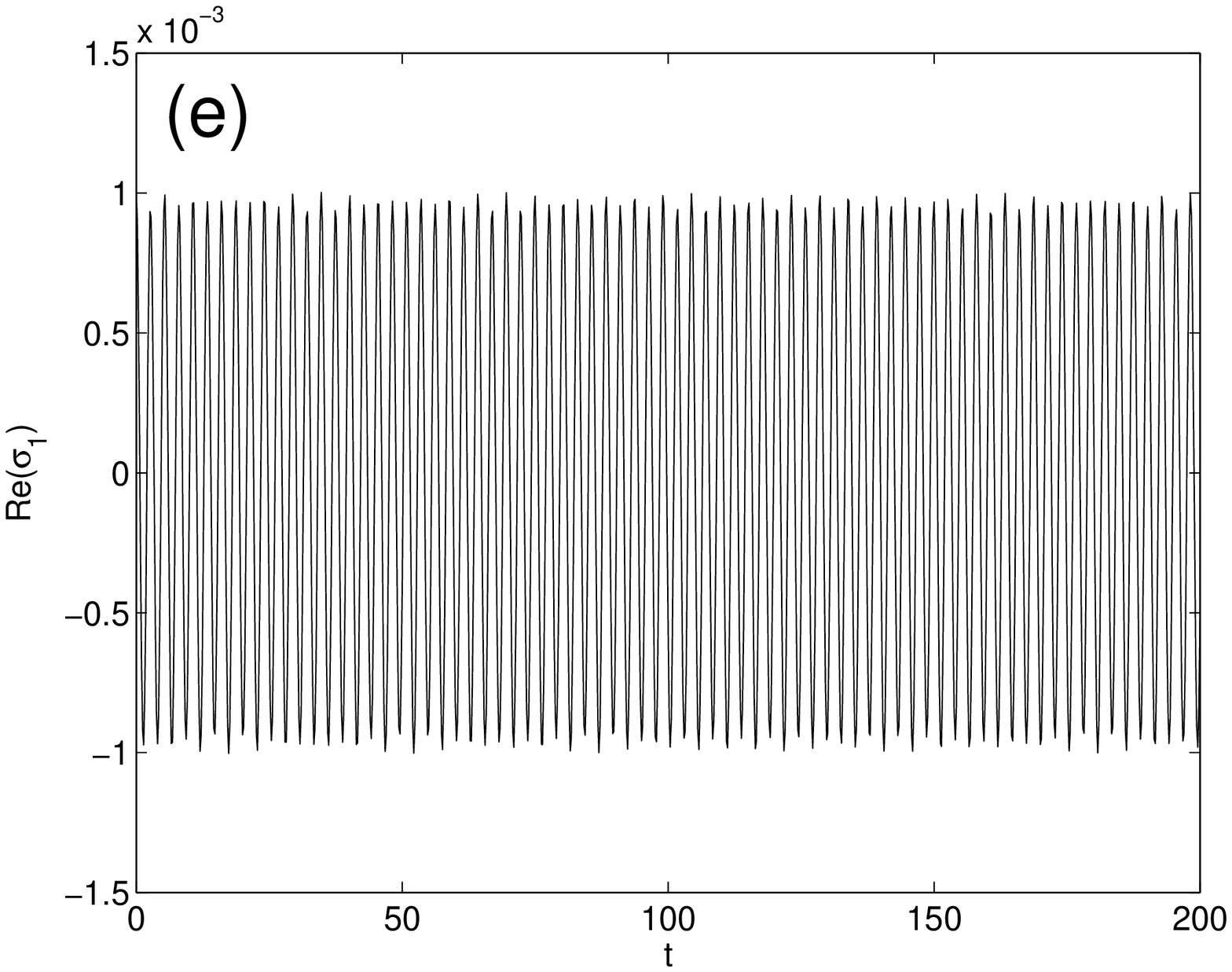}
\includegraphics[width=\mywidth]{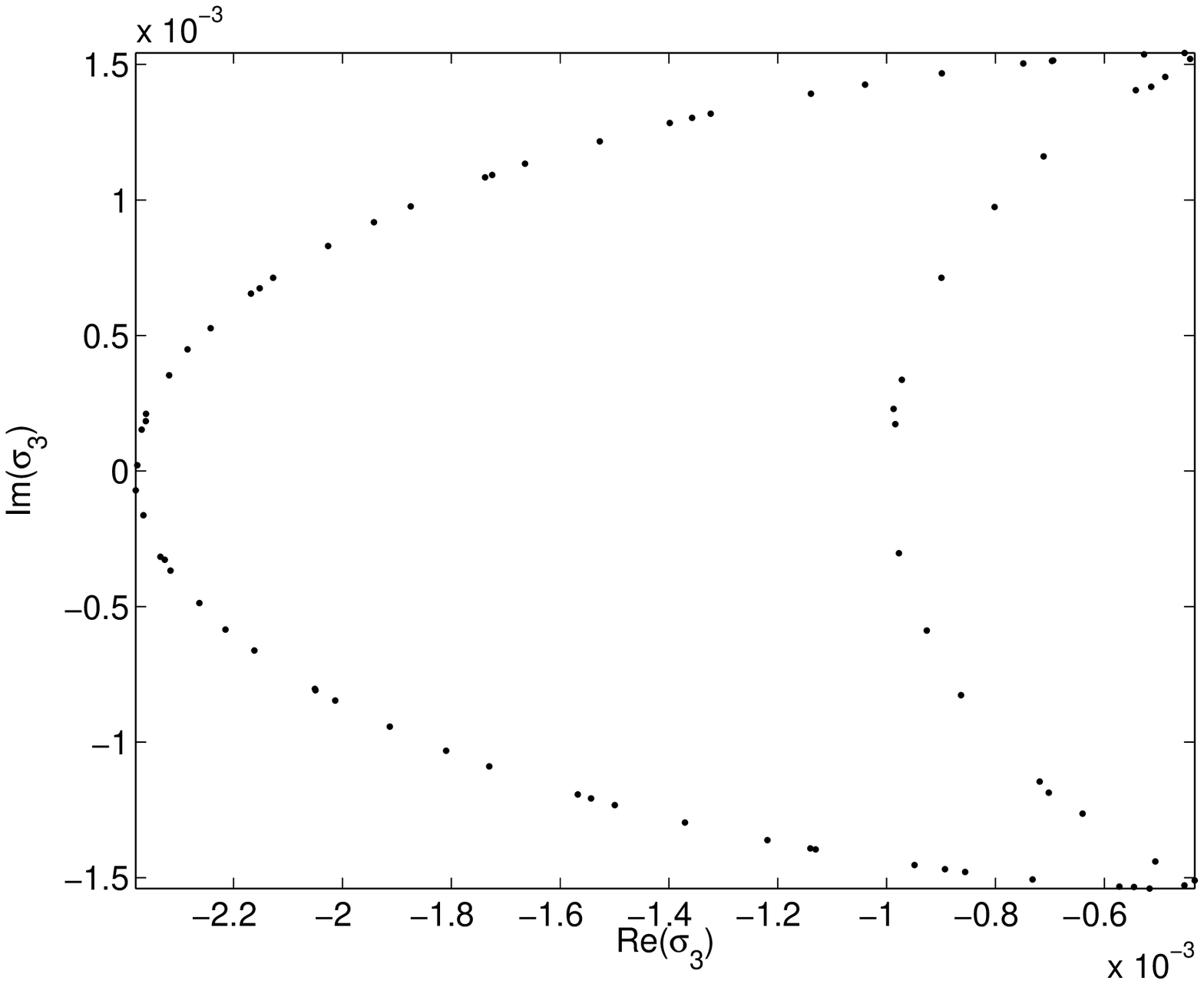}
\includegraphics[width=\mywidth]{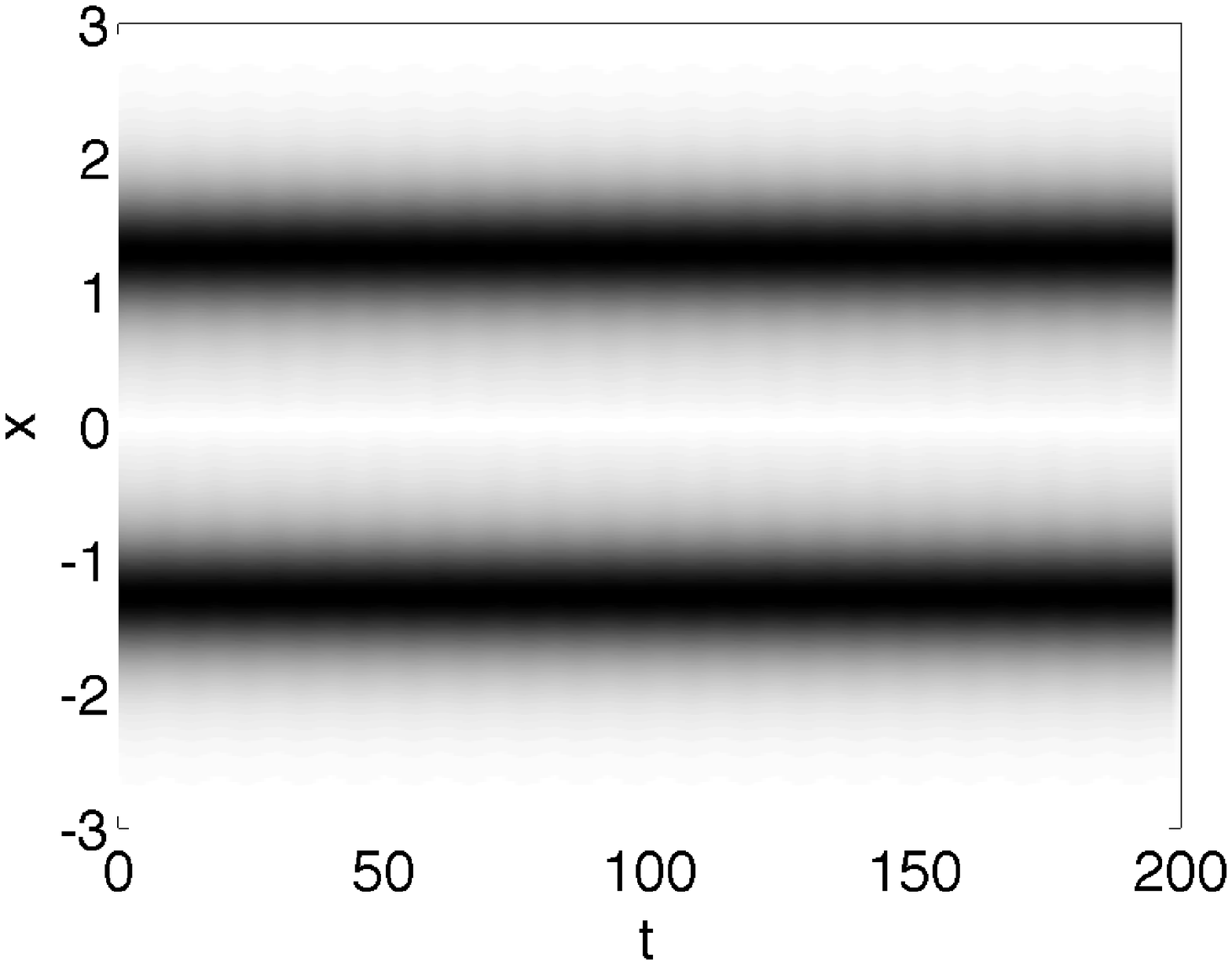}
   \caption{Time-dependent simulations of PDE system~\eqref{NLS}.  The rows, labeled (a)-(e) correspond to the values of $\cN$ indicated on figure~\ref{fig:ODE_PDE_Hopf}.  Column 1 shows $\Re{\s_1}(t)$.  Column 2 shows the intersection of the solution with Poincar\'{e} section $\Sigma_{\cN,H}$.  Column 3 shows a reconstruction of $\abs{u(t)}$ using ansatz~\eqref{complexChange} (darker areas indicated larger values).  This shows, as is predicted by figure~\ref{fig:ODE_PDE_Hopf}, that cases (a) and (e) are stable, and that instabilities, and even chaos, exist in the other three cases.  The agreement with the reduced ODE is uncanny. The initial conditions used are $u(x) = 0.001 \psi_1(x) + \psi_2(x) + 0.001 \psi_3(x)$.}
   \label{fig:PDEsolutions}
\end{figure}

As initial conditions, we use linear combinations of the three linear modes, and, to compare the solutions with those of the ODE system, we compute the projection of the solution onto the span of the localized linear modes, giving us, essentially, the parameters $c_j(t)$. Dividing $c_j$ by the phase of $c_2(t)$ gives a value analogous to $\s_j$ and $\rho(t)$. PDE simulations are shown in figure~\ref{fig:PDEsolutions}, again for $\cN \in \{0.35, 0.6, 1, 3, 5.5\}$.  The behavior is remarkably similar to the behavior found for the ODE solutions.

\section{Further reduction of the ODE}
\label{sec:further}

We now perform further analysis with the goal of understanding the dynamics observed at amplitudes above the critical value for HH bifurcation. We will formally apply the von Zeipel averaging procedure, which applies in the case of a resonance between eigenvalues~\cite{Arnold:1997,Wiggins:2003}.  Because the systems are Hamiltonian, the averaged equations will preserve some but not all features of the full system of equations---for example hyperbolic fixed points and their local un/stable manifolds will be preserved, but homoclinic orbits will not.   The averaged system will be completely integrable, but we have already seen evidence in figure~\ref{fig:ODEsolutions} that the full system is not.

The standard reference for the HH bifurcation is the monograph of van der Meer~\cite{Meer:1985}, but the analysis presented there does not apply to system~\eqref{complexform}.  In the generic HH bifurcation, the matrix of the linearization is non-semisimple (i.e.\ it has non-trivial Jordan blocks) whereas in this case, it is semisimple (diagonalizable over $\mathbb C$). This particular case is analyzed by Chow and Kim~\cite{Chow:1988}. These methods are based on a Lyapunov-Schmidt reduction and require a more involved calculation which we defer to a later study.

To reduce the number of degrees of freedom, we first make the change of variables to canonical polar coordinates
$$ 
\s_j \to \sqrt{\r_j}e^{i \th_j}; j=1,3; 
$$
 yielding a Hamiltonian:
\begin{equation}
\begin{split}
H_{\rm polar}&=
 (N(a_{1122} \cos{2 \th_1} +2 a_{1122} -a_{2222}) + W-\epsilon )\r_1\\
&\phantom{=}+ (N(a_{2233} \cos{2 \th_3}-a_{2222} +2 a_{2233})-W-\epsilon )\r_3\\
&\phantom{=}  +2 a_{1223} N  (2\cos{(\th_1-\th_3)}+\cos{(\th_1+\th_3)})\sqrt{\r_1} \sqrt{\r_3}\\
&\phantom{=}+\frac{N}{2} \r_1^2 (-2 a_{1122} \cos{2 \th_1}+a_{1111}-4 a_{1122}+a_{2222})\\
&\phantom{=}-N \r_1 \r_3 (a_{1122}\cos{2 \th_1}-a_{1133}\cos{2(\th_1- \th_3)}+a_{2233}\cos{2 \th_3}+2 a_{1122}-2 a_{1133}-a_{2222}+2 a_{2233}) \\
&\phantom{=}  +\frac{N}{2} \r_3^2 (-2 a_{2233}\cos{2 \th_3}+a_{2222}-4 a_{2233}+a_{3333})\\
&\phantom{=}+2 N \sqrt{\r_3} \r_1^{3/2} ((a_{1113}-2 a_{1223})\cos{(\th_1-\th_3)}
-a_{1223}\cos{(\th_1+\th_3)}) \\
&\phantom{=} -2 N \r_3^{3/2} \sqrt{\r_1}  (2 a_{1223}\cos{(\th_1-\th_3)}+a_{1223}\cos{(\th_1+\th_3)}-a_{1333}\cos{(\th_1-\th_3)}).
\end{split}
\label{polar}
\end{equation}
Naively, one would hope to make near-identity changes of variables that have the effect of averaging out all of the mean-zero (i.e.\ cosine) terms.  Note this would also eliminate the terms of fractional power in the $\rho_j$.  The formal equations necessary to remove some of these terms, however, will in some cases lead to zero denominators, that is, those terms are resonant.
Were it not for such resonances between the eigenvalues, one could make near-identity changes of variables to remove all terms containing trigonometric functions and fractional powers of $\r_1$ and $\r_3$, putting the system in the so-called Birkhoff normal form.  Resonances of higher order terms become a problem precisely when the linear part of the equations contains a resonance of the type defined in equation~\eqref{resonanceGeneral}.  In this case, the linear part of the Hamiltonian in the linear limit $N\to0$
$$
H = (W-\epsilon) \r_1 + (-W-\epsilon)\r_3 \equiv \w_1 \r_1 + \w_2 \r_2
$$
satisfies the near resonance of order 2:
$$
k_1 \w_1 + k_3 \w_3 = 2 \epsilon \ll 1
$$
where $k_1=k_3=1$.  In this case, one cannot completely average the system and is forced to consider the Gustavson normal form.  For more information see Wiggins~\cite[\S19.10, \S20.9]{Wiggins:2003}. To put the system in normal form, we find $(l_1,l_3)\in {\mathbb Z}^2$ satisfying $k_1 l_3 - k_3 l_1=1$ and make the symplectic change of variables:
\begin{align*}
\th_1 &= \phantom{-}l_3 \psi_1 - k_3 \psi_3; & \r_1 & = k_1 J_1 + l_1 J_3; \\
\th_3 &= -l_1 \psi_1 + k_1 \psi_3; & \r_3 & = k_3 J_1 + l_3 J_3.
\end{align*}

We also make the assumption that the nonlinearity is small: $N=\epsilon \nu$.
In particular, we could choose $(l_1,l_3)=(0,1)$.  This change of variables  explicitly separates the fast motion with $\Or{(1)}$ time scales from the slower motion with time scales of $\Or{(\epsilon^{-1})}$.  Such a change of variables would allow us to eliminate the pair $(J_3,\psi_3)$ from the Hamiltonian.  Because in figures~\ref{fig:ODEsolutions} and~\ref{fig:PDEsolutions} we show the Poincar\'e map that eliminates the pair $(J_1,\psi_1)$, we choose to make the equivalent canonical change of variables
$$
\th_1 = \psi_1, \, \th_3 = -\psi_1 +\psi_3,\, \rho_1 = J_1 + J_3, \,\rho_3 = J_3.
$$
This puts the Hamiltonian in the form
$$
H_{\rm reduced} = H_0(J_1) + \epsilon H_1(J_1,J_3,\psi_1,\psi_3)
$$
where
$$
H_0(J_1) = W J_1
$$
and $H_1$ has period $\pi$ in  $\psi_1$ and $2\pi$ in $\psi_3$ (and which we will not write out here).

Thus, on a level set of the Hamiltonian $H_{\rm reduced}=W h$, we may solve for 
$J_1$ as a function of the other three variables, which gives
$$
J_1 = hW + \epsilon L_1(h,J_3,\psi_1,\psi_3) + \Or{(\epsilon^2)}.
$$
where
$$
L_1 = -\frac{1}{W} H_1\left(\frac{h}{W},J_3,\psi_1,\psi_3\right)
$$
This indicates that to leading order in $\epsilon$ and for times of $\Or{(\epsilon^{-1})}$,  $J_1 \approx h$ is a conserved quantity and allows us to use $\psi_1$ as a time-like variable.  Renaming $\psi_1=\tau$  gives Hamiltonian~\cite{GH:83}:
\begin{equation}
H_{\rm reduced} = - \epsilon L_1 = \frac{\epsilon}{W} \tilde{H}_1(J_3,\psi_3,h) + \frac{\epsilon}{W} \hat{H}_1(J_3,\psi_3,-h,\tau)
\label{Hultimate}
\end{equation}
where
$$
 \tilde{H}_1(J_3,\psi_3;h)  =\g_1 J_3 + \g_2 J_3^2 + \g_3 \sqrt{J_3} \sqrt{J_3+h} \left(2 J_3+h-1\right) \cos{\psi_3}
$$
and $\hat{H}_1$, the details of which will not be important, satisfies
$$
\int_0^{2\pi} \hat{H}_1(J_3,\psi_3,-h,\tau) d\tau= 0
$$
with coefficients
\begin{align*}
\g_1 &= 2\mfs +\left(-a_{1111} h  +2 a_{1122} (3 h-1)  -2 a_{1133} h  -2 a_{2222} (h-1)  +2 a_{2233} (h-1) \right)\nu \\
\g_2 &= \frac{\nu}{2} \left(-a_{1111}+8 a_{1122}-4 a_{1133}-4 a_{2222}+8 a_{2233}-a_{3333}\right) \\
\g_3 &= 2 \nu a_{1223}.
\end{align*}
Standard averaging techniques~\cite{GH:83} now show that there exists a near-identity change of variables
$$ J=J_3 + \Or{(\epsilon}), \psi=\psi_3 + \Or{(\epsilon})$$
such that the solution to the averaged system with Hamiltonian
\begin{equation}
H_{\rm average}= \frac{\epsilon}{W} \tilde{H}_1(J,\psi;h) 
\label{Haverage}
\end{equation}
agrees with solutions to system~\eqref{Hultimate} with error of order $\epsilon$ for times of order $\epsilon^{-1}$.  Further, for sufficiently small $\epsilon$, fixed points and their local invariant manifolds of system~\eqref{Haverage} will correspond to periodic orbits and their local invariant manifolds of system~\eqref{Hultimate}.

By the conservation of the total intensity, equation~\eqref{l2}, $\r_1$ and $\r_3$ in system~\eqref{polar} are confined to the triangle
$$
0\le \r_1 \le 1; \; 0\le \r_2 \le 1; \; 0 \le \r_1 + \r_2 \le 1.
$$
In the reduced system, this becomes a constraint on the conserved parameter $h$ and the variable $J$,
$$
-1 \le h \le 1; \;  \min{(-h,0)} \le J \le \frac{1-h}{2}.
$$
We will consider the case $0 \le h \le 1$.  For the case $-1 \le h \le 0$, it is more convenient to eliminate $(\r_3,\theta_3)$ and work in the $(J_1,\psi_1)$ space.  In this case, the phase space is the disk $J \le \tfrac{1-h}{2}$.

A short word on this reduction is in order.  The level sets of $H$ which are manifolds of dimension $2n-1$. When the linear part of a Hamiltonian system of the form
$$H_{\rm linear} = \sum_{j=1}^n \w_j \rho_j$$
has no resonances of the form~\eqref{resonanceGeneral} and the full system has no additional conserved quantities, as discussed in section~\ref{sec:language}.  But in the near-resonance gives rise, at small nonlinearities, to the nearly-conserved quantity $h$, which allows for the dimension-reduction via averaging.

Since the resonance above is not exact, the additional conservation laws are only approximate, and the the quantity $h$ is not precisely conserved.  Formally, one may perform a countable sequence of  changes of variables that transform the system into a form that is completely integrable.  In the limit, this corresponds to defining a change of variables given as a power series in $\epsilon$.  Generally, this power series has radius of convergence zero, because the full system is not itself integrable, which we can see from the chaotic dynamics in the numerical solution given in figure~\ref{fig:ODEsolutions}c.  This analysis suggests that the solution~\ref{fig:ODEsolutions}b is also very weakly chaotic, but with a much smaller chaotic region and a longer chaotic timescale.

\begin{figure}[htb] 
   \centering
\includegraphics[width=2.5in]{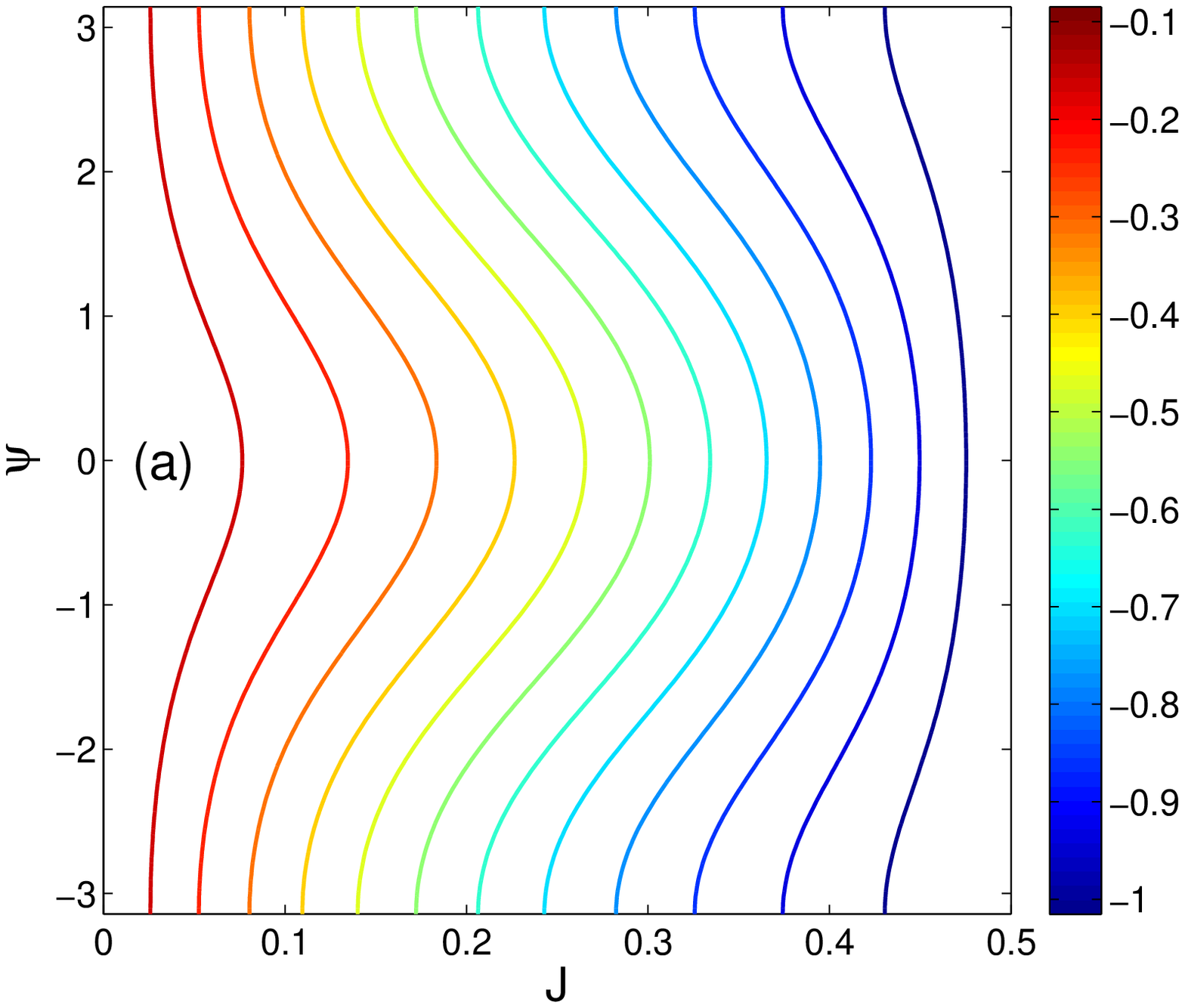}   
\includegraphics[width=2.5in]{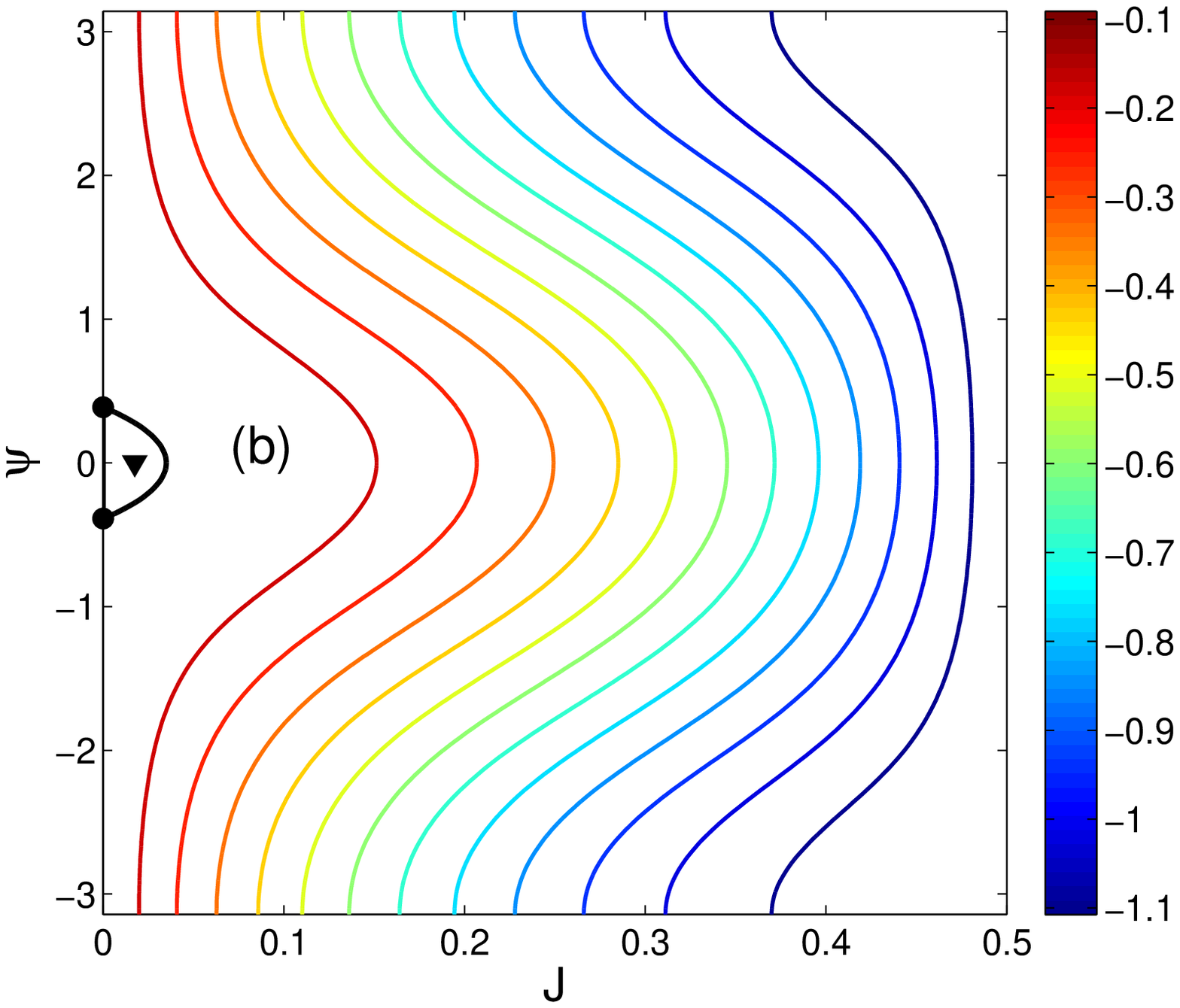}
\includegraphics[width=2.5in]{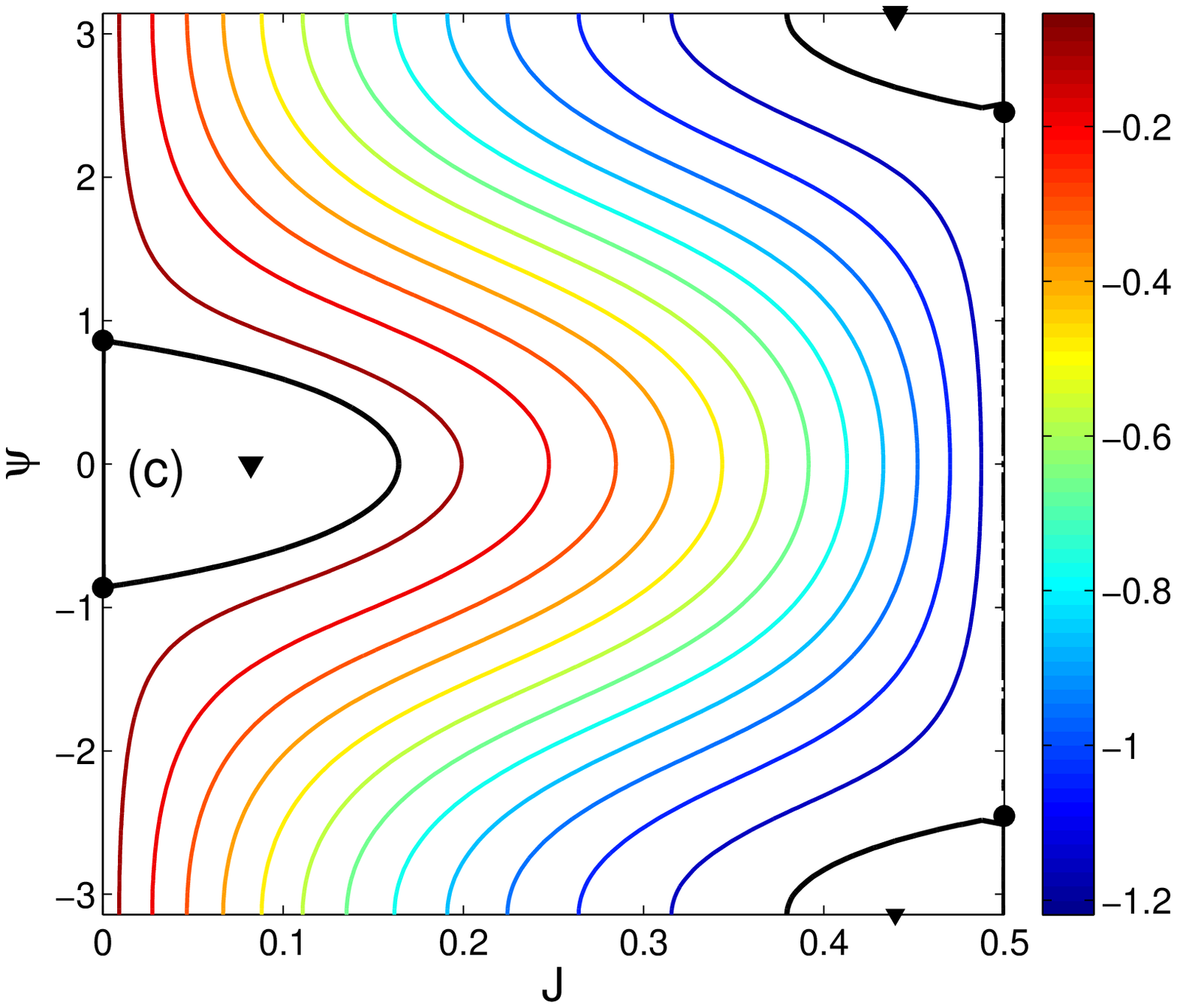}
   \caption{The averaged $(J,\psi)$ phase plane with $h=0$, and varying values of $N$. In (a), the phase changes monotonically.  In (b), a new fixed point arises due to the HH bifurcation, and with it an orbit homoclinic to the origin $(J=0)$.  In (c), three new fixed points bifurcate from the line $J=\thalf$. The colors represent the level sets of $H_{\rm average}$.}
        \label{fig:averaged}
\end{figure}

We are interested in the stability of the trivial solution of system~\eqref{polar}: $(\rho_1,\rho_3)=(0,0)$. This initial condition lies on the level set $h=0$ in $H_{\rm reduced}$.   Thus, solutions to equation~\eqref{polar} whose initial conditions satisfy $h \neq 0$ cannot  approach the origin and it suffices to set $h=0$ in system~\eqref{Haverage} when studying the stability of the trivial solution.  Any stable or unstable manifolds to the origin must also lie in this level set.
We observe what appears to be a near-homoclinic orbit in the numerical experiments presented in figure~\ref{fig:ODEsolutions}, most clearly in row B, and by the above reasoning, any homoclinic orbit to the reduced system must be on the set $h=0$.
Looking at the level set $\tilde{H}_1(J,\psi,;0)=0$ gives the following algebraic equation for the level set containing the origin.
$$
\g_1 J+ \g_2 J^2 + \g_3(2J^2-J)\cos\psi  =0.
$$
This has the trivial solution $J=0$ as well as those that satisfy
\begin{equation}
 \cos{\psi} = \frac{\g_1 + \g_2 J}{\g_3(1-2J)}
\label{averaged_heterocline}
\end{equation}
The origin will have stable and unstable manifolds if this equation has a solution with $J=0$.  After some algebra, this simplifies to
$$
\cos{\psi} = \frac{\g_1}{\g_3}=-\frac{\left(a_{1122}-a_{2222}+a_{2233}\right) \nu -\mfs}{a_{1223} \nu }
$$
which may only happen if
$$
\abs{\frac{\g_1}{\g_3}}=\abs{\frac{\left(a_{1122}-a_{2222}+a_{2233}\right) \nu -\mfs}{a_{1223} \nu }}\le 1.
$$
Thus, there exist bifurcations at 
\begin{equation}
\nu_{\rm HH} = \frac{\mfs}{\pm a_{1223} +a_{1122}-a_{2222}+a_{2233}}
\label{bifurcation_reduced}
\end{equation}
which is simply a recapitulation of the bifurcation condition found by another method in equation~\eqref{Ncritical}.
More simply, there exist fixed points of system~\eqref{Haverage} with $\sin \psi=0$ and 
\begin{equation}
J_{\rm right} \equiv \frac{2 \left(a_{1122}  \pm a_{1223}  -a_{2222}  +a_{2233}  -  \mfs/ \nu \right) }
{-a_{1111}-8 a_{1122}
-4  a_{1133} \pm 4 a_{1223}+4 a_{2222}-8 a_{2233}+a_{3333}}.
\label{J}
\end{equation}
Since $J>0$ by definition, these fixed points bifurcate from the origin exactly when the numerator vanishes, i.e.\ when the coefficients satisfy condition~\eqref{bifurcation_reduced}.
For unstable values of $\nu$, the origin is always a multiple root, and thus is a nonhyperbolic fixed point of the averaged system.
When $h\neq0$, the origin is no longer a fixed point, but the fixed point defined by equation~\eqref{J} persists, for $\nu>\nu_{\rm HH}(h)$, and this critical value now depends on $h$.  For $\nu$ below this bifurcation, the dynamics is described by monotonically decreasing angle $\psi$ (determined from the sign of the numerically calculated $\g_1$) and oscillating amplitude $J$.  For $\nu>\nu_{\rm HH}$, there exists a new fixed point, surrounded by a family of periodic orbits for which $\psi$ oscillates.  This region is separated from the region of monotonic $\psi$ by a heteroclinic orbit connecting the line $J=0$ to itself. This difference can be seen by comparing parts (a) and (b) of figure~\ref{fig:averaged}

\subsection*{Additional structure}

In addition, we note that the set $\Lambda_{\rm even} =\{(J=\thalf,\psi)\}$ is invariant under system~\eqref{Haverage} when $h=0$.  Additional fixed points exist where
\begin{equation}
\cos{\psi_{\rm F}} =  \frac{\g_1+\g_2}{\g_3}.
\label{fixedpointsOnCircumference}
\end{equation}
This corresponds to two fixed points $\sigma_\pm$ on $\Lambda_{\rm even}$. As in the case of the heteroclinic orbit given by~\eqref{averaged_heterocline}, these exist only if the right hand side has magnitude less than one.  This, then, gives a necessary condition on the amplitude $\nu \ge \nu_{\rm F}$, for their existence, where
\begin{equation}
\nu_{\rm F}= \frac{\mfs}{\frac{1}{4} a_{1111} - a_{1122} + a_{1133} \mp a_{1223} - a_{2233} + \frac{1}{4} a_{3333}}.
\label{nuF}
\end{equation}

For $\nu> \nu_{\rm F}$, equation~\eqref{fixedpointsOnCircumference} will have two solutions of saddle type connected by three heteroclinic orbits--two of them contained in $\Lambda_{\rm even}$--and an additional fixed point $J_{\rm left}$ of elliptic type with $\psi=\pi$ and $J=J_{\rm left}$ near $\thalf$. The result of this bifurcation can be seen by comparing parts (b) and (c) of figure~\ref{fig:averaged}.  Note that the left boundary $J=0$ corresponds to solutions on the odd invariant subspace $c_1=c_3=0$ of system~\eqref{3modes}, while the right boundary $\Lambda_{\rm even}$ represents solutions in the even subspace $c_2=0$, and this figure shows a clear symmetry: above $\NHH$, there exists in the averaged equation an orbit homoclinic orbit to the odd subspace, and above $N = \epsilon \nu_{\rm F}$, the averaged equations possess a pair of periodic orbits on the even subspace which are connected by three heteroclinic orbits. Note  that the orbits $\s_\pm$ correspond to periodic orbits of equation~\eqref{complexform} and cannot be found by the methods of section~\ref{sec:stationary}.  

When $0<h<1$, the two fixed points on the boundary of the phase disk persist for $h$ small, with the bifurcation condition~\eqref{nuF} generalizing to
\begin{equation}
\nu_{\rm F}(h) =  \frac{\mfs}
{%
\frac{1}{4} (1+h)a_{1111} 
-(1+h)a_{1122}
+a_{1133}
\mp  \sqrt{1-h^2} a_{1223}
-(1-h)a_{2233}
+\frac{1}{4}(1-h) a_{3333}
}
\label{nuFgeneral}
\end{equation}
Note that the averaged system~\eqref{Haverage} is valid for small values of $\epsilon$ and describes the dynamics for small $\abs{N}$ demonstrated numerically in figure~\ref{fig:ODE_PDE_Hopf}.  More concretely, the averaged system possesses the bifurcations described by equation~\eqref{Ncritical}, but not the other two roots of equation~\eqref{discriminant} that may exist for $N=\Or{(1)}$.

\begin{figure}[htb] 
   \centering
   \includegraphics[width=3.5in]{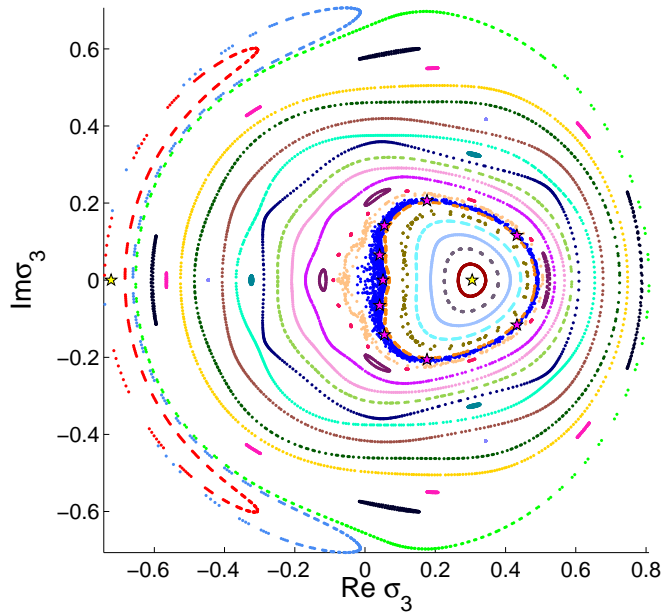} 
   \caption{Poincar\'e map showing the chaotic region of figure~\ref{fig:ODEsolutions}c (blue), as well as many other features of the dynamics.  The fixed points $\s_{\rm left, right}$ are given by yellow stars and a period-9 orbit inside the chaotic region in pink stars.  The presence of a separatrix connecting $\sigma_{\pm}$ is clear  on the left, and, on careful inspection, there is evidence for several other families of periodic orbits of period 3, 4, 7, and 13.}
   \label{fig:finalPlot}
\end{figure}

We end with a numerical computation that compares the integrable averaged system with the full system that displays Hamiltonian chaos.  For the parameter values corresponding to figure~\ref{fig:ODEsolutions}c, column 2, the averaged system has a phase space structure as in figure~\ref{fig:averaged}c.  Returning to $\sigma_j$ coordinates, the averaged system has four fixed points: the $\s_{\rm right} = \sqrt{J_{\rm right}}$ and $\s_{\rm left} = -\sqrt{J_{\rm left}}$ are elliptic, and the two points $\sigma_\pm$ on the boundary are hyperbolic.  In the full system, $h$ is not conserved, but the solution is still confined to a disk.  In figure~\ref{fig:finalPlot}, we show, in blue, the Poincar\'e section of figure~\ref{fig:ODEsolutions}c, along with several other solutions with the same value of $N$ and $H$.  On the left, there is a family of regular orbits surrounding $\s_{\rm left}$---with these parameter values, there is no chaos near this fixed points.  Around the fixed point $\s_{\rm right}$, we see what appears to be typical KAM breakup into Poincar\'e-Birkhoff islands interspersed with the preserved KAM tori.   As the parameter $N$ increases, more and more of the quasiperiodic orbits are destroyed, producing the fully-developed chaose, seen in figure~\ref{fig:ODEsolutions}d.  The chaotic dynamics near the HH bifurcation have been analyzed recently in~\cite{Dullin:2005,Gaivao:2010}.  The dynamics produced by the HH bifurcation violate a ``twist'' condition assumed by the KAM theorem, so the structure of the system is not exactly the same as in the usual KAM setup.

\subsection*{Comparison with other approaches}
Lahiri and Roy~\cite{Lahiri:2001} considered the case of the \emph{non-semisimple} HH bifurcation, so that the results they cite are not directly applicable.   Using formal averaging of a different type, they find two types of bifurcations, depending on certain coefficients in the cubic and quartic terms in the Hamiltonian.  In their \emph{Type 1} bifurcation, they find that there exists, for $\nu > \nu_{\rm HH}$, a new fixed point of the averaged equations a distance $d \propto \sqrt{\nu - \nu_{\rm HH}}$ from the origin.  Converting the solution~\eqref{J} to Cartesian coordinates, this is exactly what we find.  

In their \emph{Type II} bifurcation, there exists no nonzero periodic orbit near zero on the unstable side of the bifurcation.  This is the case for the bifurcations with $N=\Or{(1)}$ that takes place between figure~\ref{fig:ODEsolutions}, rows~\textbf{(d)} and~\textbf{(e)}.  Johansson makes the same observation in his study of the NLS trimer~\cite{Johansson:2004} but does not comment on the difference between the semisimple and non-semisimple bifurcations.

\section{Discussion and Conclusions}
\label{sec:conclusions}
While there have been a large number of papers, discussed in the introduction, examining the HH bifurcation and the onset of oscillatory instabilities in nonlinear wave equations, we believe this is the first to try to analyze the nonlinear dynamics that arise in such a system.  In so doing, we discovered the homoclinic orbit structure in the averaged equations, as well as a new family of relative periodic solutions to the full system that exists when $\nu>\nu_{\rm F}(h)$ in equation~\eqref{nuFgeneral}.

While the analysis in the present paper is purely formal, we believe that the pieces are in place to make rigorous this paper's conclusions, as has been done in~\cite{KirKevShl:08}.  In that paper, for a two-mode potential, it is shown that under suitable assumptions, that below some critical amplitude $\cN_{\rm SB}=\Or{(\epsilon})$, there exists a steady solution with even symmetry which is stable to perturbations, and that for $\cN > \cN_{\rm SB}$, there exist two new branches of asymmetric solutions which are orbitally Lyapunov stable.  In~\cite{Marzuola:2010}, it is further shown that the time-dependent dynamics of the PDE solution are well-modeled, for long but finite times, by a finite-dimensional system that is equivalent to that of a particle moving in a potential which has just one well when $\cN<\cN_{\rm SB}$ but two wells when~$\cN>\cN_{\rm SB}$.

In the finite dimensional system of approximate equations near a symmetry-breaking bifurcation, derived in~\cite{KirKevShl:08,Marzuola:2010}, it takes one line of algebra to show the existence of the two new asymmetric solutions that are born when the bifurcation occurs.  In system~\eqref{3modes}, the analysis is not so simple, and the new solution arising from the bifurcation appears, in its simplest form, as the fixed point~\eqref{J} of Hamiltonian system~\eqref{Haverage} that corresponds to a periodic orbit of system with Hamiltonian~\eqref{Hc}.  Proving the existence of this periodic orbit is a straightforward application of  a paper from the late 1980's by Chow and Kim~\cite{Chow:1988} and will constitute the first step of a planned program to put the results of the present paper on a more rigorous footing.  The chaotic dynamics near the HH bifurcation in a finite-dimensional system are rigorously demonstrated in~\cite{Dullin:2005,Gaivao:2010}.  An attempt to rigorously demonstrate complex dynamics in NLS~\eqref{NLS} must start with an understanding how these results apply to the finite dimensional model~\eqref{complexform}.

It should be noted that while in this system, it is possible to observe Hamiltonian chaotic motion, the underlying dynamics, given by system~\eqref{Hc} are essentially two degree-of-freedom.  Motion of such a system occurs on level sets of the Hamiltonian $H$ which are three-dimensional manifolds in the four-dimensional phase space.  Invariant tori in this system are two-dimensional subsets of these manifolds.  The KAM theorem (or something very similar, see~\cite{Dullin:2005}) implies that most of these tori persist when $\cN-\cNHH$ is small and positive.  A two-dimensional torus separates the three-dimensional manifold, so that trajectories cannot cross from one side of the torus to the other. This implies that solutions starting near the odd-symmetric relative fixed points must remain near that point.  If the linear system~\eqref{linearEig} is assumed to support a fourth eigenmode, with similar assumptions on the spacing of the eigenvalues, then in this weakly unstable regime, with six-dimensional phase space, solutions no longer need stay close to the fixed point, a process known as Arnol'd diffusion~\cite{Arnold:1997}.  Further studies are planned to investigate this possibility.

We have assumed throughout this paper that the potential $V(x)$ enjoys even spatial symmetry.
The HH bifurcation phenomenon discussed in this paper depends only on assumptions~\eqref{A1}-\eqref{2ndDifference} and not on this symmetry.  Lacking such a symmetry, the finite-dimensional model~\eqref{Hc}  and its relative equilibria given in section~\ref{sec:stationary} would be significantly more complicated, and the normal form for the HH bifurcation might no longer be semisimple.  An interesting question would be to see how the dynamics change in the face of such asymmetry.

Finally, when considered as a model for an optical waveguide, the system studied here should be straightforward to implement in a laboratory setting.  Discussions are underway to make this happen and will form the basis of an experimental line of research.

\section*{Acknowledgments}
Thanks to Denis Blackmore, Eduard Kirr, Elie Shlizerman, David Trubatch, and Michael Weinstein for useful discussions and Richard Kollar and Arnd Scheel for useful comments in response to a presentation.  The code used to simulate PDE solutions was written and graciously shared by Tom\'{a}\v{s} Dohnal.\   RG was supported by NSF-DMS-0807284.  This work was completed while the author was on sabbatical at Technion, the Israel Institute of Technology.  He thanks them for their hospitality.

\appendix
\section{Some 3-soliton formulas}
\label{sec:appendix}
For any three real numbers $\k_j$ satisfying  $\k_1>\k_2>\k_3>0$, there exists a three-soliton potential given by:
$$u(x) = \cN_u(x)/\cD(x)$$
and the three modes are given by
$$\psi_j(x) =\cN_j(x)/\cD(x)$$
where
\begin{equation}
\begin{split}\cN_u(x)=
2(-\k_1^6\k_2^2 + 2\k_1^4\k_2^4 - \k_1^2\k_2^6 - \k_1^6\k_3^2 - 
         \k_2^6\k_3^2 + 2\k_1^4\k_3^4 + 2\k_2^4\k_3^4 - \k_1^2\k_3^6 - \k_2^2\k_3^6)\\
- 2(\k_1^2 - \k_2^2)(\k_1^2 - \k_3^2)(\k_2 - \k_3)^2(\k_2 + \k_3)^2\cosh{2\k_1x}\\
- 2(\k_1^2 - \k_2^2)(\k_2^2 - \k_3^2)(\k_1 - \k_3)^2(\k_1 + \k_3)^2\cosh{2\k_2x}\\
- 2(\k_1^2 - \k_3^2)(\k_2^2 - \k_3^2)(\k_1 - \k_2)^2(\k_1 + \k_2)^2\cosh{2\k_3x}\\
-(\k_1 + \k_2)^2(\k_1 - \k_3)(\k_2 - \k_3)\k_3^2(\k_1 + \k_3)(\k_2 + \k_3)\cosh{2(\k_1-\k_2)x}\\
-(\k_1 - \k_2)\k_2^2(\k_1 + \k_2)(\k_2 - \k_3)(\k_1 + \k_3)^2(\k_2 + \k_3)\cosh{2(\k_1 - \k_3)x}\\
-\k_1^2(\k_1 - \k_2)(\k_1 + \k_2)(\k_1 - \k_3)(\k_1 + \k_3)(\k_2 + \k_3)^2\cosh{2(\k_2 -\k_3)x}\\
-(\k_1 - \k_2)^2(\k_1 - \k_3)(\k_2 - \k_3)\k_3^2(\k_1 + \k_3)(\k_2 + \k_3)\cosh{2(\k_1 + \k_2)x}\\
-(\k_1 - \k_2)\k_2^2(\k_1 + \k_2)(\k_1 - \k_3)^2(\k_2 - \k_3)(\k_2 + \k_3)\cosh{2(\k_1 + \k_3)x}\\
-\k_1^2(\k_1 - \k_2)(\k_1 + \k_2)(\k_1 - \k_3)(\k_2 - \k_3)^2(\k_1 + \k_3)\cosh{2(\k_2 + \k_3)x},
\end{split}
\nonumber
\end{equation}

$$
\cN_1(x)=(\k_2+\k_3)\cosh{(\k_2-\k_3)x} + (\k_2-\k_3)\cosh{(\k_2+\k_3)x},
$$
$$
\cN_2(x)=(\k_1+\k_3)\sinh{(\k_1-\k_3)x} + (\k_1-\k_3)\sinh{(\k_1+\k_3)x},
$$
$$
\cN_3(x)= (\k_1-\k_2)\cosh{(\k_1+\k_2)x} -(\k_1+\k_2)\cosh{(\k_1-\k_2)x},
$$
and
\begin{align*}\cD(x)=&\phantom{+}
(\k_1+\k_2)(\k_1+\k_3)(\k_2-\k_3)\cosh{(\k_1-\k_2-\k_3)x}\\
&+(\k_1-\k_2)(\k_1+\k_3)(\k_2+\k_3)\cosh{(\k_1+\k_2-\k_3)x}\\
&+(\k_1+\k_2)(\k_1-\k_3)(\k_2+\k_3)\cosh{(\k_1-\k_2+\k_3)x}\\
&+(\k_1-\k_2)(\k_1-\k_3)(\k_2-\k_3)\cosh{(\k_1+\k_2+\k_3)x}
\end{align*}
The three discrete eigenvalues are given by $\W_j = -\k_j^2$.

\section{Remainder terms}
\label{sec:projection_appendix}
Equation~\eqref{3modesPlus} depends on remainder terms~$R_1$, $R_2$, $R_3$, and $R_{\rm cont}$ which we define here. The remainder terms in equations~\eqref{c1}-\eqref{c3} are given by 
$$R_j = -\cN \cdot \Pi_j F$$
where $\Pi_j$ is given in equation~\eqref{proj_j} and
$$
F = \abs{c_1 \Psi_1 + c_2 \Psi_2 + c_3 \Psi_3+\eta}^2(c_1 \Psi_1 + c_2 \Psi_2 + c_3 \Psi_3+\eta)-\abs{c_1 \Psi_1 + c_2 \Psi_2 + c_3 \Psi_3}^2(c_1 \Psi_1 + c_2 \Psi_2 + c_3 \Psi_3)
$$
The remainder term for the $\eta(x,t)$ equation~\eqref{eta_eqn} is given by
$$R_{\rm cont} = -\cN \cdot \Pi_{\rm cont}  G $$
where $\Pi_{\rm cont}$ is given in~\eqref{proj_cont} and
$$
G = \abs{c_1 \Psi_1 + c_2 \Psi_2 + c_3 \Psi_3+\eta}^2(c_1 \Psi_1 + c_2 \Psi_2 + c_3 \Psi_3+\eta).
$$


\begin{thebibliography}{10}

\bibitem{Boyd:2008}
R.~W. Boyd.
\newblock {\em Nonlinear Optics}.
\newblock Academic Press, 3rd edition, 2008.

\bibitem{Newell:2003}
A.C. Newell and J.V. Moloney.
\newblock {\em {Nonlinear optics}}.
\newblock Advanced Book Program. Westview Press, 2003.

\bibitem{Pitaevskii:2003}
L.~Pitaevskii and S.~Stringari.
\newblock {\em Bose {E}instein Condensation}.
\newblock Oxford University Press, 2003.

\bibitem{KirKevShl:08}
E.~W. Kirr, P.~G. Kevrekidis, E.~Shlizerman, and M.~I. Weinstein.
\newblock Symmetry-breaking bifurcation in nonlinear
  {S}chr\"{o}dinger/{G}ross-{P}itaevskii equations.
\newblock {\em SIAM J. Math. Anal.}, 40:566--604, 2008.

\bibitem{KapKevChe:06}
T.~Kapitula, P.~G. Kevrekidis, and Z.~Chen.
\newblock Three is a crowd: {S}olitary waves in photorefractive media with
  three potential wells.
\newblock {\em SIAM J. Appl. Dyn. Syst.}, 5:598--633, 2006.

\bibitem{Marzuola:2010}
J.~L. Marzuola and M.~I. Weinstein.
\newblock Long time dynamics near the symmetry bwreaking bifurcation for
  nonlinear {S}chr\"{o}dinger/{G}ross-{P}itaevskii equations.
\newblock {\em DCDS-A}, 28:1505--1554, 2010.

\bibitem{Mayteevarunyoo:2008}
T.~Mayteevarunyoo, B.A. Malomed, and G.~Dong.
\newblock Spontaneous symmetry breaking in a nonlinear double-well structure.
\newblock {\em Phys. Rev. A}, 78:53601, 2008.

\bibitem{Pelinovsky:2011}
D.~Pelinovsky and T.~Phan.
\newblock Normal form for the symmetry-breaking bifurcation in the nonlinear
  schrodinger equation.
\newblock {\em arXiv:1101.5402v1}, 2011.

\bibitem{Kapitula:2005}
T.~Kapitula and P.~G. Kevrekidis.
\newblock Bose-{E}instein condensates in the presence of a magnetic trap and
  optical lattice.
\newblock {\em Chaos}, 15:037114, 2005.

\bibitem{Johansson:2004}
M.~Johansson.
\newblock {H}amiltonian {H}opf bifurcations in the discrete nonlinear
  {S}chr{\"o}dinger trimer: oscillatory instabilities, quasi-periodic solutions
  and a `new' type of self-trapping transition.
\newblock {\em J. Phys. A}, 37:2201, 2004.

\bibitem{Kapitula:2001}
T.~Kapitula, P.~G. Kevrekidis, and B.~A. Malomed.
\newblock Stability of multiple pulses in discrete systems.
\newblock {\em Phys. Rev. E}, 63:036604, 2001.

\bibitem{Morgante:2000}
A.M. Morgante, M.~Johansson, G.~Kopidakis, and S.~Aubry.
\newblock Oscillatory instabilities of standing waves in one-dimensional
  nonlinear lattices.
\newblock {\em Phys. Rev. Lett.}, 85:550--553, 2000.

\bibitem{Panda:2005}
S.~Panda, A.~Lahiri, T.~K. Roy, and A.~Lahiri.
\newblock Standing waves in a non-linear 1d lattice: {F}loquet multipliers,
  {K}rein signatures, and stability.
\newblock {\em Phys. D.}, 210:262 -- 283, 2005.

\bibitem{Kevrekidis:2003}
P.~G. Kevrekidis, D.J. Frantzeskakis, B.~A. Malomed, A.~Bishop, and
  I.~Kevrekidis.
\newblock Dark-in-bright solitons in {B}ose-{E}instein condensates with
  attractive interactions.
\newblock {\em New J. Phys.}, 5:64, 2003.

\bibitem{Kevrekidis:2005}
P.G. Kevrekidis, B.A. Malomed, D.J. Frantzeskakis, A.R. Bishop, HE~Nistazakis,
  and R~Carretero-Gonz{\'a}lez.
\newblock Domain walls of single-component {B}ose-{E}instein condensates in
  external potentials.
\newblock {\em Math. Comput. Simulat.}, 69:334--345, 2005.

\bibitem{Li:2005}
L.~Li, Z.~Li, B.A. Malomed, and D.~Mihalache.
\newblock Exact soliton solutions and nonlinear modulation instability in
  spinor {B}ose-{E}instein condensates.
\newblock {\em Phys. Rev. A}, 72:033611, 2005.

\bibitem{Nistazakis:2007}
H.~Nistazakis, D.J. Frantzeskakis, and P.~G. Kevrekidis.
\newblock Polarized states and domain walls in spinor {B}ose-{E}instein
  condensates.
\newblock {\em Phys. Rev. A}, 76:063603, 2007.

\bibitem{Theocharis:2010}
G.~Theocharis, A.~Weller, J.P. Ronzheimer, C.~Gross, M.K. Oberthaler, P.G.
  Kevrekidis, and D.J. Frantzeskakis.
\newblock Multiple atomic dark solitons in cigar-shaped {B}ose-{E}instein
  condensates.
\newblock {\em Phys. Rev. A}, 81:063604, 2010.

\bibitem{Kapitula:2004}
T.~Kapitula, P.~G. Kevrekidis, and B.~Sandstede.
\newblock Counting eigenvalues via the {K}rein signature in
  infinite-dimensional {H}amiltonian systems.
\newblock {\em Phys. D}, 195:263--282, 2004.

\bibitem{Kapitula:2005a}
T.~Kapitula, P.~G. Kevrekidis, and B.~Sandstede.
\newblock Addendum: Counting eigenvalues via the {K}rein signature in
  infinite-dimensional {H}amiltonian systems.
\newblock {\em Phys. D}, 201:199 -- 201, 2005.

\bibitem{Kapitula:2007}
T.~Kapitula, P.G. Kevrekidis, and R.~Carretero-Gonz{\'a}lez.
\newblock Rotating matter waves in {B}ose-{E}instein condensates.
\newblock {\em Phys. D}, 233:112 -- 137, 2007.

\bibitem{Goodman:08}
R.~H. Goodman and M.~I. Weinstein.
\newblock Stability and instability of nonlinear defect states in the coupled
  mode equations---analytical and numerical study.
\newblock {\em Phys. D.}, 237:2731--2760, 2008.

\bibitem{Luzzatto:2010}
P.~Luzzatto-Fegiz and C.~H.~K. Williamson.
\newblock Resonant instability in two-dimensional vortex arrays.
\newblock {\em P. Roy. Soc. A-Math. Phy.}, TBA, 2010.

\bibitem{Har:80}
E.~Harrell.
\newblock Double wells.
\newblock {\em Comm. Math. Phys.}, 75:239--261, 1980.

\bibitem{Yukon:1980}
S.P. Yukon and B.~Bendow.
\newblock Design of waveguides with prescribed propagation constants.
\newblock {\em J. Opt. Soc. Amer.}, 70:172--179, 1980.

\bibitem{Drazin:1993}
P.~G. Drazin and R.~S. Johnson.
\newblock {\em Solitons: An Introduction}.
\newblock Cambridge University Press, 1993.

\bibitem{Ablowitz:2004}
M.~J. Ablowitz, B.~Prinari, and A.~D. Trubatch.
\newblock {\em Discrete and continuous nonlinear {S}chr\"odinger systems},
  volume 302 of {\em London Mathematical Society Lecture Note Series}.
\newblock Cambridge University Press, Cambridge, 2004.

\bibitem{MatSal:91}
V.~B. Matveev and M.~A. Salle.
\newblock {\em Darboux transformations and solitons}.
\newblock Springer Series in Nonlinear Dynamics. Springer-Verlag, Berlin, 1991.

\bibitem{Hirsh:2009}
I.~Hirsh, M.~Horowitz, and A.~Rosenthal.
\newblock Design of planar waveguides with prescribed mode-profile using
  inverse scattering theory.
\newblock {\em IEEE J. Quantum Elect.}, 45:1133 -- 1141, 2009.

\bibitem{MacKay:1987}
R.~Mac{K}ay.
\newblock Stability of equilibria of {H}amiltonian systems.
\newblock In J.~Meiss R.~MacKay, editor, {\em Hamiltonian Dynamical Systems},
  pages 137--153. Adam Hilger, 1987.

\bibitem{Brugnano:2009a}
L.~Brugnano, F.~Iavernaro, and T.~Susca.
\newblock Hamiltonian {BVM}s ({HBVM}s): {I}mplementation details and
  applications.
\newblock {\em Proceedings of ICNAAM}, pages 723--726, 2009.

\bibitem{Brugnano:2009}
L.~Brugnano, F.~Iavernaro, and D.~Trigiante.
\newblock Analysis of {H}amiltonian boundary value methods ({{HBVM}}s): A class
  of energy-preserving {R}unge-{K}utta methods for the numerical solution of
  polynomial {H}amiltonian systems.
\newblock {\em arXiv:0909.5659v2}, 2009.

\bibitem{Brugnano:2009b}
L.~Brugnano, F.~Iavernaro, and D.~Trigiante.
\newblock Hamiltonian {BVM}s ({HBVM}s): {a} family of ``drift free'' methods
  for integrating polynomial {H}amiltonian problems.
\newblock {\em AIP Conf. Proc}, 2009.

\bibitem{Kennedy:2003}
C.~A. Kennedy and M.~H. Carpenter.
\newblock Additive {R}unge-{K}utta schemes for convection-diffusion-reaction
  equations.
\newblock {\em Appl. Numer. Math.}, 44:139--181, 2003.

\bibitem{Dohnal:2007}
T.~Dohnal and T.~Hagstrom.
\newblock Perfectly matched layers in photonics computations: 1{D} and 2{D}
  nonlinear coupled mode equations.
\newblock {\em J. Comput. Phys.}, 223:690--710, 2007.

\bibitem{Arnold:1997}
V.~I. Arnol'd, V.~V. Kozlov, and A.~I. Neishtadt.
\newblock {\em Mathmatical Aspects of Classical and Celestial Mechanics}.
\newblock Springer, 2nd edition, 1997.

\bibitem{Wiggins:2003}
S.~Wiggins.
\newblock {\em Introduction to applied nonlinear dynamical systems and chaos}.
\newblock Texts in Applied Mathematics. Springer-Verlag, New York, 2nd edition,
  2003.

\bibitem{Meer:1985}
J.-C. van~der Meer.
\newblock {\em The {H}amiltonian {H}opf bifurcation}, volume 1160 of {\em
  Lecture Notes in Mathematics}.
\newblock Springer-Verlag, Berlin, 1985.

\bibitem{Chow:1988}
S.-N. Chow and Y.-I. Kim.
\newblock Bifurcation of periodic orbits for non-positive definite
  {H}amiltonian systems.
\newblock {\em Applicable. Anal.}, 31:163--199, 1988.

\bibitem{GH:83}
J.~Guckenheimer and P.~Holmes.
\newblock {\em Nonlinear oscillations, dynamical systems, and bifurcations of
  vector fields}.
\newblock Springer-Verlag, New York, 1983.

\bibitem{Dullin:2005}
H.R. Dullin and A.V. Ivanov.
\newblock Vanishing twist in the {H}amiltonian {H}opf bifurcation.
\newblock {\em Phys. D}, 201:27--44, 2005.

\bibitem{Gaivao:2010}
J.~P. Gaiv{\~a}o and V.~Gelfreich.
\newblock Splitting of separatrices for the {H}amiltonian-{H}opf bifurcation
  with the {S}wift-{H}ohenberg equation as an example.
\newblock {\em arXiv:1004.2054v1}, 2010.

\bibitem{Lahiri:2001}
A.~Lahiri and M.S. Roy.
\newblock The {H}amiltonian {H}opf bifurcation: an elementary perturbative
  approach.
\newblock {\em Internat. J. Non-Linear Mech.}, 36:787--802, 2001.

\end{thebibliography}

\end{document}